\documentclass[floats,floatfix,showpacs,amssymb,prd,twocolumn,superscriptaddress,nofootinbib,nolongbibliography,reprint]{revtex4-1}

\usepackage{amssymb,amsmath,verbatim,mathtools,needspace,enumitem,etoolbox,graphicx,physics,microtype,afterpage,dcolumn,diagbox}

\usepackage[dvipsnames, usenames]{xcolor}

\definecolor{linkcolor}{rgb}{0.0,0.3,0.5}
\usepackage[unicode, colorlinks=true, linkcolor=linkcolor, citecolor=linkcolor, filecolor=linkcolor,urlcolor=linkcolor, pdfusetitle]{hyperref}
\usepackage[all]{hypcap}
\usepackage[T1]{fontenc}
\usepackage[utf8]{inputenc}
\usepackage{tabularx}
\allowdisplaybreaks

\usepackage{lipsum}

\usepackage{diagbox}
\newcommand{\rmi}{\mathrm{i}}
\newcommand{\e}{\mathrm{e}}

\newcounter{count}

\begin{document}

\title{Core collapse in massive scalar-tensor gravity}

\author{Roxana Rosca-Mead}
\email{rr417@cam.ac.uk}
\affiliation{DAMTP, Centre for Mathematical Sciences, University of Cambridge, Wilberforce Road, Cambridge CB3 0WA, United Kingdom}

\author{Ulrich Sperhake}
\email{u.sperhake@damtp.cam.ac.uk}
\affiliation{DAMTP, Centre for Mathematical Sciences, University of Cambridge, Wilberforce Road, Cambridge CB3 0WA, United Kingdom}
\affiliation{TAPIR 350-17, California Institute of Technology, 1200 E. California Blvd., Pasadena, California 91125, USA}

\author{Christopher J. Moore}
\email{cmoore@star.sr.bham.ac.uk}
\affiliation{School of Physics and Astronomy \& Institute for Gravitational Wave Astronomy, \\ University of Birmingham, Birmingham B15 2TT, UK}

\author{Michalis Agathos}
\email{magathos@damtp.cam.ac.uk}
\affiliation{DAMTP, Centre for Mathematical Sciences, University of Cambridge, Wilberforce Road, Cambridge CB3 0WA, United Kingdom}
\affiliation{Kavli Institute for Cosmology Cambridge, Madingley Road CB3 0HA, Cambridge, UK}

\author{Davide Gerosa}
\email{d.gerosa@bham.ac.uk}
\affiliation{School of Physics and Astronomy \& Institute for Gravitational Wave Astronomy, \\ University of Birmingham, Birmingham B15 2TT, UK}

\author{Christian D. Ott }
\email{christian.d.ott@gmail.com}
\affiliation{OCS Labs LLC, Pasadena, California 91104, USA}

\date{\today}

\begin{abstract}

This paper provides an extended exploration of the inverse-chirp
gravitational-wave signals from stellar collapse in massive
scalar-tensor gravity reported in [Phys.~Rev.~Lett.~{\bf 119},
201103].  We systematically explore the parameter space that
characterizes the progenitor stars, the equation of state, and the
scalar-tensor theory of the core collapse events. We identify a
remarkably simple and straightforward classification scheme of the
resulting collapse events. For any given set of parameters, the
collapse leads to one of three end states: a weakly scalarized
neutron star, a strongly scalarized neutron star, or a black hole,
possibly formed in multiple stages.  The latter two end states can lead
to strong gravitational-wave signals that may be detectable in
present continuous-wave searches with ground-based detectors. We
identify a very sharp boundary in the parameter space that separates
events with strong gravitational-wave emission from those with
negligible radiation.
\end{abstract}

\maketitle

%=============================================================================
\section{Introduction}

Black holes (BHs) and neutron stars (NSs) populate the graveyard of massive stars. As the star's iron core exceeds its effective Chandrasekhar mass, gravitational instability causes collapse to a NS. Collapse is initially halted by the repulsive character of nuclear interactions, causing the inner core to bounce. This bounce may liberate a hydrodynamical shock that will propagate through the star's envelope and eventually result in a supernova. For some progenitors, further accretion from the star's outer layers, can then turn the NS into a BH.

The formation of BHs and NSs via stellar collapse naturally involves strong, dynamical gravitational fields, thus constituting a precious tool to investigate the nature of gravity~\cite{Berti:2015itd}. In particular, core collapse is ideal to constrain those generalizations of Einstein's general relativity (GR) where compact objects present a substantially different structure. Examples of these are spontaneously scalarized NSs \cite{Damour:1993hw,Ramazanoglu:2016kul,Doneva:2018ouu,Andreou:2019ikc} and BHs  \cite{Silva:2017uqg,Doneva:2017bvd} in some classes of scalar-tensor (ST) theories, universal horizons in theories with Lorentz violation \cite{Barausse:2011pu}, or the spontaneous growth
of vector or tensor fields around compact objects in modified gravity
\cite{Ramazanoglu:2017xbl,Ramazanoglu:2019gbz,Annulli:2019fzq}.

Probing the dynamics and gravitational-wave (GW) emission of
compact objects undergoing
such dynamic processes
requires a well-posed formulation of the underlying
theory that allows for implementation in numerical
evolution codes.
The demonstration of the well-posedness of GR by Choquet-Bruhat
\cite{FouresBruhat:1952zz,ChoquetBruhat:1969cb}
represents a milestone in the mathematical understanding of Einstein's
theory and the corresponding problem
is now being tackled for some of the most popular alternative theories of gravity~\cite{Salgado:2008xh,Delsate:2014hba,Papallo:2017qvl,Papallo:2017ddx,Papallo:2019lrl}.

ST theories, where gravity is mediated by the usual graviton and  an additional scalar field, are arguably the simplest and most intensively studied generalization of GR. Extending early seminal work by \citeauthor{Brans:1961sx} \cite{Brans:1961sx}, the theory's most general formulation was first written down by \citeauthor{Horndeski:1974wa} \cite{Horndeski:1974wa}.
These theories have been strongly tested in the weak-field regime by the Cassini mission \cite{Bertotti:2003rm}, Lunar Laser Ranging \cite{Williams:2005rv}, and binary pulsars \cite{Wex:2014nva}. ST theories of gravity are now being severely constrained by GW observations~\cite{Yunes:2016jcc,LIGOScientific:2019fpa}. In particular, the multimessenger observation of GW170817 \cite{GBM:2017lvd} has ruled out all variants of Horndeski theory where the speed of photons and gravitons differs by more than $\sim 5\times10^{-16}$~\cite{Ezquiaga:2017ekz,Sakstein:2017xjx,Creminelli:2017sry}. For some Horndeski theories, gravity has a dispersion relation (i.e.\ waves with different frequencies travel at different speeds) which provides a further handle to constrain the nature of gravity with GW signals.

In this paper, we study BH and NS formation in a particular subclass of massive scalar-tensor (MST) gravity and explore its consequences for current and future GW observations.
We note in this context that the above mentioned constraints on the
propagation of GWs apply to the spin-two modes but do not, as yet,
constrain the propagation speed and, hence, the mass of scalar
degrees of freedom.
In particular, the ST formulation by Refs.~\cite{Bergmann:1968ve,Wagoner:1970vr,Damour:1993hw} with the addition of a mass term (e.g.~\cite{Ramazanoglu:2016kul}) constitute an ideal playground for probing additional physics with stellar collapse~\cite{Novak:1998rk,Novak:1997hw,Novak:1999jg,Gerosa:2016fri,Mendes:2016fby,Sperhake:2017itk,Cheong:2018gzn,Rosca-Mead:2019seq,Geng:2020slq}. This class of ST theories presents three crucial features:
\begin{enumerate}
\item The Einstein frame reduction (see, e.g.,~\cite{Salgado:2005hx})
immediately proves that the theory is well-posed and thus suitable to be tackled by numerical integration.
\item A new family of stationary NS solutions is present, which are macroscopically different from their GR counterparts \cite{Damour:1993hw}.
\item The presence of a nonzero scalar-field mass introduces a dispersion relation, with a consequent new phenomenology for the emitted GW signal.
\end{enumerate}
With these ingredients in the blender, our previous contribution \cite{Gerosa:2016fri,Sperhake:2017itk,Rosca-Mead:2019seq} has presented a limited suite of simulations of NS and BH
formation from realistic presupernova stellar density profiles and highlighted the presence of characteristic ``inverse GW chirp'' signals. Encoded in the oscillation of the scalar field, high-frequency GW signals reach the detector sooner compared to low-frequency modes. Signals might still be present for decades, or even centuries, after the core collapse event, thus providing us with the tantalizing possibility of testing massive ST theories with GW observations of historic supernovae.

In this paper, we extend our previous work by presenting a systematic exploration of the phenomenology of core collapse in massive ST gravity. In Sec.~\ref{sec:formalism_and_equations}, we review the complete formalism used in this study, including equations of motions in flux-conservative form and details on the equation of state. In Sec.~\ref{sec:framework}, we summarize our numerical implementation, including initial data and the evolution scheme.  Section~\ref{sec:wavesignals} presents a complete taxonomy of the collapse process and its end points.

A surprisingly simple picture emerges: despite the large dimensionality of the problem, the collapse dynamics can always be classified as one of only five possible scenarios. These are the following: (i) single-stage collapse to GR-like NSs, (ii) collapse to a BH following one accretion episode, (iii) collapse to a BH following multiple accretion and proto-NS stages, (iv) collapse to a strongly scalarized NS via accretion onto a GR-like proto-NS, and (v) direct collapse to a strongly scalarized NS.

We then proceed by analyzing the GW consequences of our findings. Section~\ref{propagation} provides a careful derivation and analysis of the inverse-chirp signal morphology. In particular, we argue that the features depend only on the mass of the scalar field and not the details of the source dynamics.
Moreover, the main characteristics of the GW signal,
its frequency and amplitude as functions of time,
depend (to good accuracy) on the scalar mass only through
a remarkably simple rescaling. In Sec.~\ref{sec:LIGO_obs} we present the relevance of our simulations to current and future GW searches. Finally,  in Sec.~\ref{concl} we draw our conclusions. To streamline the flow of the paper, several details are postponed to the appendices. In particular, Appendix~ \ref{app:scenarios} provides a more detailed description of each collapse scenario through the analysis of a representative example. Appendix~\ref{appeos} illustrates more results on the impact of the equation of state and progenitor model on the degree of scalarization. The accuracy of the stationary-phase approximation in describing the propagation of massive scalar waves is verified through a numerical test in Appendix~\ref{app:comparing_the_methods}, and Appendix~\ref{app:App_SNR_estimates_LIGO} provides more results on the LIGO detectability of the inverse-chirp signal.

Overall, this paper contains the results of
$\mathcal{O}(4\times 10^3)$
one-dimensional (1D) core-collapse simulations for a total computational time of
$\mathcal{O}(2\times 10^6)$
CPU hours. Throughout this paper we use geometric units $c=G=1$.

%=============================================================================
\section{Scalar-tensor theory}
\label{sec:formalism_and_equations}

In this work we consider the class of scalar-tensor theories of
gravity first studied by \citeauthor{Bergmann:1968ve}~\cite{Bergmann:1968ve} and \citeauthor{Wagoner:1970vr}~\cite{Wagoner:1970vr}, which satisfy the following assumptions:

\begin{enumerate}

\item
The equations of motion are derived from the variation of an action
$S=S_{\rm G}+S_{\rm M}$ where $S_{\rm G}$ consists exclusively of
the gravitational fields and $S_{\rm M}$ represents the interaction
of gravity with all matter fields.

\item All long-range forces
are mediated by the three lowest-spin bosons. Electromagnetism is
the only spin one interaction and the spin zero contribution is
described by a single real scalar field.

\item Variation
of the action results in at most two-derivative field equations,
i.e.~terms linear in second derivatives or quadratic in first
derivatives or of lower order.

\item The theory is diffeomorphism invariant, i.e.~formulated
in terms of tensorial equations.
\item The weak equivalence principle is satisfied.
\end{enumerate}

Using the above principles, we can formulate the action in the Jordan-Fierz frame~\cite{Berti:2015itd}:

\begin{eqnarray}\label{eq:Jordan_action}
  S &=& \int \mathrm{d}x^4 \sqrt{-g} \left[ \frac{F(\phi)}{16\pi} R
  - \frac{1}{2}g^{\mu \nu} (\partial_{\mu} \phi)(\partial_{\nu}\phi)
  - W(\phi) \right] \nonumber \\[5pt]
 &&+ S_M\left[\psi_m,g_{\mu \nu}\right]\, ,
\end{eqnarray}
where $g_{\mu\nu}$ represents the metric (from now on referred to as the Jordan metric), $g$ is its determinant, $R$ is the Ricci scalar corresponding to $g_{\mu\nu}$, $\phi$ represents the scalar field, $F$ and $W$ are functions of $\phi$, and $S_M$ represents the action of the matter fields $\psi_m$. A particularly convenient
(and in some instances preferable \cite{Faraoni:1999hp}) formulation
of this class of theories is
obtained in the so-called {\em Einstein frame}. This is achieved through a conformal transformation
\begin{equation} \label{eq:conformal_metric_Ffactor}
  \bar{g}_{\mu\nu}   \equiv F(\phi) g_{\mu\nu}\,,
\end{equation}
and a redefinition of the scalar field according to
\begin{equation}\label{eq: varphiofphi}
  \frac{\partial \varphi}{\partial \phi} = \sqrt{\frac{3}{4} \frac{F_{,\phi}{}^2}{F^2} + \frac{4\pi}{F}} \,;
\end{equation}
for an exploration of the regime of viability of this
transformation see \cite{Geng:2020ftu}.
The action of Bergmann-Wagoner scalar tensor theory is then
given by \cite{Fujii:2003pa,Berti:2015itd}
\begin{eqnarray}
  S&=&\int \mathrm{d}^{4}x\; \frac{\sqrt{-\bar{g}}}{16\pi}
    \left[ \bar{R}-2\bar{g}^{\mu\nu}\partial_{\mu}\varphi\,
    \partial_{\nu}\varphi - 4V(\varphi)\right] \nonumber \\[5pt]
    && + S_M\left[\psi_m,\frac{\bar{g}_{\mu \nu}}{F}\right]\,,
  \label{eq:actionE}
\end{eqnarray}
where $V(\varphi)$ is the scalar potential and $\bar{R}$ and
$\bar{g}$, respectively, denote the Ricci scalar and determinant
constructed from the conformal metric. Note that we recover Brans-Dicke
theory \cite{Brans:1961sx} with the choice
$F=\exp({-2\varphi/\sqrt{3+2\omega_{\rm BD}}})$ while general relativity
corresponds to the trivial case $\varphi = \mathrm{const}$.

In this work we choose the matter part of the action $S_M$
such that the physical energy momentum tensor describes a perfect
fluid with baryon density $\rho$, pressure $P$, internal energy
$\epsilon$, enthalpy $H$ and 4-velocity $u^{\alpha}$,
\begin{equation}
  T^{\mu\nu} \equiv \frac{2}{\sqrt{-g}}
    \frac{\delta S_M}{\delta g_{\mu\nu}}
    = \rho H u^{\mu}u^{\nu}+Pg^{\mu\nu}\,.
  \label{eq:EMtensor}
\end{equation}
The equations of motion are obtained through variation of the action
(\ref{eq:actionE}) with respect to the metric, the scalar and the
matter fields, as well as the continuity equation for
baryon conservation in the physical frame,
\begin{align}
  &\bar{G}_{\alpha\beta} = 2\partial_{\alpha}\varphi
    \partial_{\beta} \varphi-\bar{g}_{\alpha\beta}
    \partial^{\mu}\varphi\,
    \partial_{\mu} \varphi + 8\pi \bar{T}_{\alpha\beta}
    -2V \bar{g}_{\mu\nu}\,,
  \label{eq:Einstein}\\
  &\bar{\nabla}^{\mu} \bar{\nabla}_{\mu} \varphi
    = 2\pi \frac{F_{,\varphi}}{F}\bar{T}+V_{,\varphi}\,,
  \label{eq:scalareq}\\
  &\bar{\nabla}_{\mu} \bar{T}^{\mu\alpha} = -\frac{1}{2}
    \frac{F_{,\varphi}}{F}\,\bar{T}\,\bar{g}^{\alpha\mu}
    \bar{\nabla}_{\mu} \varphi\,,
    \\
    &\nabla_{\mu}(\rho u^{\mu}) = 0\,.
  \label{eq:baryoncon}
\end{align}
Here $\bar{T}_{\alpha\beta}=T_{\alpha \beta}/F$ is the conformal
energy momentum tensor, $\bar{\nabla}$ and $\nabla$ are the covariant
derivatives associated with $\bar{g}_{\mu\nu}$ and $g_{\mu\nu}$,
respectively, and the subscript $,\varphi$ denotes differentiation
with respect to $\varphi$.

The specific scalar-tensor theory of gravity is determined by the choice
of the potential function $V(\varphi)$ and the conformal factor
$F(\varphi)$. Here we consider a noninteracting scalar field with mass
parameter $\mu$, so that the potential is given by
\begin{equation}
  V(\varphi) = \frac{\mu^2\varphi^2}{2\hbar^2}\,.
\end{equation}
The scalar mass introduces a characteristic frequency
\begin{equation}
  \omega_{*} = 2\pi f_{*} = \frac{\mu}{\hbar}\,.
\end{equation}
Finally, we write the conformal factor as
\begin{equation}
  F({\varphi}) = e^{-2\alpha_0 \varphi-\beta_0 \varphi^2}\,,
  \label{eq:F}
\end{equation}
where $\alpha_0$ and $\beta_0$ are dimensionless parameters. This
choice for the conformal factor (sometimes also written as
$A\equiv F^{-1/2}$; cf.~\cite{Damour:1992we})
is very common in the literature
and motivated by the fact that in this form $\alpha_0$ and $\beta_0$
completely determine all modifications of gravity at first
post-Newtonian order \cite{Damour:1992we,Damour:1996ke,Chiba:1997ms}.

Henceforth, we consider spherical symmetry and impose polar slicing
and radial gauge \cite{Bardeen:1983}
in the Einstein frame, so that the line element
takes on the form
\begin{equation}
  \mathrm{d}\bar{s}^2 = \bar{g}_{\mu\nu}\mathrm{d}x^{\mu}\mathrm{d}x^{\nu} =
    -F\alpha^2 \mathrm{d}t^2+FX^2 \mathrm{d}r^2 + r^2\mathrm{d}\Omega^2\,,
  \label{eq:lineelement}
\end{equation}
where $\alpha$ and $X$ are functions of $(t,r)$. Following common
practice, we introduce for convenience the potential $\Phi(t,r)$
and the mass function $m(t,r)$ through
\begin{equation}
  F\alpha^2 = e^{2\Phi}\,,~~~~~FX^2 = \left(1-\frac{2m}{r}
    \right)^{-1}\,. \label{eq:Pot_mass}
\end{equation}
\medskip
The four velocity in spherical symmetry is
\begin{equation}
  u^{\mu} = \frac{1}{\sqrt{1-v^2}} \left[
    \frac{1}{\alpha},\,\frac{v}{X},\,0,\,0\right]\,,
    \label{eq:4velocity}
\end{equation}
where the velocity field $v$ as well as the matter variables $\rho$,
$P$, $H$, $\epsilon$ of Eq.~(\ref{eq:EMtensor}) are functions of
$(t,r)$. By inserting the expressions of Eqs.~(\ref{eq:EMtensor}) and
(\ref{eq:lineelement})-(\ref{eq:4velocity}) into the field equations
(\ref{eq:Einstein})-(\ref{eq:baryoncon}), we obtain the set of
equations that govern the dynamics of spherically symmetric fluid
configurations in Bergmann-Wagoner ST theory of gravity. In order
to accurately model discontinuities arising through shock formation
in the fluid profiles, however, we require high resolution shock
capturing and, hence, a flux conservative form of the matter
equations. This is achieved by converting the {\em primitive}
variables $(\rho,\,v,\,H)$ to their flux conservative counterparts
\cite{O'Connor:2009vw,Gerosa:2016fri},
\begin{equation}
  D=\frac{\rho X F^{-3/2}}{\sqrt{1-v^2}}\,,~S^r
    = \frac{\rho Hv F^{-2}}{1-v^2}\,,~
    \tau = \frac{S^r}{v}-\frac{P}{F^2}-D\,.
\end{equation}
Finally, we convert the wave equation (\ref{eq:scalareq}) for the
scalar field into a first order system by defining
\begin{equation}
  \eta = \frac{1}{X}\partial_r \varphi\,,~~~~~\psi
    = \frac{1}{\alpha}\partial_t \varphi\,.
\end{equation}
The final set of equations can then be written in the form
\begin{align}
 &\partial_r \Phi = X^2 F\left[\frac{m}{r^2}+4\pi r
    \left(S^r v+\frac{P}{F^2}\right)+\frac{r}{2F}(\eta^2+\psi^2)
    \right]
    \notag\\&\qquad
    -rFX^2V\,,
  \label{eq:Phir}\\
  &\partial_r m = 4\pi r^2 (\tau+D)+\frac{r^2}{2F}(\eta^2+\psi^2)
    +r^2V\,,
  \label{eq:mr}\\
&  \partial_t \varphi = \alpha \psi\,,
  \label{eq:varphit}\\
  &\partial_t \eta = \frac{1}{X}\partial_r(\alpha \psi)
    -rX\alpha \eta\,(\eta\psi-4\pi F\,S^r)
    +\frac{F_{,\varphi}}{2F}\alpha\eta\psi\,,
  \label{eq:etat}\\
 & \partial_t \psi = \frac{1}{r^2X}\partial_r(r^2\alpha\eta)
    -rX\alpha\psi(\eta\psi-4\pi F\,S^r)
    +\frac{F_{,\varphi}}{2F}\alpha\psi^2
      \notag\\&\qquad
   +2\pi \alpha\left(\tau-S^rv +D-3\frac{P}{F^2}\right)
    F_{,\varphi}-\alpha FV_{,\varphi}\,,
  \label{eq:psit}\\
 & \partial_t \begin{pmatrix} D \\[5pt]S^r \\[5pt] \tau \end{pmatrix}
    = \frac{1}{r^2}\partial_r \left[ r^2\frac{\alpha}{X}
    \begin{pmatrix} \boldsymbol{f}_D \\[5pt]
                    \boldsymbol{f}_{S^r} \\[5pt]
                    \boldsymbol{f}_{\tau} \end{pmatrix}
    \right] =
    \begin{pmatrix} \boldsymbol{s}_D \\[5pt]
                    \boldsymbol{s}_{S^r}\\[5pt]
                    \boldsymbol{s}_{\tau}\end{pmatrix}\,,
  \label{eq:mattert}
\end{align}
with fluxes and sources given by
\begin{align}
 & \boldsymbol{f}_D = Dv\,,
  \\
 &   \boldsymbol{f}_{S^r} = S^r v+ \frac{P}{F^2}\,,
    \\
&    \boldsymbol{f}_{\tau} = S^r-Dv\,,
  \label{eq:f}\\
&  \boldsymbol{s}_D = -D\frac{F_{,\varphi}}{2F}\alpha
    (\psi+v\eta)\,,
  \label{eq:sD}\\
 & \boldsymbol{s}_{S^r} = (S^r v-\tau-D)\alpha X F
    \bigg(8\pi r\frac{P}{F^2}+\frac{m}{r^2}
    -\frac{F_{,\varphi}}{2F^2X}\eta
             \notag\\&\qquad
    -rV\bigg)
    + \frac{\alpha X}{F}P\frac{m}{r^2}
    + 2\frac{\alpha P}{rXF^2}-r\alpha X \frac{P}{F}V
         \notag\\&\qquad
- 2r\alpha XS^r\eta\psi-\frac{3}{2}\alpha\frac{P}{F^2}
    \frac{F_{,\varphi}}{F}\eta
               \notag\\&\qquad
    -\frac{r}{2}\alpha X(\eta^2+\psi^2)
    \left(\tau+\frac{P}{F^2}+D\right)(1+v^2)\,,
  \label{eq:sSr}\\
 & \boldsymbol{s}_{\tau} = -\left(\tau+\frac{P}{F^2}+D\right)
    r\alpha X \,[(1+v^2)\eta\psi+v(\eta^2+\psi^2)]\,
            \notag\\&\qquad
 +\frac{\alpha}{2}\frac{F_{,\varphi}}{F}
    \left[ Dv\eta+\left(S^r v-\tau+3\frac{P}{F^2}\right)\psi
    \right]\,.
  \label{eq:stau}
\end{align}
Note that these equations differ from Eqs.~(2.21), (2.22), (2.26)-(2.28),
and (2.33)-(2.39) in Ref.~\cite{Gerosa:2016fri} through the
presence of the potential terms involving $V$ in our Eqs.~(\ref{eq:Phir}),
(\ref{eq:mr}), (\ref{eq:psit}), and (\ref{eq:sSr}). In particular,
the principal part and the characteristic structure of the equations
are identical to those in the case of a massless scalar field,
and we consequently inherit the well-posed character of the
evolution equations of the massless case.

In order to close the system of differential equations
(\ref{eq:Phir})-(\ref{eq:stau}), we need to prescribe an equation
of state (EOS) that provides the pressure as a function of $\rho$
and $\epsilon$. Here we use a so-called hybrid EOS introduced in
Ref.~\cite{Janka:1993} that captures in closed analytic form the
stiffening of the matter at nuclear densities and models the response
of shocked material through a thermal pressure component; see also Refs.~\cite{Zwerger:1997sq,Dimmelmeier:2002bm,Dimmelmeier:2007ui,Dimmelmeier:2008iq} for comparisons with
modern finite-temperature EOSs. The hybrid EOS consists of a cold and a thermal
pressure component given by
\begin{equation}
  P = P_{\rm c} + P_{\rm th}\,.
  \label{eq:P}
\end{equation}
The cold component has piecewise polytropic form
\begin{equation}
  P_{\rm c} = \left\{ \begin{array}{l} K_1 \rho^{\Gamma_1}~~~~~
    \text{if}~~~~~\rho \leq \rho_{\rm nuc} \\[5pt]
    K_2 \rho^{\Gamma_2}~~~~~\text{if}~~~~~\rho >\rho_{\rm nuc}
    \end{array} \right.\,,
  \label{eq:Pc}
\end{equation}
and the thermal contribution is given by
\begin{equation}
  P_{\rm th}= (\Gamma_{\rm th}-1)\,\rho\,(\epsilon-
    \epsilon_{\rm_{\rm c}})\,,
  \label{eq:Pth}
\end{equation}
where $\epsilon$ is the internal energy and $\epsilon_{\rm c}$ follows from the first law of thermodynamics
for adiabatic processes,
\begin{equation}
  \epsilon_{\rm c}=\left\{ \begin{array}{ll}
    \frac{K_1}{\Gamma_1 -1}\rho^{\Gamma_1-1} & \text{if}~~~
    \rho \leq \rho_{\rm nuc}
  \\[5pt]
  \frac{K_2}{\Gamma_2-1} \rho^{\Gamma_2-1}+E~~~&\text{if}~~~
    \rho >\rho_{\rm nuc} \end{array} \right.\,.
  \label{eq:epsc}
\end{equation}
Prior to core bounce, the flow is adiabatic
which implies $\epsilon \approx \epsilon_{\rm c}$, but at core
bounce the shocked material becomes nonadiabatical and thus subject
to a non-negligible thermal pressure component.

  We set the nuclear density
$\rho_{\rm nuc}=2\times 10^{14}~\text{g}\,\text{cm}^{-3}$
\cite{Dimmelmeier:2002bm} and $K_1=4.9345 \times 10^{14} \,[\text{cgs}]$ 
as predicted for a relativistic degenerate gas of electrons with
electron fraction $Y_{\rm e}=0.5$ \cite{Shapiro:1983du}.  The constants $K_2$ and $E$
follow from continuity at $\rho=\rho_{\rm nuc}$.  The EOS given by Eqs.~(\ref{eq:P})-(\ref{eq:epsc}) is thus determined
by the three adiabatic indices $\Gamma_1$, $\Gamma_2$, and $\Gamma_{\rm
th}$.
A gas of
relativistic electrons has an adiabatic index of $4/3$, but electron
capture during the collapse phase reduces the effective adiabatic
index $\Gamma_1$ to slightly lower values in the range $\Gamma_1\approx
1.28$ to $\Gamma_1\approx 1.32$
\cite{Dimmelmeier:2007ui,Dimmelmeier:2008iq,Shen:2011qu}. At densities
$\rho > \rho_{\rm nuc}$, however,
the repulsive core of the nuclear force stiffens the EOS
which leads to a larger adiabatic index $\Gamma_2$. Reference
\cite{Dimmelmeier:2008iq} find $\Gamma_2\approx 2.5$ and $\Gamma_2\approx
3$ to approximate well the finite-temperature EOSs of Lattimer-Swesty
\cite{Lattimer:1985zf,Lattimer:1991} and Shen {\em et al}
\cite{Shen:1998by,Shen:1998gq}, respectively.  Finally, the thermal
adiabatic index $\Gamma_{\rm th}$ models a mixture of relativistic
and nonrelativistic gas which leads to the bounds $4/3 <\Gamma_{\rm
th}<5/3$.

\begin{table}[t]
  \centering
  \begin{tabular}{c | c c c c c}
        & EOS1  & EOS3& EOS5& EOS8 & EOSa\\
    \hline
    $\Gamma_1$       & 1.30 & 1.32 & 1.30 & 1.30 & 1.28\\
    $\Gamma_2$       & 2.50 & 2.50 & 3.00 & 2.50 & 3.00\\
    $\Gamma_{\rm th}$& 1.35 & 1.35 & 1.35 & 1.50 & 1.50\\
      \end{tabular}
    \caption{Parameters for different hybrid equations of state. 
    The noncontiguous EOS labels are due to
    the fact that we have also explored collapse configurations
    with EOSs using different
    combinations of the given parameter values. These simulations,
    without exception, fit into the classification scheme
    of Sec.~\ref{sec:wavesignals} and are therefore not reported here.
    }

  \label{tab:EOS}
\end{table}
Our hybrid EOS is therefore determined by three parameters.
Motivated by the above considerations,
we select values
$\Gamma_1 \in \{1.28,~1.3,~1.32\}$,
$\Gamma_2 \in \{2.5,~3\}$, and
$\Gamma_{\rm th}\in \{1.35,~1.5\}$
with $(\Gamma_1,~\Gamma_2,~\Gamma_{\rm th})=(1.3,~2.5,~1.35)$ as
our fiducial model. In particular, we pick five different combinations of the EOS parameters as listed in Table ~\ref{tab:EOS}.

\medskip

%=============================================================================
\section{Computational framework and initial data}
\label{sec:framework}
\begin{figure*}[t]
  \includegraphics[width=0.48\textwidth,clip=true]{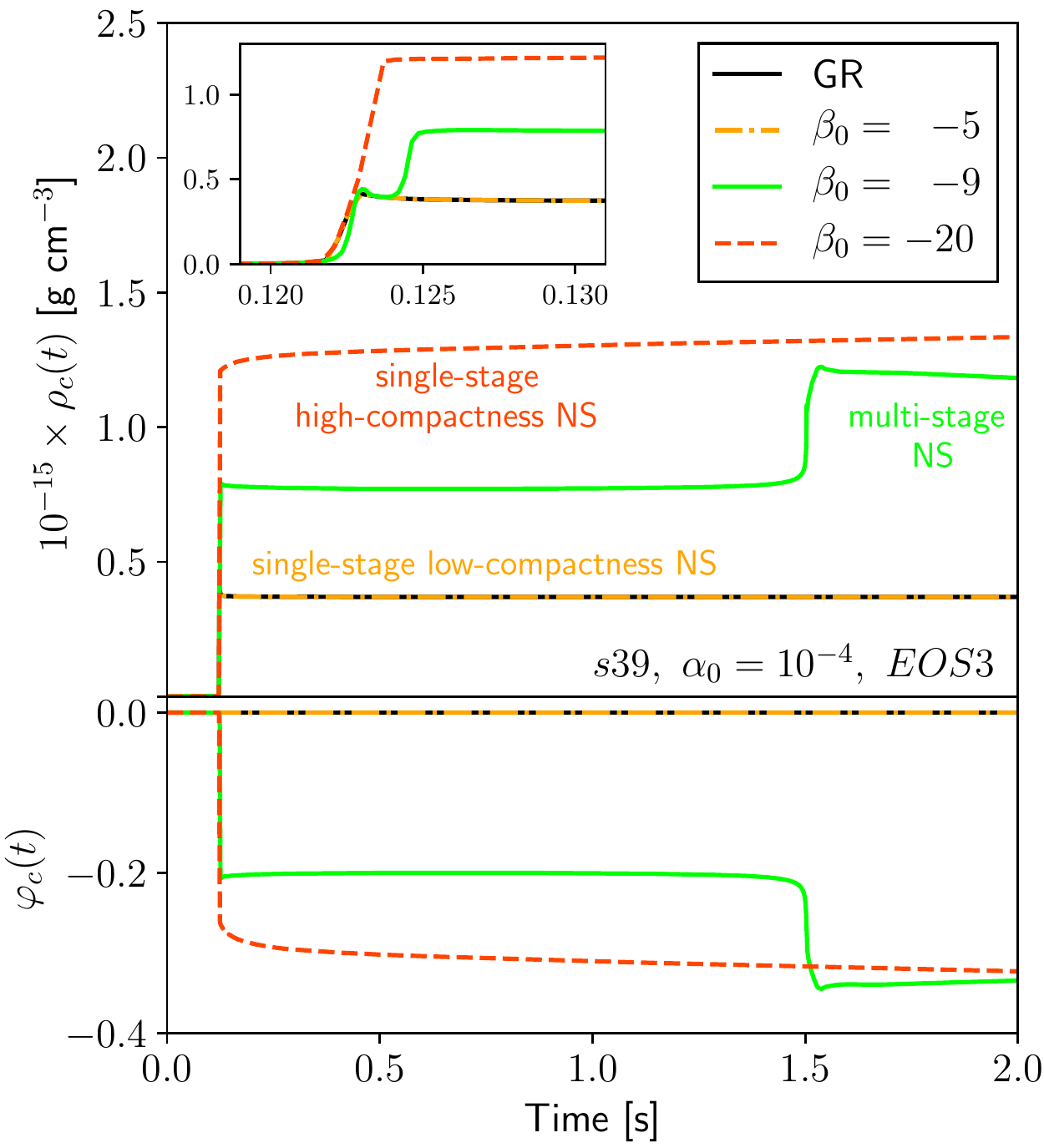}
  \includegraphics[width=0.48\textwidth,clip=true]{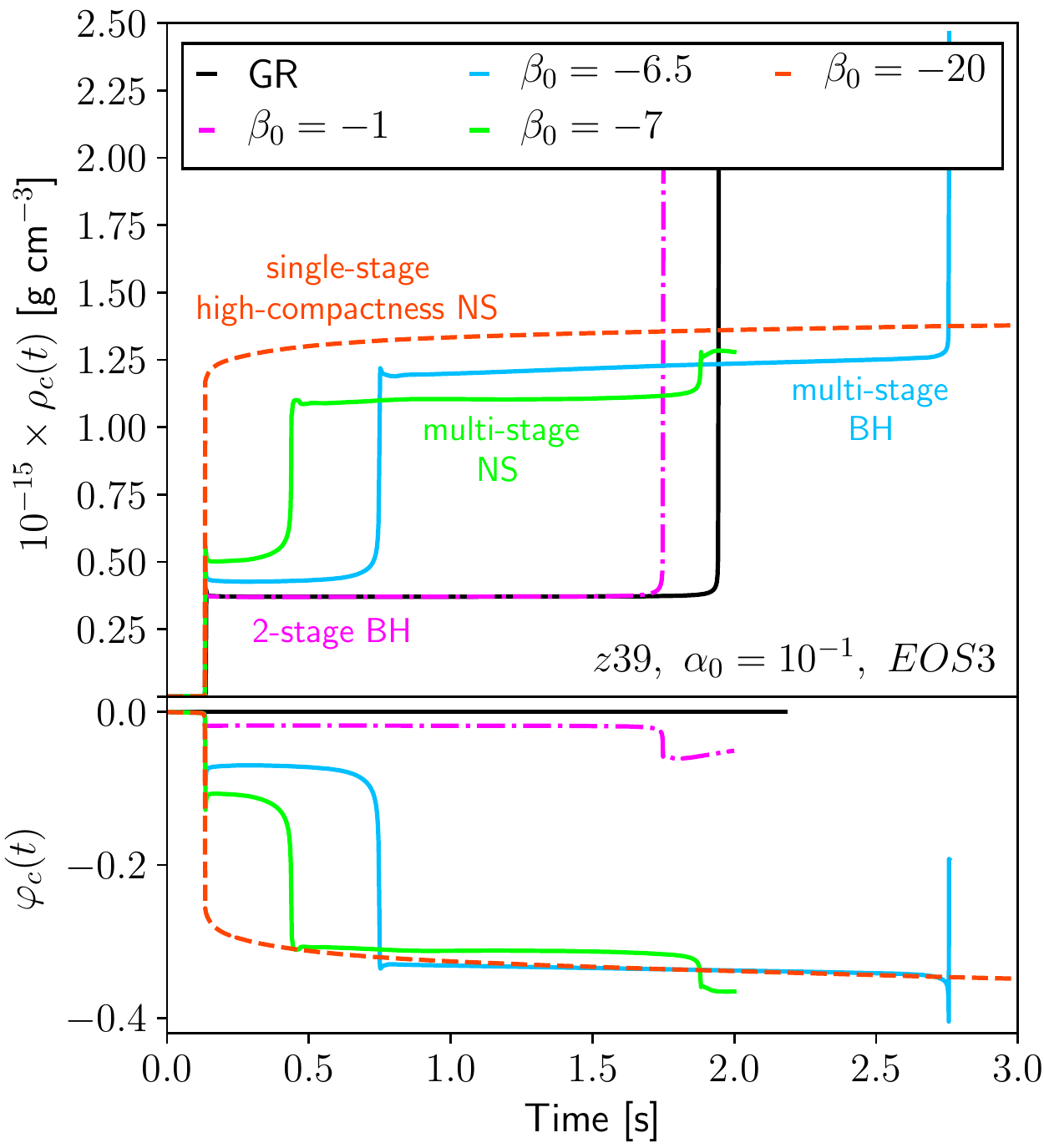}
  \caption{
           The central density (top panel) and maximal scalarization
           (bottom panel) are shown for representative
           examples of the five collapse configurations
           summarized in Sec.~\ref{sec:classification}.
           On the left, we show three evolutions of progenitor
           s39 with EOS3 for different ST parameters
           corresponding to the NS formation scenarios
           1 (single-stage low-compactness NS), 4 (multi-stage NS),
           and 5 (single-stage high-compactness NS), respectively.
           On the right, we show the evolution of progenitor
           z39 with EOS3 for different ST parameters corresponding
           to scenarios 2 (two-stage BH formation),
           3 (multi-stage BH formation), 4 (multi-stage NS),
           and 5 (single-stage high-compactness NS), respectively.
           For comparison, we display with solid black curves
           the corresponding evolution of the progenitors in GR
           which result in a NS in the left ``s39'' case
           and a BH in the right ``z39'' case. All curves have
           been obtained for a scalar mass $\mu=10^{-14}\,{\rm eV}$.
           }
  \label{fig:5examples}
\end{figure*}
We evolve the set of differential equations (\ref{eq:Phir})-(\ref{eq:mattert})
with an extended version of the open-source code {\sc gr1d}
\cite{O'Connor:2009vw} originally developed for modeling stellar collapse
in general relativity. {\sc gr1d} has been generalized to massless
scalar-tensor gravity in Ref.~\cite{Gerosa:2016fri}, and we have
merely added to this version of the code the potential terms
involving $V$ or $V_{,\varphi}$ in Eqs.~(\ref{eq:Phir})-(\ref{eq:stau}).
As mentioned above, these terms do not change the characteristics of
the differential equations and thus allow us to use the shock-capturing
scheme in the very same form as in \cite{Gerosa:2016fri}.

In order to capture the vastly different length scales encountered
in our simulations, we employ a computational grid consisting of
an inner grid with uniform resolution $\Delta r_1$ out to $r=40\,{\rm
km}$ and an outer component with logarithmic spacing up to $r=9\times
10^5\,{\rm km}$, resulting in a total of $N$ grid points. In
Ref.~\cite{Sperhake:2017itk}, some of the authors have analyzed the convergence of the
resulting core collapse simulations and found a discretization error
in the wave signal of about $4\,\%$ for a grid setup using
$\Delta r_1=250\,{\rm m}$ and $N=10\,000$. This is the minimum resolution
used for all the simulations of this work.
Finally, we have verified that the error due to extracting
the wave signal at a large but finite radius is negligible compared with the
discretization error, and we therefore estimate the total numerical
uncertainty as $\sim 4\,\%$. 

All simulations presented in this work start with the nonrotating
models of the catalog of spherically symmetric
presupernova stars provided by
\citeauthor{Woosley:2007as} \cite{Woosley:2007as}. These models
have been obtained by evolving stars in Newtonian gravity up to the
moment of iron core collapse and provide profiles for stars with
zero-age-main-sequence (ZAMS) masses from $10.8$ to $75$ solar masses
and three different metallicities: solar, $10^{-4}$ times solar,
and primordial metallicity.
Throughout this work, we denote the
progenitor models by a prefix ``$s$'', ``$u$'', or ``$z$'', respectively for the
three metallicities, followed by the ZAMS mass. With this notation,
for instance, ``u39'' denotes a progenitor with $10^{-4}$ times
solar metallicity and mass $M_{\rm ZAMS}=39\,M_{\odot}$. In the weak-gravity
regime of these low-density progenitor stars (their central density
is a factor about $10^5$ below nuclear density), the scalar field
is negligible, and we therefore set $\varphi=0$ initially. The initial
metric variables can then be computed directly from the matter
profile using quadrature in Eqs.~(\ref{eq:Phir}) and (\ref{eq:mr}).

\medskip

%=============================================================================
\section{Phenomenology of stellar collapse}
\label{sec:wavesignals}
%

%=============================================================================
\begin{figure*}[t!]
\includegraphics[width=0.8\textwidth]{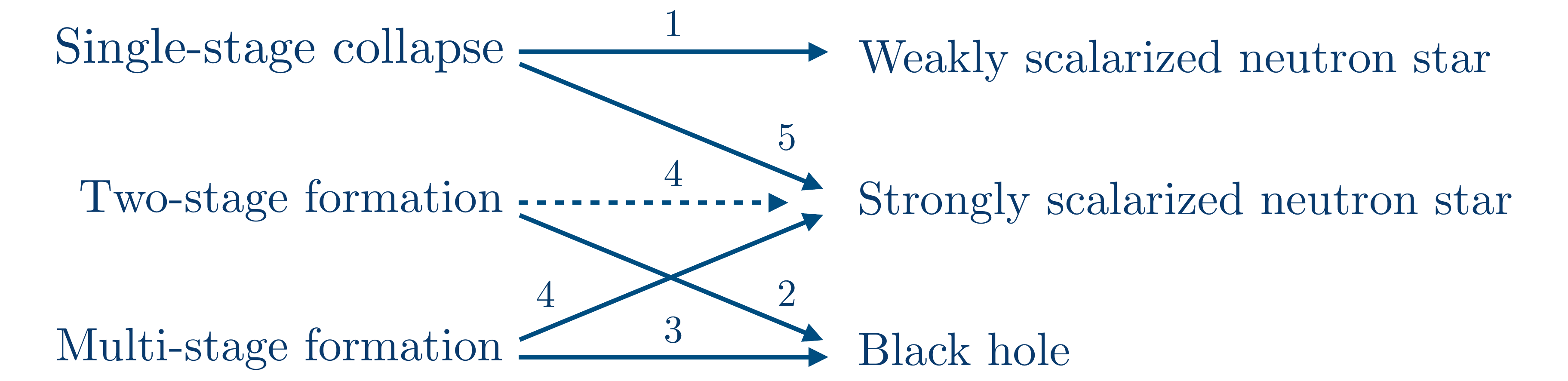}
\caption{
A graphical illustration of the main collapse scenarios identified in
our simulations. All stellar progenitors collapse into a weakly
scalarized (GR-like) NS, a strongly scalarized NS, or a BH.
As indicated in the diagram the outcomes may be reached promptly
or in two or more stages.
The two-stage formation of a strongly scalarized NS has been
marked by a dashed arrow because this case appears in our
set of simulations only a handful of times, and we suspect numerical noise to have impeded the occurrence of new stages. In the multi-stage
NS category we count the simulations where all
stages remain distinct, even if they happen on a short timescale.
}
\label{fig:arrowdiagram}
\end{figure*}

%===============================================================
\subsection{Classification}
\label{sec:classification}
\begin{figure*}[t]
  \includegraphics[width=0.49\textwidth,clip=true]{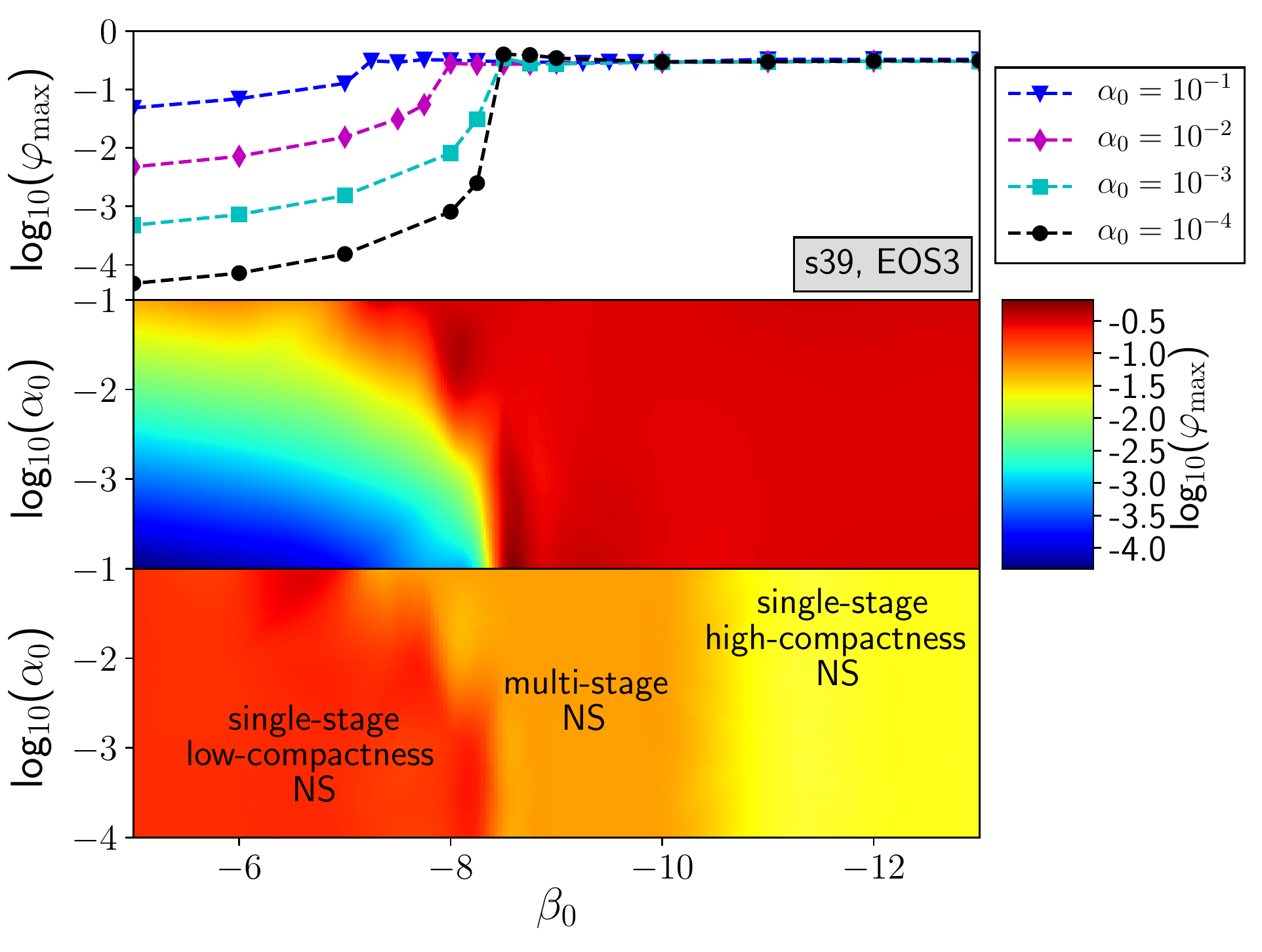}
  \includegraphics[width=0.49\textwidth,clip=true]{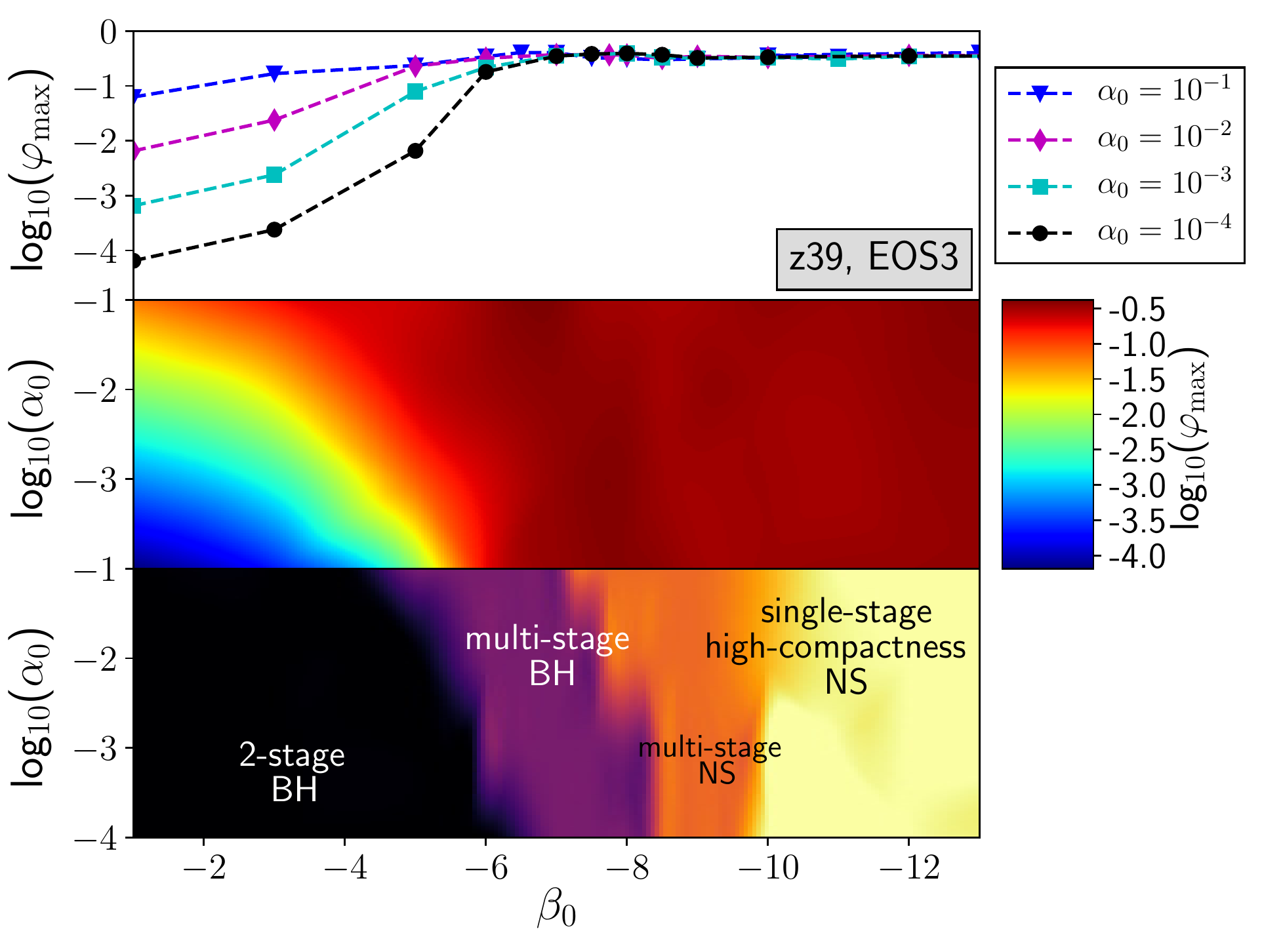}
  \caption{We consider a fixed progenitor star with ZAMS mass $39\,M_{\odot}$,
           equation of state EOS3 of Table \ref{tab:EOS}, and fix
           the scalar mass at $\mu=10^{-14}\,{\rm eV}$. The progenitor
           `s39' in the left panel has solar metallicity $Z_{\odot}$,
           and the progenitor `z39' in the right panel has primordial
           metallicity.
           Top row: For selected values of $\alpha_0$, we plot the
           maximal scalarization of the collapsing star as a function of $\beta_0$.
           The middle row provides a color (or ``heat'') map of the same
           quantity in the $(\alpha_0,\beta_0)$ plane: ``Red'' = strong
           scalarization, ``Blue'' = weak scalarization. The bottom
           row presents a color code of the five qualitatively different
           collapse scenarios listed in Sec.~\ref{sec:classification}.
           Note that the `s39' progenitor exclusively collapses to
           a neutron star, whereas 'u39' collapses to a black hole
           for $\beta_0\gtrsim -8$ and to a neutron star for
           $\beta_0\lesssim -8$. We find that every progenitor model
           results in heat maps in the $(\alpha_0,\beta_0)$ plane qualitatively
           equal to that on the left (the ``neutron star'' case) or that on the
           right (the ``black hole'' case).
          } %\dg{larger fontsize}
  \label{fig:heatmaps}
\end{figure*}

Stellar core collapse and supernova explosions are highly
complex processes, and the dynamics in numerical
simulations can depend sensitively on the level of detail
included in the modeling.
The focus of our study is an exploration of the parameter
space through a large number
[$\mathcal{O}(4\times 10^3)$] of long simulations
(several seconds). 
For computational feasibility,
we consider nonrotating
stars in spherical symmetry with piecewise polytropic
EOS and do not consider neutrino transport.
We characterize the
progenitor stars in terms of their ZAMS
mass and metallicity (the grid used in the progenitor
catalog of \cite{Woosley:2007as}), but note the
strong correlation of the outcome of a collapse
event with the compactness of the stellar core at bounce
\cite{OConnor:2010moj}.
While the qualitative picture from our simulations
is robust, some caution is advised on the quantitative details;
in particular the location of the boundaries between strongly and
scalarized configurations in Figs.~\ref{fig:heatmaps}
and \ref{fig:EOSzeta} may change under a refinement of
the modeling framework.

Within our framework,
a given stellar collapse model is characterized
by eight parameters:
\begin{itemize}
\item The EOS is characterized by two polytropic exponents
$\Gamma_1$, $\Gamma_2$, and the thermal pressure coefficient
$\Gamma_{\rm th}$.
\item The stellar progenitors are characterized
by metallicity $Z$
and zero-age-main-sequence mass $M_{\rm ZAMS}$.
\item The ST theory of gravity is determined by the mass of the scalar field $\mu$ and the coefficients $\alpha_0$ and $\beta_0$
entering the conformal factor.
\end{itemize}

Such a vast parameter space allows for an enormous phenomenology
and, through sheer numbers, represents a major challenge for a numerical
exploration; surmounting this challenge is the central goal of this
section.
More specifically, we will see that
within our modeling framework, the phenomenology of the
different collapse scenarios reveals distinct patterns and systematics
that enable us to provide a remarkably comprehensive description
of core collapse in massive ST gravity.

For this purpose, we first consider the possible end products of our
collapse simulations. There are only three qualitatively different
end states we have obtained in all of our simulations: (i) A weakly
scalarized neutron star where $\varphi= \mathcal{O}(\alpha_0)$,
(ii) a strongly scalarized neutron star with $\varphi=\mathcal{O}(1)$,
or (iii) a black hole.  The latter two end states, however, may be
reached either directly or through several stages.  This observation
leads to our main classification scheme of five qualitatively
different collapse scenarios.
\begin{list}{\rm{(\arabic{count})}}{\usecounter{count}
             \labelwidth1cm \leftmargin1.0cm \labelsep0.4cm \rightmargin0cm
             \parsep0.5ex plus0.2ex minus0.1ex \itemsep0ex plus0.2ex}
  \item Single-stage collapse to a weakly scalarized neutron star.
  \item Two-stage formation of a black hole. Here the configuration
        temporarily settles down into a weakly scalarized neutron star.
        As the continued accretion of matter exceeds a threshold mass,
        the star undergoes a second collapse phase into a BH.
  \item Formation of a black hole through multiple stages. Here
        the configuration undergoes at least two approximately stationary
        neutron star phases; the first is weakly scalarized, and later phases
        are strongly scalarized.
  \item Collapse to a strongly scalarized neutron star through multiple
        stages. Here the configuration intermittently forms one
        or more approximately
        stationary neutron star stages with ever increasing
        central density. The transition from weak to strong
        scalarization always occurs in the second collapse phase.
  \item Single-stage collapse to a strongly scalarized neutron star.
\end{list}
These five different scenarios are most conveniently visualized in terms
of the central baryon density $\rho_c$ and the central value of the
scalar field $\varphi_c$ as functions of time. We plot these quantities
for a set of representative configurations in Fig.~\ref{fig:5examples}.
A more detailed discussion of the five scenarios is given in
Appendix \ref{app:scenarios} and a diagram-style visualization in
Fig.~\ref{fig:arrowdiagram}.

The strength of the GW signal depends on the maximum
scalarization achieved during the time evolution. This is not necessarily
the degree of scalarization at the end of the simulation since black holes
will descalarize in agreement with the no-hair theorems for BHs
\cite{Thorne:1971,Hawking:1972qk}.
For\footnote{For $\alpha_0= \mathcal{O}(1)$ the scalar field will always
reach a large amplitude $\varphi_{\rm max}=\mathcal{O}(\alpha)
=\mathcal{O}(1)$ and the distinction between weak and strong scalarization
disappears. We only consider $\alpha_0\leq 0.1$.}
$\alpha_0\ll 1$,
this implies that case (1) always leads to a negligible GW signal
whereas cases (3), (4), and (5) always lead to strong signals.
For the two-stage BH formation of case (2), we find that
either weak or strong gravitational radiation is possible,
depending on the degree of scalarization that can be achieved
during the rapid collapse from a weakly scalarized neutron star to
a BH. This sensitively depends on the parameters of the configuration.

In summary, for any given set of parameters, the collapse proceeds according
to one of the five scenarios listed above. The question that remains is
to establish a mapping between the parameter space and the possible outcomes.
For this purpose we separate the parameters into two sets.
The first consists of the EOS and progenitor parameters
$(M_{\rm ZAMS},~Z,~\Gamma_1,~\Gamma_2,~\Gamma_{\rm th})$ and
the second of the ST parameters $(\alpha_0,~\beta_0,~\mu)$.
Let us then consider a given stellar progenitor with fixed ZAMS mass,
metallicity, and EOS and consider the fate of this progenitor as a function
of the ST parameters. Our first observation, which will be discussed in
further detail below in Sec.~\ref{sec:universality}, is that over
a wide range of values the scalar mass $\mu$ does not affect
the outcome qualitatively, but merely rescales the frequency of the
GW signal and modifies its amplitude by a factor of order unity.
In the remainder of this section, we set $\mu=10^{-14}\,{\rm eV}$.

This leaves $\alpha_0$ and $\beta_0$, and we now explore the
main properties of the collapse scenarios in the plane spanned
by these two parameters.

The resulting pattern is best understood by considering two examples, the
progenitors s39 and z39 for EOS3 of Table \ref{tab:EOS}. These stellar
models differ in their metallicity which
leads to a different compactness of the core
at bounce and, hence, significantly different
collapse scenarios as shown in Fig.~\ref{fig:heatmaps}.
In this figure, we display the maximal scalarization defined as
\begin{equation}  \label{eq:max_phi_c_over_time}
  \varphi_{\rm max} = \max (|\varphi_c(t)|)\,.
\end{equation}
In all of our simulations, the extremal value of the central
$\varphi_c$ is negative, hence the modulus sign in Eq.~(\ref{eq:Phir}). [The overall sign of $\varphi$ is merely a matter of convention; inspection of the action in Eq.~(\ref{eq:actionE}) reveals that it is invariant under the simultaneous redefinitions $\varphi\rightarrow-\varphi$ and $\alpha_0\rightarrow-\alpha_0$.]
In the top row of
Fig.~\ref{fig:heatmaps}, we plot $\varphi_{\rm max}$, in
logarithmic measure, as a function of
$\beta_0$ for selected values $\alpha_0$, and in the middle row it is
shown in the form of a heat map in the $(\alpha_0,\beta_0)$ plane. Note that
$\varphi_{\rm max} \propto \alpha_0$ for weakly scalarized configurations,
whereas all strongly scalarized stars reach a comparable $\varphi_{\rm max}
=\mathcal{O}(1)$. The measure $\varphi_{\rm max}$ furthermore determines
the strength of the GW signal emitted in the collapse; $\varphi_{\rm max}
=\mathcal{O}(1)$ always implies a strong GW signal and
$\varphi_{\rm max}=\mathcal{O}(\alpha_0)$ a correspondingly weaker one
by a factor $\alpha_0 \ll 1$.
Finally, we display in the bottom row of Fig.~\ref{fig:heatmaps}
in the form of an (arbitrarily chosen) color code which of the above
five collapse scenarios is realized for the `s39' or `z39' progenitor
for ST parameters $(\alpha_0,\beta_0)$. Clearly, the two progenitors
result in qualitatively different color maps.
All our simulations of progenitor `s39'
result in a neutron star: for mildly negative $\beta_0$,
a weakly scalarized NS is formed in a single stage.
For moderate $\beta_0$,
a multi-stage collapse leads to a strongly scalarized NS and for
highly negative $\beta_0$ a strongly scalarized NS forms
without intermediate stages. In contrast, we encounter
for progenitor `z39' the following scenarios as $\beta_0$ becomes
more negative: two-stage BH formation, multi-stage BH formation,
multi-stage formation of a strongly scalarized NS, single-stage formation
of a strongly scalarized NS. The parameter $\alpha_0$ only
weakly affects the respective threshold values of $\beta_0$.

In principle, we could now construct heat maps analogous to those in
Fig.~\ref{fig:heatmaps} for any possible progenitor, i.e.~for
every mass $M_{\rm ZAMS}$, metallicity $Z$, and EOS. We have done
this for
about 20 additional cases and
always obtained a set of maps qualitatively equal to either the left neutron star
case of Fig.~\ref{fig:heatmaps} or the right black hole case in the
figure. The boundaries of the different regions vary with EOS,
$M_{\rm ZAMS}$, and $Z$,
but we always get one of the two maps. In consequence,
the question which of the two qualitatively different maps
of Fig.~\ref{fig:heatmaps}, the neutron star or the black hole case,
applies to a given progenitor (and EOS) is completely determined by its fate
in GR.

%=============================================================================
\subsection{Dependency on the equation of state and progenitor model}
\label{sec:dependency}
\begin{figure*}[t]
  \includegraphics[width=0.48\textwidth,clip=true]{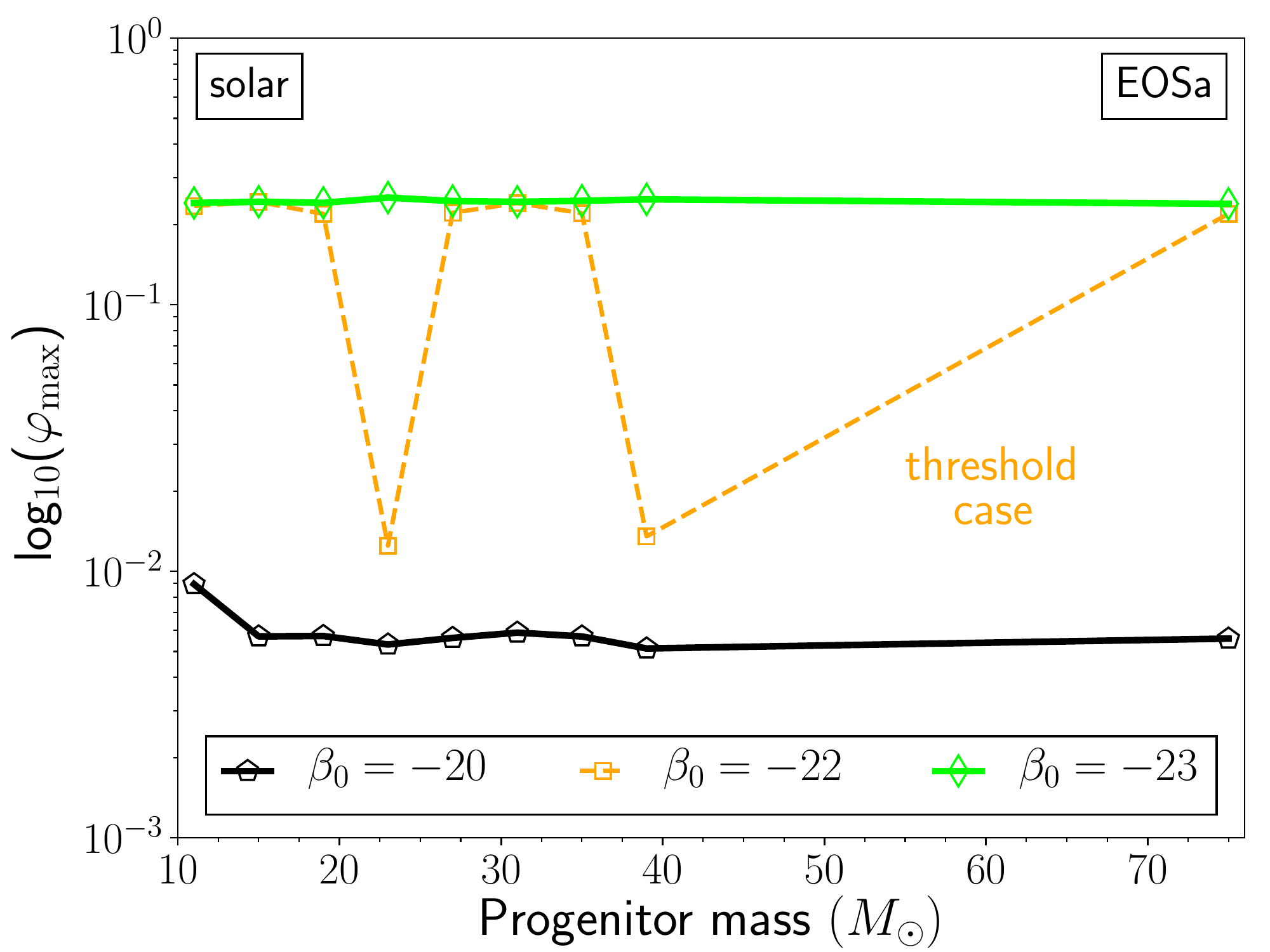}
  \includegraphics[width=0.48\textwidth,clip=true]{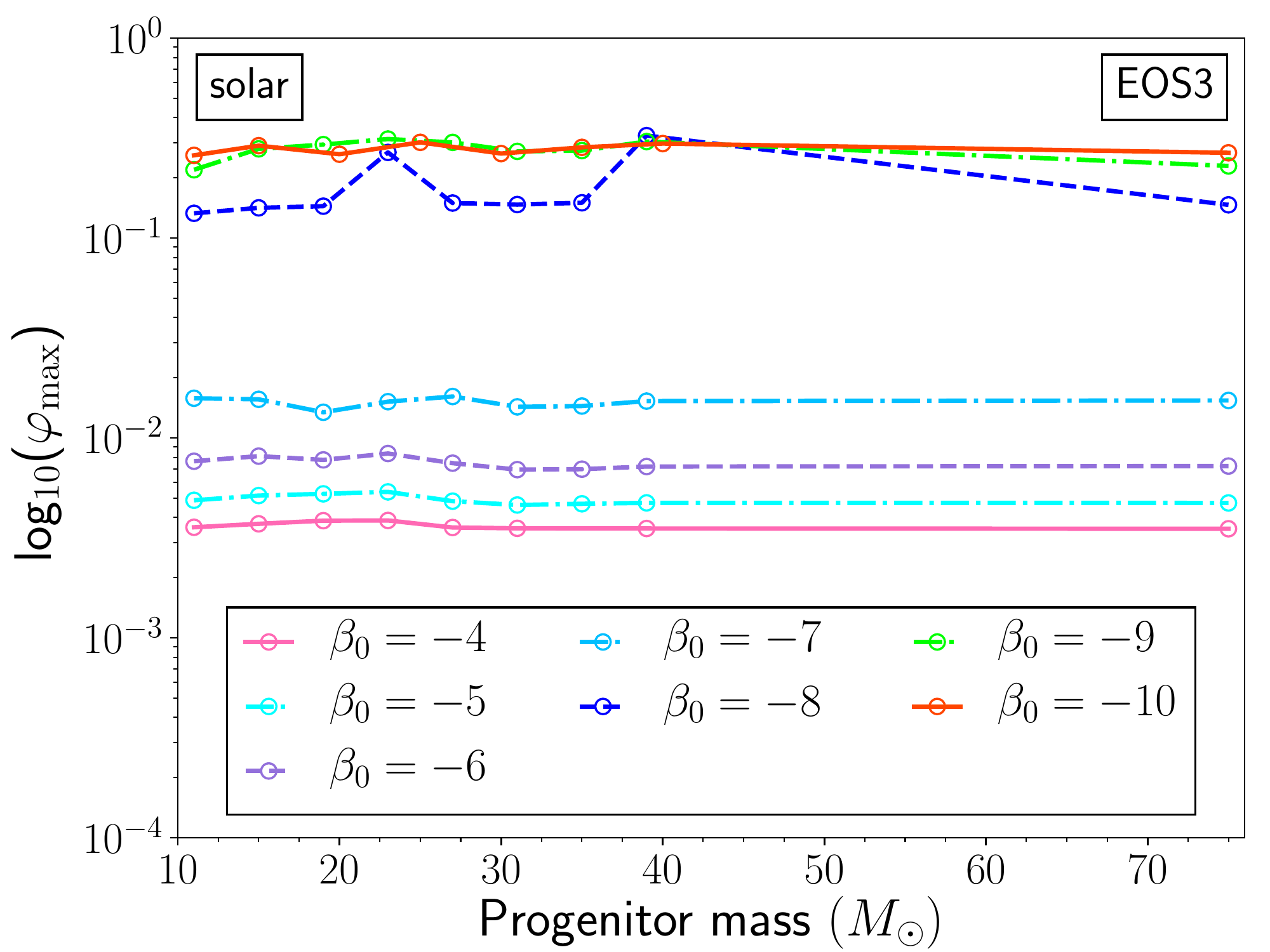}
  \includegraphics[width=0.48\textwidth,clip=true]{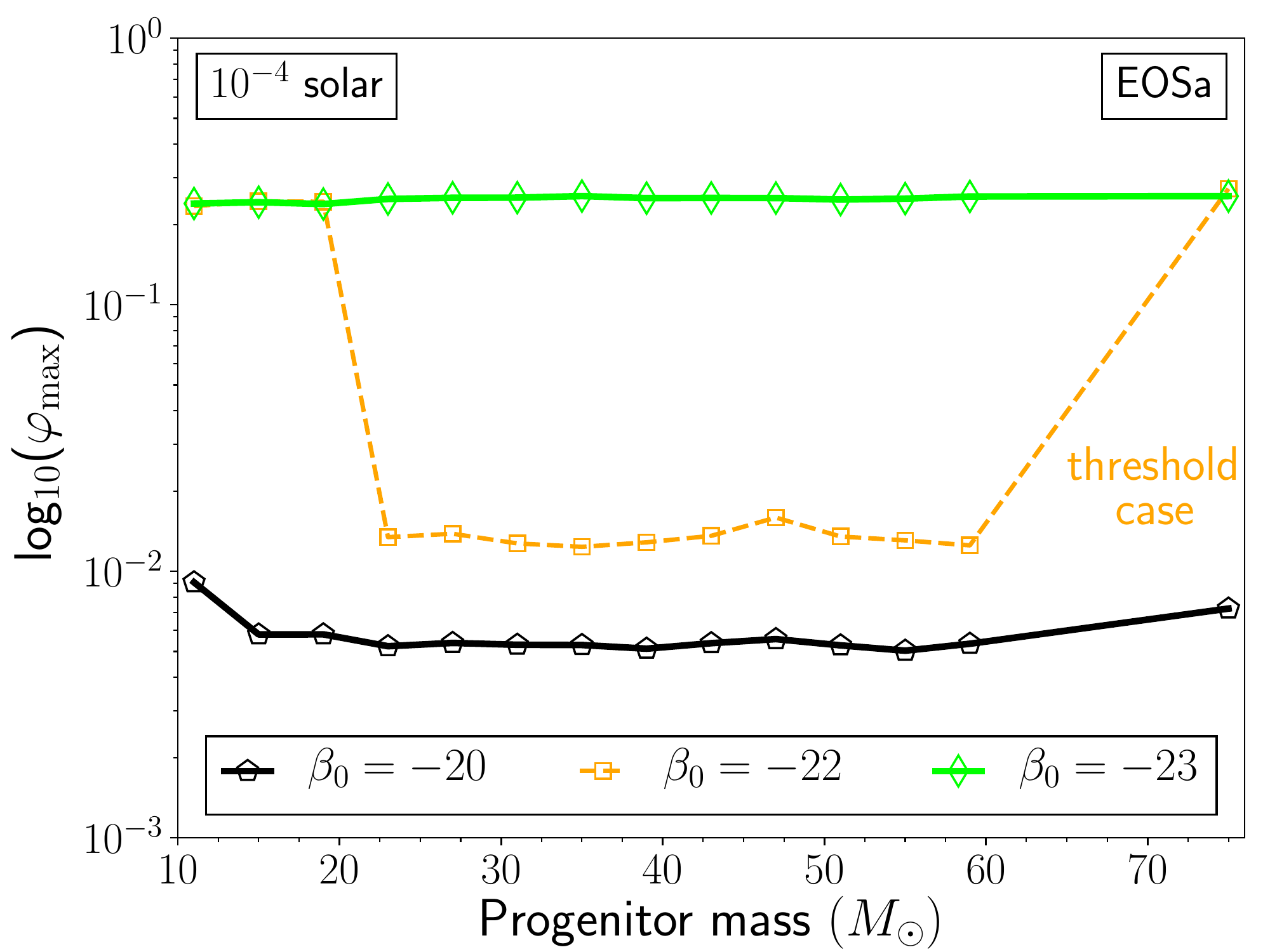}
  \includegraphics[width=0.48\textwidth,clip=true]{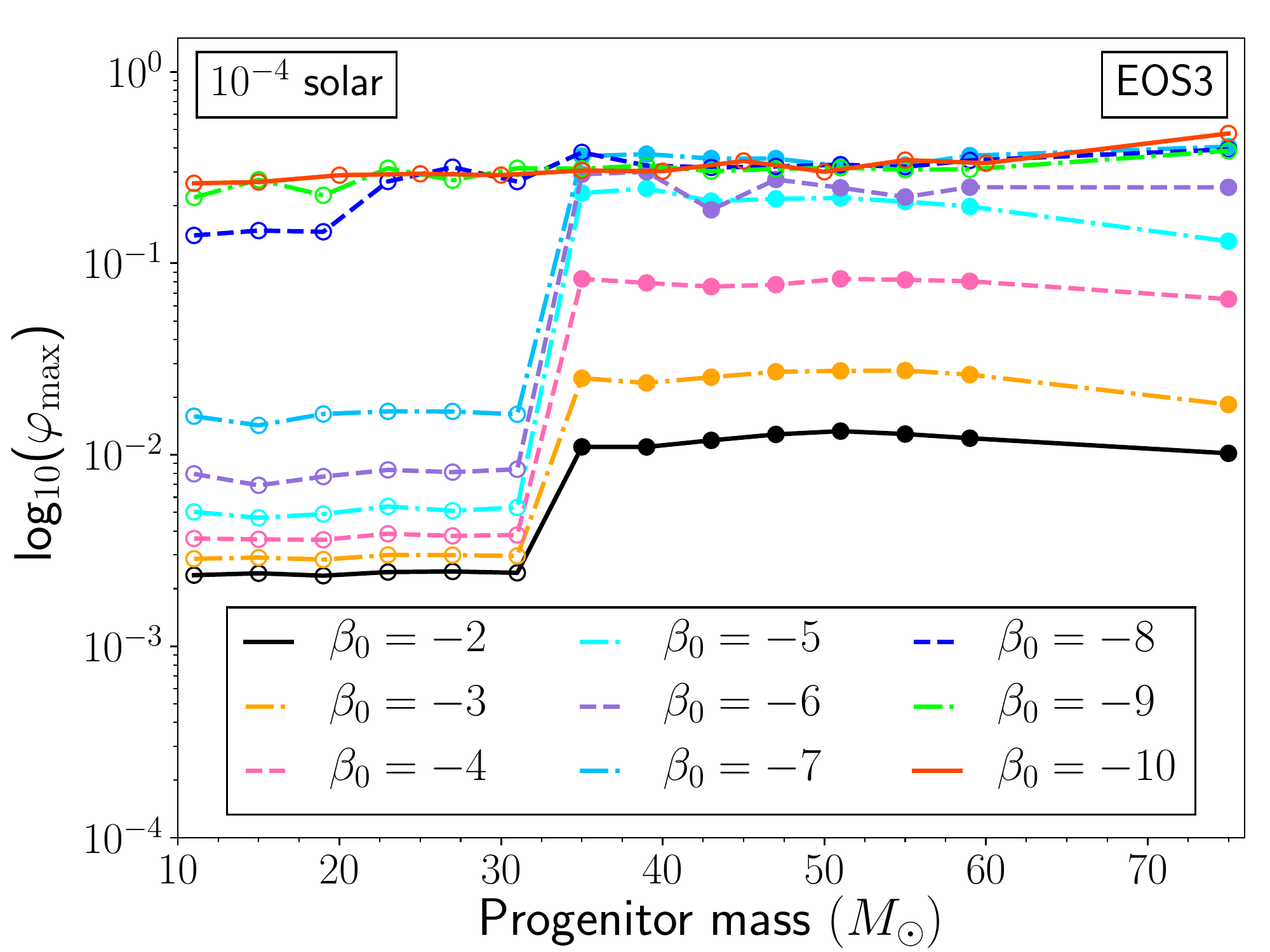}
  \includegraphics[width=0.48\textwidth,clip=true]{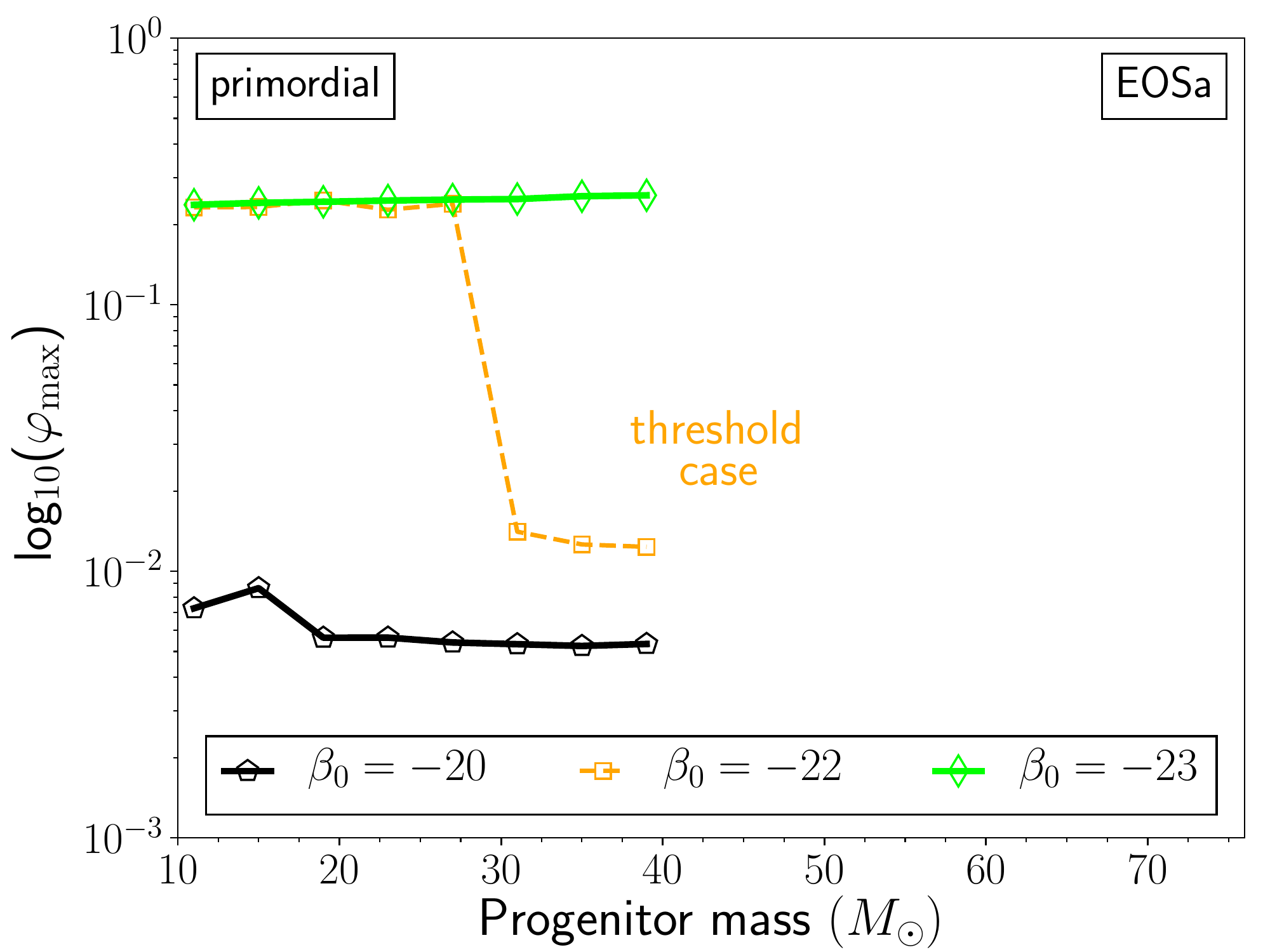}
  \includegraphics[width=0.48\textwidth,clip=true]{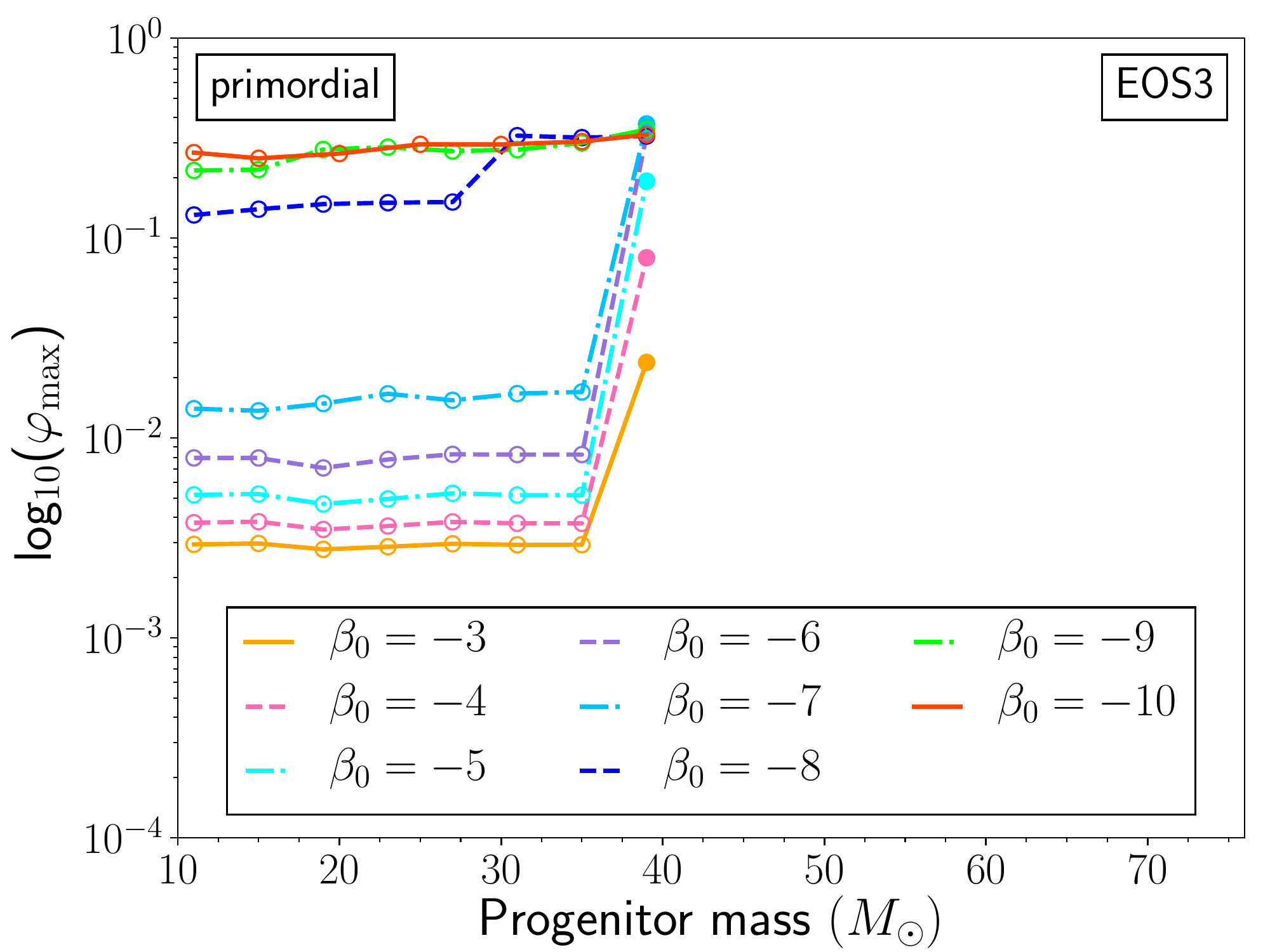}
  \caption{ Each panel shows the maximum scalarization
           of a core collapse as a function of the ZAMS progenitor mass
           for selected values of $\beta_0$ that bracket the transition
           from weak to strong scalarization. The left column represents EOSa and the right column EOS3. The rows represent a different metallicity as labeled
           in the panels. The right panels and the top-left panel
           exclusively contain collapse scenarios forming NSs.
           In the center-right and bottom-right panels, we distinguish
           NS cases from those forming BHs by using empty or filled
           symbols, respectively. Note that for primordial metallicity,
           the catalog of stellar progenitors contains models
           up to $M_{\rm ZAMS}=40~M_{\odot}$ only.
          }
  \label{fig:EOSzeta}
\end{figure*}
The classification of the collapse scenarios has given us a qualitative
picture of the possible outcomes of a stellar core collapse in ST gravity.
The main task that remains is to understand more quantitatively how
the boundaries in the diagrams of Fig.~\ref{fig:heatmaps} depend on
the choice of the EOS and the progenitor. Here we are particularly
interested in the strength of the GW signal and will therefore focus
on the sharp transition between weakly scalarized (blue) and the
strongly scalarized (red) regions in the central panels in
Fig.~\ref{fig:heatmaps}. Our strategy for this purpose is as follows.
We consider EOS3 and EOSa
from Table \ref{tab:EOS}
as representative examples of a soft and a
stiff EOS, respectively. Next, we note in Fig.~\ref{fig:heatmaps}
that the parameter $\alpha_0$ only mildly affects whether
a configuration is weakly or strongly scalarized; the corresponding
$\beta_0$ threshold in the center panels of the figure
varies by a few units but no more. Bearing in mind
this variation, we fix in our analysis $\alpha_0=10^{-2}$.

We then have six combinations with different
EOS and/or metallicity of the progenitor.
For each of these cases we plot in Fig.~\ref{fig:EOSzeta} the maximum
scalarization $\varphi_{\rm max}$ as a function
of the progenitor mass $M_{\rm ZAMS}$
for selected values of $\beta_0$; these $\beta_0$
values have been chosen such that they bracket the
threshold between weak and strong scalarization. The results
of the figure are summarized as follows.
\begin{itemize}
  \item The transition between weak and strong scalarization is
        abrupt, occurring in a brief interval around
        a threshold value $\beta_0^*$.
        Without fine-tuning $\beta_0$, we obtain
        either weakly or strongly scalarized configurations but
        rarely cases in between.
  \item For $\beta_0$ values close to the threshold, the degree
        of scalarization can be highly sensitive to the ZAMS mass.
        Such a sensitive dependence on the parameters is
        reminiscent of the critical phenomena well known in
        gravitational collapse \cite{Choptuik:1992jv}
        and is also expected from the phase-transition
        character of the spontaneous scalarization phenomenon
        \cite{Damour:1993hw}.
  \item Besides this sensitive dependency near the critical
        $\beta_0^*$, the only significant variation of the degree
        of scalarization with the ZAMS mass occurs at the onset
        of BH formation in the center-right
        and bottom-right panels of Fig.~\ref{fig:EOSzeta}.
        Here the scalarization increases visibly at
        $M_{\rm ZAMS}\approx 30~M_{\odot}$ and $35~M_{\odot}$,
        respectively. Progenitor masses below the threshold
        value result in
        weakly scalarized NSs and higher masses lead to
        BH formation and stronger scalarization. Note, however,
        the logarithmic scaling of the vertical axis, so that even
        in these cases, the strong variation of $\varphi_c$
        with $M_{\rm ZAMS}$ is restricted to $\beta_0$ values close
        to the critical threshold $\beta_0^*$.
  \item For $\beta_0$ values significantly below or above the
        threshold,
        our simulations show only a mild dependence
        of the scalarization on the
        progenitor mass $M_{\rm ZAMS}$. The same holds for the
        metallicity $Z$.
  \item Stiff EOSs result in less compact neutron stars and
        correspondingly
        more negative threshold values $\beta_0^*$ for strong scalarization.
        For soft EOSs, highly compact neutron stars can form even for
        mild $\beta_0$ values and lead to strong scalarization.
\end{itemize}
In summary, we observe strong scalarization when $\beta_0$ becomes
more negative than a threshold value $\beta_{0}^*$. This
threshold is $\approx -25$ for a stiff EOS but drops to the
well-known limit $\beta_{0,{\rm thr}}-4.35$ observed for the spontaneous
scalarization of stationary neutron-star models in massless ST theory
\cite{Damour:1993hw,Novak:1998rk}. This threshold varies only
mildly with the mass or the metallicity of the progenitor model.

Throughout this analysis, we set the scalar mass parameter
$\mu=10^{-14}\,{\rm eV}$. As it turns out, the degree of scalarization
barely changes even when we vary $\mu$ over several orders of magnitude.
This insensitivity to $\mu$
of the strong scalarization effect is not only supported by our
simulations, but can also be understood at the analytic
level. This will be done in the next section where we also discuss in
more detail the propagation of the wave signal to astrophysically
large distances.

%=============================================================================
\section{Wave extraction and propagation}
\label{propagation}

\label{sec:wave_extraction_and_propagation}
In this section, the extraction of the scalar field from the core
collapse simulations is described along with a procedure for
converting this into a prediction for the GW signal at astrophysically
large distances, potentially observable by LIGO/Virgo.  
The latter step is complicated by the dispersive nature of wave propagation
for massive fields; it will be shown how this dispersion generically
leads to the \emph{inverse chirp} described in \cite{Sperhake:2017itk}.

There are two natural length scales relevant to the problem: the
gravitational radius associated with the mass of the remnant NS, $r_{\rm
G}=GM_{\rm NS}c^{-2}$; and the reduced Compton wavelength for the massive
scalar field, $\lambdabar_{\rm C}=c/\omega_{*}$ where $\omega_{*}=\mu c^{2}\hbar^{-1}$.
The remainder of this section again uses natural units in which $G=c=1$.

At large distances from the star ($r\gg r_{\rm G}$) the dynamics
of the gravitational scalar are, to a good approximation, governed
by the flat-space Klein-Gordon equation,
\begin{equation} \label{eq:KDeqn}
    \partial_t^2 \varphi - \nabla^2\varphi + \omega_{*}^2\varphi= 0 \,.
\end{equation}
In spherical symmetry (using coordinates $\{t,\,r,\,\theta,\,\phi\}$)
the field depends only on time and radius ($\psi=\psi(t;r)$), the
Laplacian is given by
$\nabla^{2}\cdot=r^{-2}\partial_{r}(r^{2}\partial_{r}\cdot)$, and
the rescaled field $\sigma\equiv r\varphi$ satisfies a 1D wave
equation,
\begin{equation} \label{eq:KGr}
    \partial_t^2 \sigma - \partial_r^2 \sigma + \omega_{*}^2\sigma= 0\,.
\end{equation}

Consider first the behavior of a single Fourier mode, $\sigma
\propto e^{-i(\omega t-kr)}$; Eq.~\eqref{eq:KGr} gives the
\emph{dispersion relation}
\begin{equation} \label{eq:omegak}
    \omega^2 = k^2+\omega_{*}^2 \,.
\end{equation}
The wave number, $k$, is real for high frequencies ($|\omega|>\omega_{*}$)
and the solution describes a propagating wave.  For low frequencies
($|\omega|<\omega_{*}$; including the static case $\omega=0$) the
wave number is imaginary leading to solutions which decay exponentially
over a characteristic length $\lambdabar_{\rm C}$.
The critical frequency $\omega_{*}$ associated with the scalar field mass acts
as a low frequency cutoff in the GW spectrum.
For propagating solutions, the phase velocity
($v_{\textrm{phase}}=\omega/k= [1-(\omega_{*}/\omega)^{2}]^{-1/2}$)
is superluminal, while the group velocity ($v_{\textrm{group}}=
\textrm{d}\omega/\textrm{d}k=[1-(\omega_{*}/\omega)]^{+1/2}$)
is subluminal.

In the massless case ($\omega_{*}=0$), the general solution to Eq.~\eqref{eq:KGr}
can be written as the sum of ingoing and outgoing pulses traveling
at the speed of light.
This makes interpreting the output of core collapse
simulations particularly simple.  First, one extracts the field
as a function of time at a fixed \emph{extraction radius},
$\sigma(t;r_{\rm ex})$.  This radius must be sufficiently large
that (i) the flat space Eq.~\eqref{eq:KDeqn} holds, and (ii) $r_{\rm
ex}$ is in the wave zone so that the signal has decoupled from the
source and is purely outgoing.  In the massless case both (i) and
(ii) are satisfied by choosing $r_{\rm ex}\gg r_{\rm G}$.  Then,
the signal as a function of time at some larger \emph{target radius},
$\sigma(t;r)$, is simply obtained via $\sigma(t-[r-r_{\rm
ex}];r)=\sigma(t;r_{\rm ex})$.  The only change in the signal
between $r_{\rm ex}$ and $r$ is a time delay and a reduction in the amplitude
of the field $\varphi$ by a factor $(r/r_{\rm ex})$.

We seek an analogous method in the massive case ($\omega_{*}>0$) for
relating the signal at the extraction radius to the signal at the
much larger target radius.  The extraction radius is chosen to
satisfy the two conditions as before, but now (ii) requires $r_{\rm
ex}\gg\lambdabar_{\rm C}$.  This is generally a stricter condition
than $r_{\rm ex}\gg r_{\rm G}$; for $\mu=10^{-14}\,\mathrm{eV}$ the
Compton wavelength is $\lambdabar_{\rm C}\approx 10^{7}\,\mathrm{m}$,
whereas the gravitational radius for NSs is typically only
$r_{\rm G}\sim 10^{3}\,\mathrm{m}$.  In this paper the extraction
radius is taken to be $r_{\rm ex}=7.0\times10^{7}\,\mathrm{m}$.
The target radius, the distance of the supernova from Earth, is
very much large, e.g.\ $\sim 10\,\mathrm{kpc}$.

The remainder of this section describes two methods for evolving
signals from the extraction radius out to large radii.  First, a
numerical evolution of Eq.~\eqref{eq:KGr} in the time domain is
described.  This numerical method, while very accurate at a short
distance, is of limited use in practice because it struggles to
cope with the very large astrophysical distances.  Second, an
analytic method for solving Eq.~\eqref{eq:KGr} in the frequency
domain is described.  The two methods are validated by comparing
them against each other in the regime where both can be evaluated.
Finally, the analytic method is used to study the asymptotic behavior
at large distances using the stationary phase approximation (SPA).

%=============================================================================
\subsection{Numerical evolution in the time domain}\label{sec:1+1}
\begin{figure}[t]
  \includegraphics[width=0.34\textwidth]{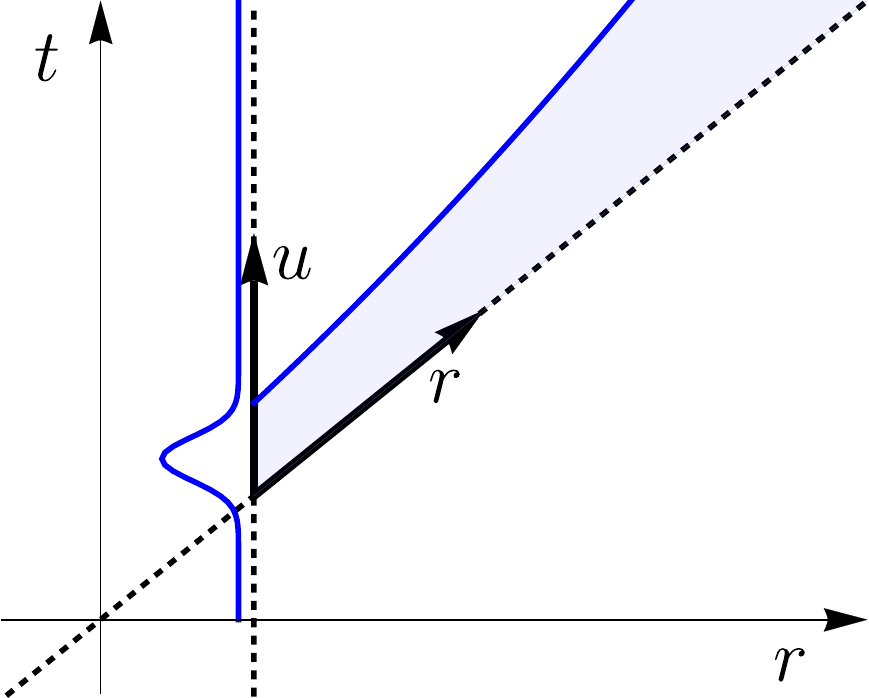}
  \caption{ \label{fig:CoordinateDiagram}
           A sketch of the coordinates used in the numerical
           evolution.  The main axes show the standard $\{t,r\}$
           coordinates and the inset arrows show the $\{u,r\}$
           coordinates.  The vertical blue line indicates the signal
           as a function of time at the extraction radius,
           $\sigma(t;r_{\rm ex})$, and the shading indicates the
           region where the signal propagates dispersively.
           A numerical grid based on the $\{u,r\}$ coordinates can
           cover the shaded region with less redundant space than
           one based on $\{t,r\}$.
          }
\end{figure}

Given suitable initial data it is possible to numerically evolve
Eq.~\eqref{eq:KGr}.  Here it is necessary to evolve some given
outgoing data on a timelike surface out to larger radii (see
Fig.~\ref{fig:CoordinateDiagram}).  Equation \eqref{eq:KGr} is
written in a manner that makes a $1+1$ dimensional split obvious
using the coordinates $\{t,r\}$.  However, these coordinates are
not well adapted for signals traveling at, or near, the speed of
light.  Alternatively, and much more efficiently, a $1+1$ split can
be implemented based on coordinates $\{u,r\}$, where $u\equiv
t-r$ is the (null) retarded time coordinate.  Using these
coordinates the wave equation becomes
\begin{align} \label{eq:wavechar}
    2\partial_{u}\partial_{r}\sigma-\partial_{r}^{2}\sigma+\omega_{*}^{2}\sigma = 0\,.
\end{align}
By defining the conjugate momentum $\Pi_{u}\equiv\partial_{u}\sigma(u;r)$,
Eq.~\eqref{eq:wavechar} can be reduced down to the first order form.

Given an initial signal on the extraction sphere, $\sigma(u;r_{\rm
ex})$, it is straightforward to solve Eq.~\eqref{eq:wavechar} using
standard techniques; in our case a {\em method of line} integration
with the {\em iterated Crank-Nicholson} scheme \cite{Teukolsky:1999rm}.
From this numerical solution, we directly extract the signal at
some larger target radius, $\sigma(u;r)$ .

%=============================================================================
\subsection{Analytic evolution in the Fourier domain} \label{sec:FTprop}
\label{sec:analytic_propagation}
We now revert to coordinates $\{t,r\}$ in Eq.~\eqref{eq:KGr}.
With the Fourier transform conventions
\begin{align} \label{BBB2}
    \tilde{\sigma}(\omega;r)=&
        \int_{-\infty}^{\infty}\textrm{d}t\;\sigma(t,r)e^{\rmi\omega t} \,, \\
    \sigma(t;r)=&
        \int_{-\infty}^{\infty}\frac{\textrm{d}\omega}{2\pi}
        \;\tilde{\sigma}(\omega,r)e^{-\rmi\omega t} \,,
\end{align}
the Fourier transform of Eq.~\eqref{eq:KGr} yields the simple
harmonic motion equation for $\tilde{\sigma}(\omega;r)$,
\begin{equation} \label{BBB3}
    \partial_{r}^{2}\,\tilde{\sigma}(\omega;r)=-\big(\omega^{2}
        -\omega_{*}^{2}\big)\tilde{\sigma}(\omega;r)\;.
\end{equation}
Defining $k^{+}\!\equiv\! +\sqrt{\omega^{2}-\omega_{*}^{2}}$ as the positive root of the dispersion relation in
Eq.~\eqref{eq:omegak}, the solution to Eq.~\eqref{BBB3} can be
written in terms of two arbitrary functions,
\begin{align} \label{BBB4}
    \tilde{\sigma}(\omega;r) &= f(\omega)\e^{\rmi k^{+} (r-r_{\rm
        ex})} + g(\omega)\e^{-\rmi k^{+} (r-r_{\rm ex})}\,.
\end{align}
The radial coordinate has been shifted to the extraction radius for
later convenience.  Taking the inverse Fourier transform to convert
back into the time domain gives
\begin{align} \label{eq:BBB_invFT}
    \sigma(t;r)=\int_{-\infty}^{\infty}  \frac{\mathrm{d}\omega}{2\pi}
        \, \Big[ & f(\omega) \e^{\rmi k^{+} ( r-r_{\rm ex} )} \\ + &
        g(\omega) \e^{-\rmi k^{+} ( r-r_{\rm ex} )} \Big]\e^{-\rmi
        \omega t} \,. \nonumber
\end{align}

The fact that the field $\varphi$ is real imposes some constraints
on the otherwise arbitrary functions $f$ and $g$:
\begin{align} \label{eq:condition1}
    & (\mathrm{a}) \quad \sigma(t;r)\in\mathbb{R} \;\Rightarrow\;
        \tilde{\sigma}(\omega;r)=\tilde{\sigma}^{*}(-\omega;r)
    \;\Rightarrow\; \\
    & \begin{cases}
    f(\omega)=g^{*}(-\omega)\,&\textrm{if }\,|\omega|>\omega_{*} \\
    f(\omega)=f^{*}(-\omega)\textrm{ and
        }g(\omega)=g^{*}(-\omega)\,&\textrm{if }\,|\omega|<\omega_{*}.
\end{cases}\nonumber
\end{align}
A further constraint on the function $g$ is obtained by imposing
boundary conditions at infinity.  The field $\varphi$ must decay
as $1/r$ (or faster) which implies that $\tilde{\sigma}(\omega;r)$
remains bounded at large radii.  From Eq.~\eqref{eq:BBB_invFT}, and
recalling that $k^{+}$ is imaginary for $|\omega|<\omega_{*}$, gives
the constraint
\begin{align} \label{eq:condition2}
     (\mathrm{b}) \quad g(\omega)=0 \quad \textrm{if }|\omega|<\omega_{*} \,.
\end{align}

The constraints (a) and (b) can be used to eliminate $g(\omega)$
in favor of $f(\omega)$.  Furthermore, the symmetries implied by
the constraint (a) allow the Fourier integral in Eq.~\eqref{eq:BBB_invFT}
to be written over positive frequencies; the general solution in
Eq.~\eqref{eq:BBB_invFT} now becomes
\begin{align} \label{BBB5}
    \sigma &(t;r)  =  2 \Re \bigg\{
        \int_{0}^{\omega_{*}}\!\frac{\mathrm{d}\omega}{2\pi}\,f(\omega)\e^{\rmi
        k^{+} (r-r_{\rm ex})}\e^{-\rmi \omega t} + \\
    & \int_{\omega_{*}}^{\infty}\!\frac{\mathrm{d}\omega}{2\pi}\,\Big[
        f(\omega)\e^{\rmi k^{+} (r-r_{\rm ex})} + f^{*}(-\omega)\e^{-\rmi
         k^{+} (r-r_{\rm ex})} \Big]\e^{-\rmi \omega t} \bigg\}. \nonumber
\end{align}
From Eq.~\eqref{BBB5}, and considering the sign of $k^{+}$, it can
be seen that the high frequencies $f(\omega>\omega_{*})$ represent
outgoing modes, the large negative frequencies $f(\omega<-\omega_{*})$
represent ingoing modes, and the intermediate frequencies
$f(|\omega|<\omega_{*})$ represent nonpropagating modes.

It only remains to relate the unknown function $f(\omega)$ to the
(purely outgoing) scalar profile at the extraction radius obtained
from the core collapse simulation, $\sigma(t;r_{\rm ex})$.  The
function $f(\omega)$ is given by
\begin{equation} \label{BBB6}
    f(\omega)=\begin{cases} 0 & \textrm{if }\omega\leq-\omega_{*} \\
    \tilde{\sigma}(\omega;r_{\rm ex}) & \textrm{if }\omega>-\omega_{*}
    \end{cases}\Bigg\}\;.
\end{equation}
Substituting into Eq.~\eqref{BBB5}, and returning to writing the
integral over both positive and negative frequencies, gives
\begin{align} \label{BBB6point5}
    \sigma(t;r)= & \int\frac{\textrm{d}\omega}{2\pi}\,
        \tilde{\sigma}(\omega;r_{\rm ex})\,\times \\ &\begin{cases}\e^{-\rmi
         k^{+} (r-r_{\rm ex})} & \textrm{if } \omega\leq-\omega_{*} \\ \e^{+\rmi
        k^{+}(r-r_{\rm ex})} & \textrm{if } \omega >-\omega_{*}
        \end{cases}\Bigg\}\e^{-\rmi\omega t} \,. \nonumber
\end{align}
This shows that the frequency domain signal at the target radius
is related to that at the extraction radius via
\begin{equation} \label{BBB7}
    \tilde{\sigma}(\omega;r) \!=\! \tilde{\sigma}(\omega;r_{\rm ex})
        \!\times\! \begin{cases}\e^{-\rmi k^{+} (r\textrm{--}r_{\rm ex})}
        &\textrm{if } \omega\!\leq\! -\omega_{*} \\ \e^{+\rmi
        k^{+}(r\textrm{--}r_{\rm ex})} &\textrm{if } \omega \!>\!
        -\omega_{*} \end{cases} \Bigg\} .
\end{equation}
Note that the effect of the dispersion enters only in the complex
phase of the Fourier transform.  Therefore, the effect of the
dispersion is to disperse the signal, rearranging the frequency
components in time, while leaving the overall power spectrum invariant
for all $|\omega|>\omega_*$.
Lower frequencies, $|\omega|<\omega_*$,
are exponentially suppressed during propagation and are not observable
at large distances.

We now have a prescription for analytically propagating signals out
to larger radii.  First, numerically evaluate the fast Fourier
transform of the scalar profile on the extraction sphere,
$\tilde{\sigma}(\omega;r_{\rm ex})$.  Second, use Eq.~\eqref{BBB7}
to obtain the Fourier domain signal at the target radius,
$\tilde{\sigma}(\omega;r)$.  Finally, numerically evaluate the
inverse Fourier transform to obtain the desired signal, $\sigma(t;r)$.
In Appendix \ref{app:comparing_the_methods} we compare the results of analytically propagating signals in this way with the results obtained via numerical evolution described in Sec.~\ref{sec:1+1} and find good agreement.

Unfortunately, neither of the methods described (in their current
form) is suitable for propagating the signal to astrophysically
large distances (e.g.\ $r_{\rm ex}=10\,\mathrm{kpc}$).  The unavoidable
problem is that as the signal propagates further, the longer (i.e. containing more
cycles) it becomes due to the dispersive stretching.
 This poses
two problems for the time domain numerical integration: first,
the evolution becomes increasingly expensive due to the large
numerical grids required; and second the numerical errors tend
to grow as the signal is propagated over greater distances.  The analytic
frequency domain method can be pushed to somewhat larger radii;
however, even this fails when the signal eventually becomes longer
than the largest array for which the fast Fourier transform can be
numerically evaluated.
The next section describes how the behavior of the
scalar field at very large distances may be studied.

%=============================================================================
\subsection{Asymptotic behaviour: the inverse chirp}\label{sec:asymp}
As the signal is stretched out it becomes ever more oscillatory,
and the amplitude varies more slowly relative to the phase.  Therefore,
in the large distance limit the \emph{stationary phase approximation}
(SPA) may be used to evaluate the inverse Fourier transform in
Eq.~\eqref{BBB6point5}.  It should be noted that the SPA \emph{becomes}
valid at large radii regardless of whether it was initially
valid for the signal at the extraction radius.  As will be shown
below, dispersive signals tend to ``forget'' the details of their
initial profile as they propagate over large distances and always
tend to a generic ``inverse chirp'' profile.

\begin{figure}[t]
\centering
\includegraphics[width=0.34\textwidth]{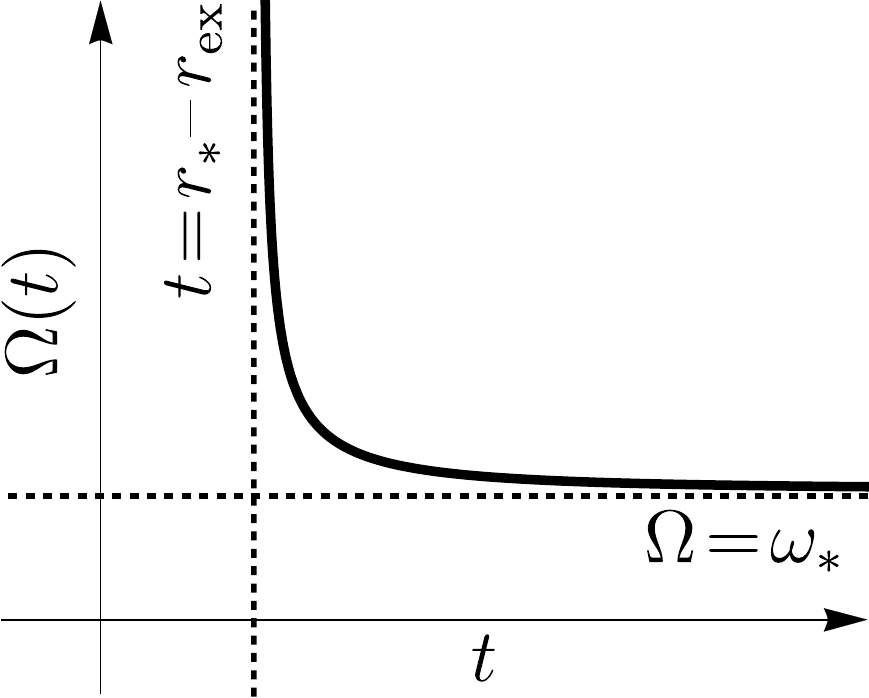}
\caption{ \label{fig:InverseChirp}
        A sketch plot showing the time-frequency structure of the
        ``inverse chirp'' in Eq.~\eqref{eq:omegastar}. The frequency
        decays over time; the high frequency components (traveling
        at almost the speed of light) arrive first, followed by the
        slower low frequency components. Frequencies below $\omega_{*}$
        are exponentially suppressed and never reach large radii.
        }
\end{figure}

The initial Fourier domain signal on the extraction sphere may be decomposed into
its amplitude and phase:
\begin{align}
	\tilde{\sigma}(\omega;r_{\rm ex}) = \mathcal{A}(\omega;r_{\rm
ex})\e^{\rmi\Psi(\omega)} \,.
\end{align}
As noted above, at large radii frequencies $|\omega|<\omega_{*}$
do not contribute to the signal because they decay exponentially with $r$.
It will be convenient to write the time domain solution at large radii in
Eq.~\eqref{BBB6point5} as an integral over positive frequencies
only:
\begin{align} \label{eq:intstatphaseA}
    \sigma(t;r)= 2\Re\bigg\{\int_{\omega_{*}}^{\infty}
        \frac{\textrm{d}\omega}{2\pi}\;\mathcal{A}(\omega)\e^{\rmi
        \psi(\omega,t)}\bigg\}\,.
\end{align}
where the modified complex phase is defined
as $\psi(\omega,t)\equiv\Psi(\omega)+k^{+}(r-r_{\rm ex})-\omega t$.
This phase has a stationary point when
$\partial\psi(\omega,t)/\partial\omega=0$ which is satisfied by
\begin{equation} \label{eq:eqnforomegastar}
    t = \frac{\mathrm{d}\Psi(\omega)}{\mathrm{d}\omega}
        +\frac{\omega(r-r_{\rm ex})}{\sqrt{\omega^{2}
        -\omega_{*}^{2}}}\,.
\end{equation}
Note that the final term in Eq.~\eqref{eq:eqnforomegastar} can be written as $(r-r_{\rm
ex})/v_{\textrm{group}}$.  In the limit $r\gg r_{\rm ex}$ the final
term in Eq.~\eqref{eq:eqnforomegastar} becomes dominant and the
$\mathrm{d}\Psi/\mathrm{d}\omega$ term can be neglected.  In this
approximation, it is straightforward to invert
Eq.~\eqref{eq:eqnforomegastar}, which gives us the frequency of the
signal at $r$ as a function of time, $\omega=\Omega(t)$, where
\begin{equation} \label{eq:omegastar}
    \Omega(t)=\frac{\omega_{*}t}{
	    \sqrt{   t^2 - \left(r-r_{\rm ex} \right)^{2}   }
        } \,,\;\textrm{ for }\,t>r-r_{\rm ex}\,.
\end{equation}
This frequency varies as an \emph{inverse chirp} (see
Fig.~\ref{fig:InverseChirp}) with low frequencies arriving after
high frequencies.  The origin of the inverse chirp is easily
understood as the modes of each frequency arriving at a time
corresponding to the group velocity of that frequency.

All that remains is to evaluate the amplitude as a function of time.
This can also be done via the SPA.  The integrand in
Eq.~\eqref{eq:intstatphaseA} is highly oscillatory when $r-r_{\rm
ex}$ is large, except for frequencies near $\Omega(t)$ which therefore
dominate the result.  Expanding the amplitude to zeroth order, and
the phase to quadratic order, about $\omega=\Omega(t)$ and substituting
into Eq.~\eqref{eq:intstatphaseA} gives
\begin{align} \label{eq:intstatphase2}
    \sigma(t;r)=2\Re\bigg\{&\mathcal{A}\big[\Omega\big]\e^{\rmi
        \psi\left(\Omega,t\right)} \,\times \\
    & \int_{\omega_{*}}^{\infty}\frac{\textrm{d}\omega}{2\pi}\,
        \e^{\frac{\rmi}{2}\left(\omega-\Omega\right)^{2}\psi''}\bigg\}\,,
    \nonumber
\end{align}
where
$\psi''\equiv\partial^{2}\psi/\partial\omega^{2}|_{\omega=\Omega}$.
The integrand in Eq.~\eqref{eq:intstatphase2} is dominated by
frequencies near $\omega=\Omega$; at the current approximation
order, the integration limits can be changed to
$\int_{\Omega-a}^{\Omega+b}\textrm{d}\omega$ for any $a,b>0$.
Choosing $a,b\rightarrow\infty$, and changing variables to
$u^{2}=\left(\omega-\Omega\right)\psi''$ gives
\begin{align} \label{eq:intstatphase3}
  \sigma(t;r) = \Re\bigg\{&\frac{ \mathcal{A}(\Omega)
        \e^{\rmi \psi\left(\Omega,t\right)}}
        {\sqrt{\pi^{2}\left|\psi''\right|}}\int_{-\infty}^{\infty}\textrm{d}u\;\e^{\frac{\rmi}{2}u^{2}
        \textrm{sign}(\psi'')}\bigg\}\,.
\end{align}
The integral in Eq.~\eqref{eq:intstatphase3} is a standard Gaussian
integral which may be readily evaluated to give
\begin{align}
    \sigma(t;r)=\Re\bigg\{ A (t;r)\,\e ^{\rmi
        \phi(t;r) } \bigg\} \,,
\end{align}
where the amplitude and phase are given by
\begin{align} \label{eq:intstatamp4}
    A(t;r) &=
    \sqrt{\frac{2\big[\Omega^{2}-\omega_{*}^{2}\big]^{3/2}}
        {\pi\omega_{*}^{2}(r-r_{\rm ex})}}
    \mathcal{A}(\Omega) \,, \\
    \label{eq:intstatphase4}
    \phi(t;r) &=
    \Psi(\Omega) +
        \sqrt{\Omega^{2}-\omega_{*}^{2}}(r-r_{\rm ex})  - \Omega t - \frac{\pi}{4}\,,
\end{align}
where $\Omega(t)$ is given in Eq.~\eqref{eq:omegastar} [cf. Eq.~(11) in \cite{Sperhake:2017itk}].
At each instant the signal is quasimonochromatic with a frequency
$\Omega(t)$ and an amplitude, $A(t;r)$, proportional to
the square root of the power spectrum of the initial (extraction
radius) signal evaluated at that frequency divided by a factor to
account for the dispersive stretching of the signal.

The inverse chirp profile described by Eq.~\eqref{eq:intstatphase4}
(see Fig.~\ref{fig:InverseChirp}) is an extremely robust prediction
for the signal observed at large distances.  The signal frequency
as a function of time depends only on the distance to the source
and the mass of the scalar field (and there is a near universal
scaling behavior with the scalar mass, as described in the next
section). The frequency as a function of time is completely
independent of the details of the original signal near the source.
The signal amplitude as a function of time does retain some information
about the original source, through its dependence on the spectrum
$\mathcal{A}(\omega)$, although even this gets highly smeared out by the
dispersion.  The inverse chirp waveforms can be extremely long and
highly oscillatory; for the scalar field masses and distances of
interest here (i.e.\ $\mu\approx 10^{-14}\,\mathrm{eV}$ and $r_{\rm
ex}\approx 10\,\mathrm{kpc}$) the signals can retain frequencies
and amplitudes potentially detectable by LIGO/Virgo for centuries.
These signals are best visualized by plotting the amplitude and
frequency separately as functions of time (see Fig.~2 and the
accompanying discussion in \cite{Sperhake:2017itk}).

%=============================================================================
\subsection{Approximate universality under changes of the scalar mass}
\label{sec:universality}
%

%=============================================================================
\subsubsection{Theoretical considerations}
The asymptotic behavior of the wave signal under its dispersive
propagation is determined by Eq.~(\ref{eq:omegastar}) for the
frequency and Eq.~(\ref{eq:intstatamp4}) for the amplitude of the
signal. The dependence of the propagated signal on the scalar mass
$\mu$ through its associated frequency $\omega_*$ becomes clearer if
we rewrite the solution in terms of dimensionless quantities. For
this purpose, we define the rescaled frequency, radius, and time by
%\us{In the previous section we call this $r_{\rm ex}$ instead of $r_0$.}
%
\begin{eqnarray}
  \bar{\Omega} = \frac{\Omega}{\omega_*}\,,~~~~~&&\bar{r}_{\rm ex}
    = \omega_* r_{\rm ex}\,, \nonumber \\
  \bar{t}=\omega_* t\,,~~~~~&&\bar{r}=\omega_* r\,.
  \label{eq:resc}
\end{eqnarray}
In this notation, Eqs.~(\ref{eq:omegastar}),
(\ref{eq:intstatphase4}), and (\ref{eq:intstatamp4}) become
\begin{eqnarray}
  \bar{\Omega}(\bar{t},\bar{r}) &=&
     \frac{\bar{t}}{\sqrt{\bar{t}^{2}-(\bar{r}-\bar{r}_{\rm ex})^2}}\,,
      \nonumber\\[5pt]
  \phi(\bar{t},\bar{r}) &=& \sqrt{\bar{\Omega}^2-1}
    (\bar{r}-\bar{r}_{\rm ex})-\bar{\Omega}\bar{t}
    -\frac{\pi}{4}
    +{\rm Arg}[\tilde{\sigma}(\Omega;r_{\rm ex})]\,,
    \nonumber \\[5pt]
  A(\bar{t},\bar{r})&=&\sqrt{\frac{2}{\pi}}
    \frac{\omega_*(\bar{\Omega}^2-1)^{3/4}}
         {(\bar{r}-\bar{r}_{\rm ex})^{1/2}}
  {\rm Abs}[\tilde{\sigma}(\Omega;r_{\rm ex})]
  \label{eq:rescA1}\,.
\end{eqnarray}
We have thus been able to absorb much of the dependence on the
scalar mass in terms of a simple rescaling of radius, time, and
frequency. But two issues remain: (i) a factor of $\omega_*$ is
present in the amplitude $A(\bar{t},\bar{r})$, and (ii) the phase
and amplitude implicitly depend on the scalar mass through the phase
and amplitude of the Fourier transform $\tilde{\sigma}(\Omega,r_{\rm ex})$.
Further progress requires information about the signal at $r_{\rm ex}$.
More specifically, we can exploit two features that we find to be
satisfied approximately in the generation of scalar radiation in
stellar collapse in ST theory.

The first observation is that the scalar field at the center of the
star evolves largely independently of the scalar mass.  Likewise,
the scalar profile $\varphi(r)$ at late stages in the evolution is
independent of the scalar mass (always assuming that the other
parameters of the configuration are held fixed).  This suggests
that in the region of wave generation $\sigma(t,r)$ [rather than
$\sigma(\bar{t},\bar{r})]$ is approximately independent of the
scalar mass. Let us take this as a working hypothesis and compute
its implications.

From the definition of the Fourier transform we obtain
\begin{eqnarray}
  \tilde{\sigma}(\Omega;r_{\rm ex}) &=& \int_{-\infty}^{\infty}
  \sigma(t;r_{\rm ex})e^{i\Omega t} dt
  \nonumber \\[5pt]
  &=& \frac{1}{\omega_*}\int_{-\infty}^{\infty}
    \sigma(\bar{t}/\omega_*;r_{\rm ex})e^{i\bar{\Omega}\bar{t}}
    d\bar{t}\,.
    \label{eq:tildesigma1}
\end{eqnarray}
Now we employ the second empirical observation.  Near the star, the
dynamics in the scalar field are dominated by the sudden transition
from weak (or zero) to strong scalarization. The time dependence
of the scalar field at a given radius is therefore approximated by a
Heaviside function, $\sigma(t,r_{\rm ex})\sim f(r_{\rm ex})H(t)$. The Heaviside
function satisfies $H(t)=H(at)$ for a real constant $a$,
and we can use $\sigma(\bar{t}/\omega_*;r_{\rm ex}) =\sigma(\bar{t};r_{\rm ex})$
in Eq.~(\ref{eq:tildesigma1}), so that
\begin{equation}
  \tilde{\sigma}(\Omega;r_{\rm ex})=\frac{1}{\omega_*}
  \int_{-\infty}^{\infty}\sigma(\bar{t};r_{\rm ex})
    e^{i\bar{\Omega}\bar{t}}d\bar{t}
    =\frac{1}{\omega_*}\tilde{\sigma}(\bar{\Omega};r_{\rm ex})\,.
\end{equation}
We thus acquire a factor $1/\omega_*$ in the amplitude of
$\tilde{\sigma}(\Omega;r_{\rm ex})$ and no change in its phase and
Eq.~(\ref{eq:rescA1}) becomes
\begin{eqnarray}
  \bar{\Omega}(\bar{t},\bar{r}) &=&
     \frac{\bar{t}}{\sqrt{\bar{t}^2-(\bar{r}-\bar{r}_{\rm ex})^2}}\,,
     \nonumber \\[5pt]
  \phi(\bar{t},\bar{r}) &=& \sqrt{\bar{\Omega}^2-1}
    (\bar{r}-\bar{r}_{\rm ex})-\bar{\Omega}\bar{t}
    -\frac{\pi}{4}
    +{\rm Arg}[\tilde{\sigma}(\bar{\Omega};r_{\rm ex})]\,,
    \nonumber \\[5pt]
  A(\bar{t},\bar{r})&=&\sqrt{\frac{2}{\pi}}
    \frac{(\bar{\Omega}^2-1)^{3/4}}
         {(\bar{r}-\bar{r}_{\rm ex})^{1/2}}
  {\rm Abs}[\tilde{\sigma}(\bar{\Omega};r_{\rm ex})]
  \label{eq:rescA}\,.
\end{eqnarray}
This gives us a universal expression for the wave signal which
depends on the scalar mass $\omega_*$ only through the rescaling
of time, radius, and frequency according to Eq.~(\ref{eq:resc}).  In
other words, if we know the signal $[\Omega(t,r),~ \phi(t,r),~A(t,r)]$
of a configuration with mass parameter $\omega_{*,1}$, we obtain
the signal for the same configuration in ST theory with $\omega_{*,2}$
by replacing $t\rightarrow \lambda t$, $r\rightarrow \lambda r$,
$\Omega \rightarrow \Omega/\lambda$, $(\phi,A)\rightarrow(\phi,A)$
with $\lambda =\omega_{*,1}/\omega_{*,2}$.

%================================================================
\subsubsection{Results}
The universality under changes in the scalar mass $\omega_*$ will
only hold approximately for a number of reasons: (i) At least at
small radii, the wave propagation will be governed by the field
equations (\ref{eq:Phir})-(\ref{eq:psit}) rather than the Klein-Gordon
equation underlying the calculations of this section. (ii) The time
dependence of the scalar field near the source is only approximately
of Heaviside shape. (iii) Especially for large scalar mass parameters,
we expect the function $\sigma(t,r)$ no longer to be independent
of the value $\omega_*$ as the Compton wavelength approaches the
size of the stellar core. For example, a reduced Compton wavelength
$\lambdabar_{\rm c}<100~{\rm km}$ corresponds to a scalar mass $\mu
>1.97\times 10^{-12}~{\rm eV}$ and frequency $\omega_*>3\,000~{\rm s}^{-1}$.
(Note that such large values of the scalar mass are no
longer ideal for tests with GW observations as the contributions
relevant for LIGO-Virgo partially fall inside the exponentially
suppressed regime $\omega < \omega_*$.)

\begin{figure}
    \includegraphics[width=0.48\textwidth]{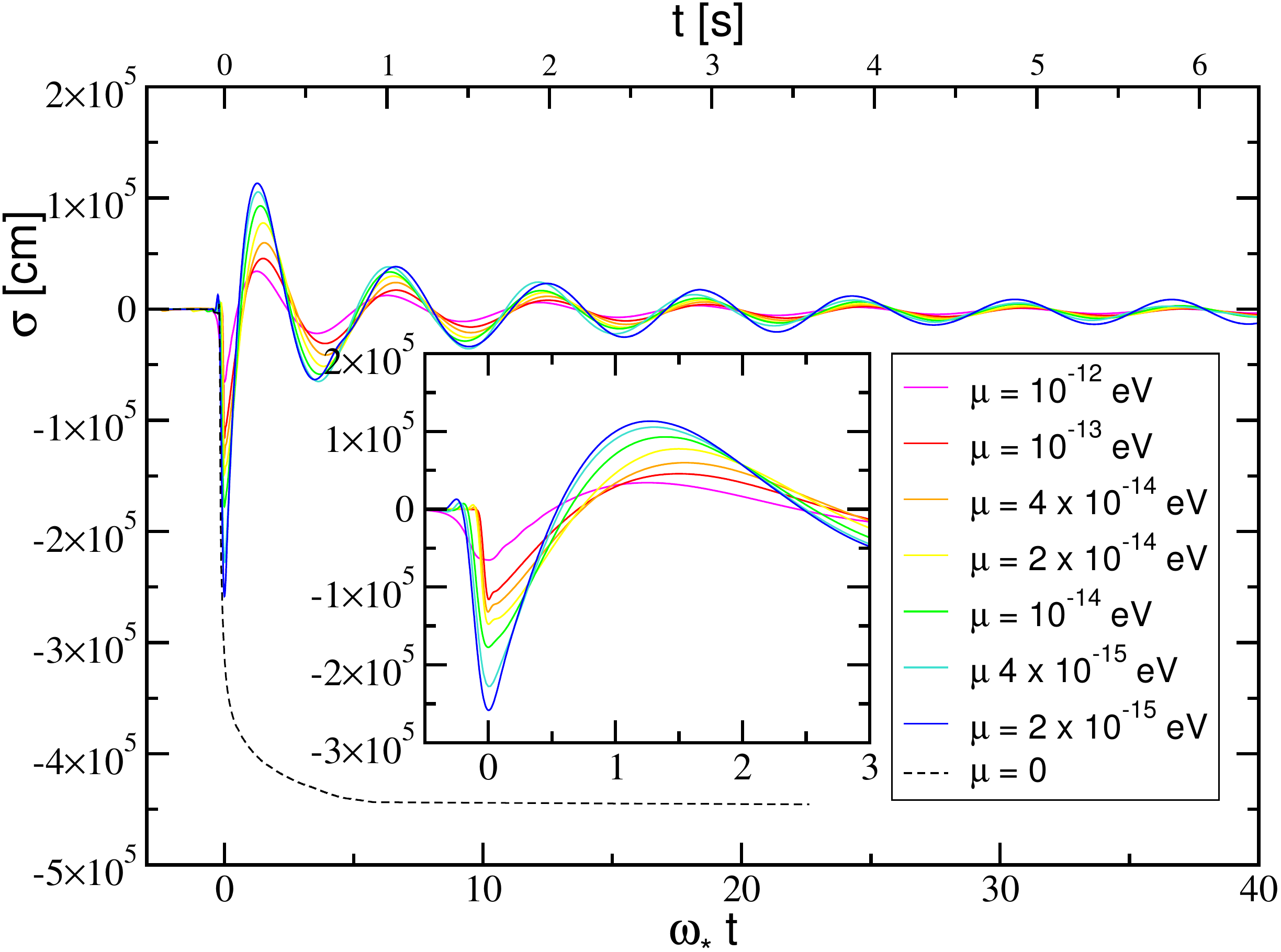}
    \caption{GW signal $\sigma(t,r_{\rm ex})$ extracted from the
             collapse of an s12 (i.e.~$12~M_{\odot}$, solar
             metallicity) progenitor model with $\alpha_0=10^{-2}$,
             $\beta_0=20$, using EOS5
             at $\omega_*
             r_{\rm ex}=5.07$ for different values of the scalar
             mass $\mu\in[2\times 10^{-15} ~\mathrm{eV},~10^{-12}~{\rm
             eV}]$. The overall amplitude increases monotonically
             with decreasing $\mu$. For reference, we also show the
             wave signal obtained for $\mu=0$ (dashed curve).  In
             this case, we cannot rescale the time with $\omega_*$
             and instead measure time in seconds as labeled on the
             upper horizontal axis.
            }
    \label{fig:rphi_a1e02_b20_EOS5}
\end{figure}

\begin{figure}
    \includegraphics[width=0.48\textwidth]{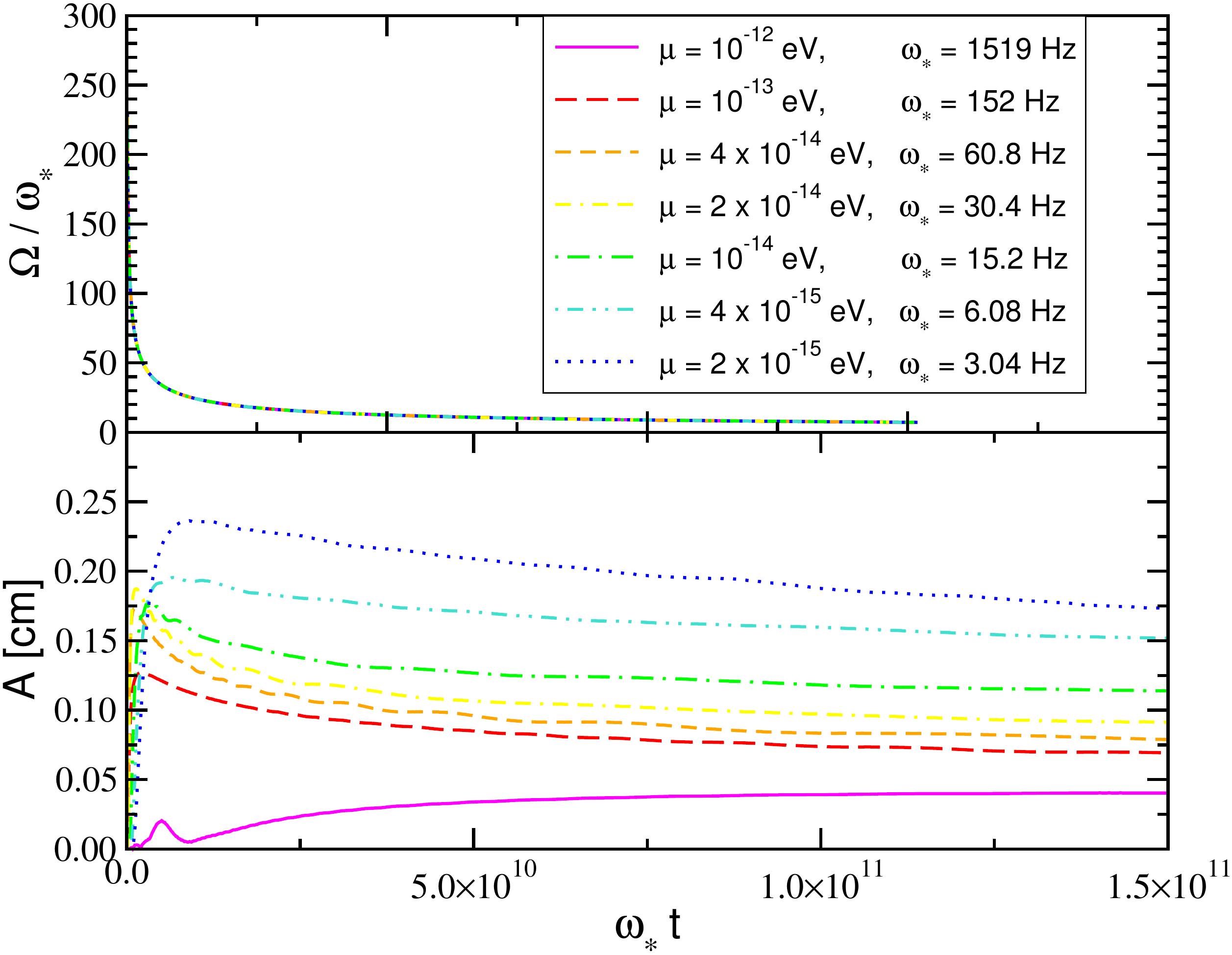}
    \caption{The wave signals of Fig.~\ref{fig:rphi_a1e02_b20_EOS5}
    propagated according to Eqs.~(\ref{eq:omegastar}) and
    (\ref{eq:intstatamp4}) to $\omega_* r =1.56\times 10^{13}
    = (\mu/10^{-14}\,\mathrm{eV})^{-1}~10\,\mathrm{kpc}$.
    As expected, the curves for the rescaled frequency
    $\Omega/\omega_*$ overlap in the upper panel. The amplitude
    in the lower panel shows a mild increase as we decrease
    the scalar mass $\mu$.}
    \label{fig:AOM_a1e02_b20_EOS5}
\end{figure}

So how well is the universality predicted by Eq.~(\ref{eq:rescA})
satisfied in practice?  To address this question we have numerically
explored a range of configurations. For each of these, we have fixed
$\alpha_0$, $\beta_0$, the EOS,
and the progenitor model
and then
performed a one-parameter study varying $\mu$ in the 
range $2\times
10^{-15}~{\rm eV}\le \mu \le 10^{-12}~{\rm eV}$.  All of these cases
exhibit the characteristic behavior we illustrate in
Figs.~\ref{fig:rphi_a1e02_b20_EOS5} and \ref{fig:AOM_a1e02_b20_EOS5}
for the specific case of an s12 progenitor star, EOS5, and ST
parameters $\alpha_0=10^{-2},~\beta_0=-20$.

The wave amplitude $\sigma$ in Fig.~\ref{fig:rphi_a1e02_b20_EOS5}
has been extracted from the core collapse simulations at rescaled
extraction radius
$\bar{r}_{\rm ex}=5.07=(\mu/10^{-14}~{\rm eV})^{-1}\times 10^5~{\rm
km}$.
We have shifted the signals in time
such that their peaks align at $\bar{t}=0$. The main difference of
the signals is a monotonic drop in amplitude as $\mu$ increases;
the strongest signal (for $\mu =2\times10^{-15}~\mathrm{eV}$) exceeds
the weakest one (for $\mu=10^{-12}~\mathrm{eV}$) by a factor of
about 5.  For scalar mass values $\mu < 2\times 10^{-15}~{\rm
eV}$, simulations over several wave cycles become prohibitively
costly (recall that the corresponding physical timescales $\propto
1/\mu$). We have, however, performed short simulations up to the
first strong peak in the signal. This peak, shifted to $\bar{t}=0$
in Fig.~\ref{fig:rphi_a1e02_b20_EOS5}, corresponds to the core
bounce at $t=\mathcal{O}(0.1)~{\rm s}$ and can be computed in shorter
simulations lasting up to about $t\approx r_{\rm ex}$.  We find the
monotonic trend in the amplitude to continue with an upper bound
given by the limiting case $\mu=0$. The wave signal $\sigma(t)$
resulting from this limit can no longer be rescaled according to
Eq.~(\ref{eq:resc}) since $\omega_*=0$; instead, we have included
it in Fig.~\ref{fig:rphi_a1e02_b20_EOS5} (black dashed curve) as a
function of physical time $t$ denoted on the upper horizontal axis.

Amplitude and frequency of the corresponding waveforms propagated
to $\omega_* r=1.56\times 10^{13}=(\mu/10^{-14}\,\mathrm{eV})^{-1}\,10~{\rm
kpc}$ are shown in Fig.~\ref{fig:AOM_a1e02_b20_EOS5}. We find the
same monotonic increase of the wave amplitude as $\mu$ decreases
from $10^{-12}~{\rm eV}$ to $2\times 10^{-15}~{\rm eV}$ with, again,
an overall factor of about 5 between the extreme cases.
We furthermore notice an additional reduction in the
high-frequency contributions for $\mu=10^{-12}\,\mathrm{eV}$
which manifests itself in the reduced signal strength at early 
times in Fig.~\ref{fig:AOM_a1e02_b20_EOS5}.
As expected
from Eq.~(\ref{eq:rescA}), the rescaled frequencies $\bar{\Omega}(\bar{r})$
agree exactly.  We have explored in the same way other configurations
differing from this case in the ST or EOS parameters or the mass
of the stellar progenitor model.  All cases show the same behavior:
the rescaled frequency is independent of the scalar mass $\mu$ when
plotted as a function of rescaled time $\bar{t}$, whereas the amplitude
shows a monotonic increase by an overall factor of about $5$
as $\mu$ decreases from $2\times 10^{-15}~{\rm eV}$ to
$10^{-12}~{\rm eV}$.

Finally, we have explored whether the onset of strong scalarization
as shown in the heat maps in Fig.~\ref{fig:heatmaps} depends on the
scalar mass $\mu$. The answer is {\em no} for all configurations
we have tested; while the degree of strong scalarization mildly
weakens for larger $\mu$, the transition occurs at the same $\beta_0$
independent of the value of $\mu$.

In summary, once we have computed a wave signal from a configuration
for some value of $\mu$, the signal for the (otherwise) identical
configuration with a different scalar mass $\hat{\mu}$ can be
obtained by a linear rescaling of the argument and result of the
frequency $\Omega(t)$ while an approximate estimate of the amplitude
$A(t)$ can be obtained by a rescaling of the time (but not of $A$).
The frequency scaling is exact within the SPA, whereas the amplitude
scaling is approximate to within an order of magnitude and we cannot
rigorously exclude exceptions from its rule.

%=============================================================================
\section{GW Observations} \label{sec:LIGO_obs}
\begin{figure}[th]
        \centering
        \includegraphics[width=\columnwidth]{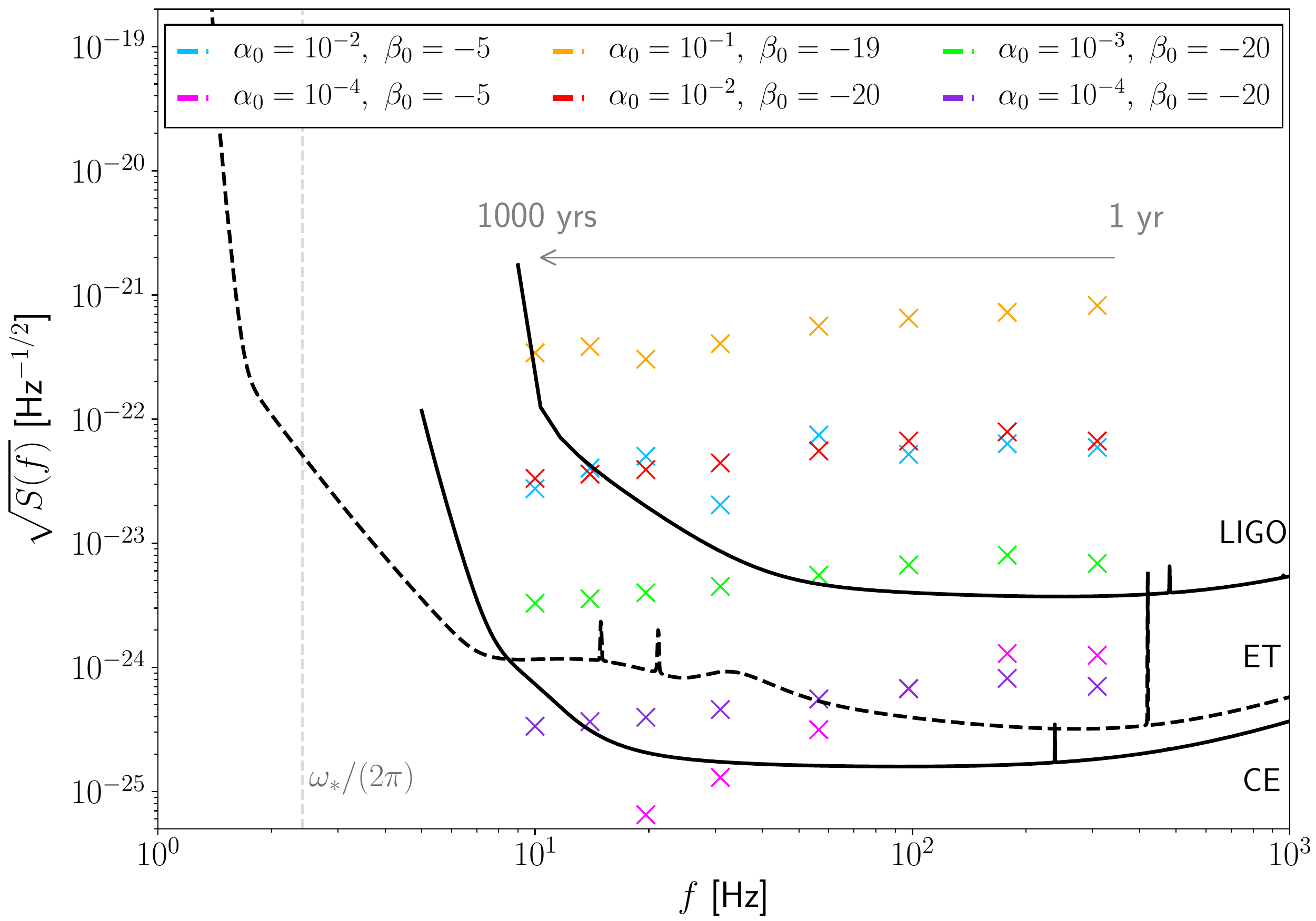}
        \\[10pt]
        \includegraphics[width=\columnwidth]{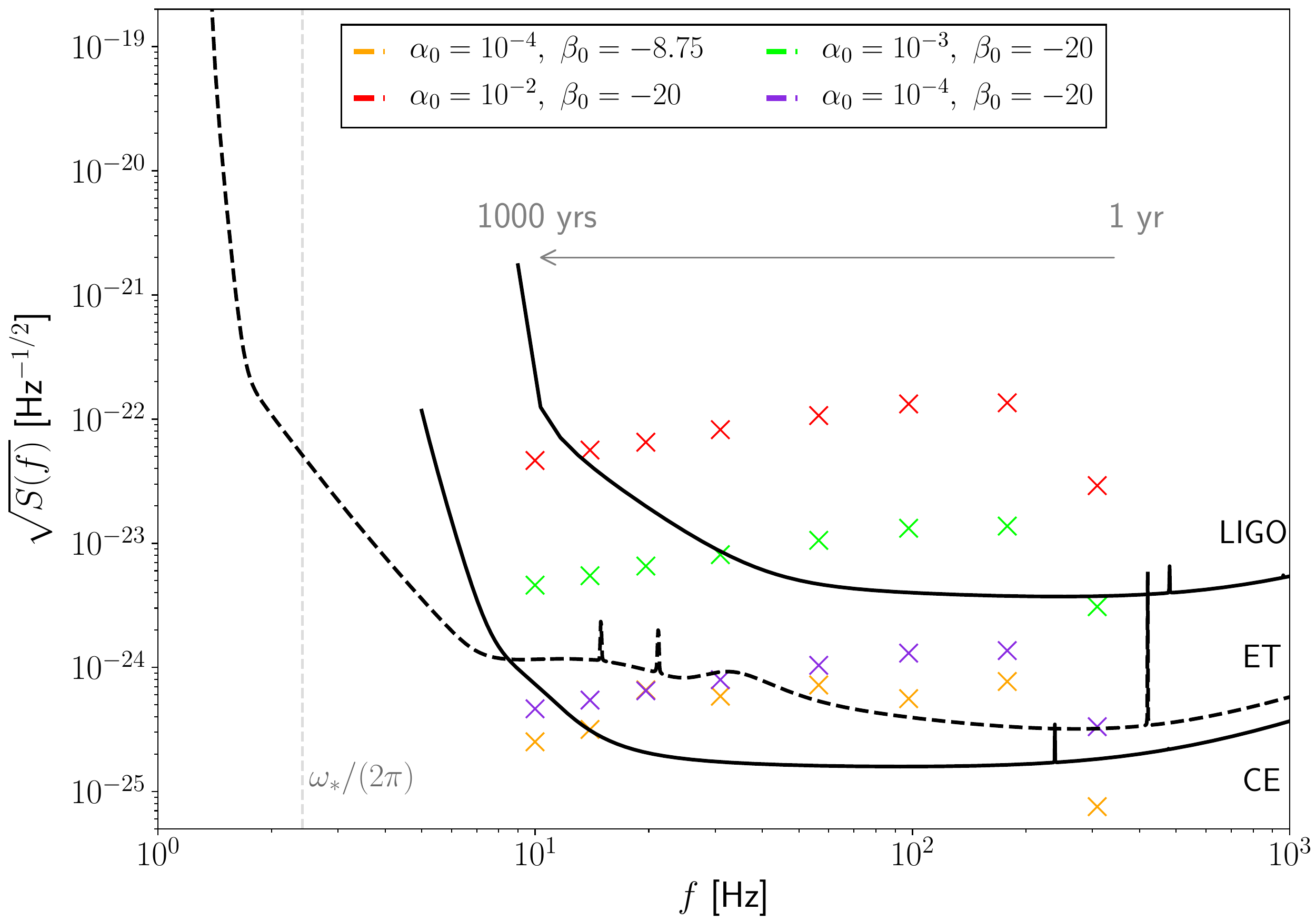}
        \caption{
        Signal amplitudes $\sqrt{S_o}$ for quasimonochromatic GWs emitted during stellar collapse for several $(\alpha_0,\,\beta_0)$ values with $\mu=10^{-14}$ eV using progenitor $s39$ with EOS1 (top) and EOS3 (bottom).
        These are compared against the expected noise curves $\sqrt{S_n(f)}$ of LIGO \cite{Aasi:2013wya}, the Einstein Telescope, and the Cosmic Explorer \cite{Evans:2016mbw}.
        The ratio of $\sqrt{S_o}$ to $\sqrt{S_n(f)}$ gives the SNR which is calculated assuming a 2 month period of observation.
        Results are shown for observations performed different times after the original supernova: $t=1$, 3, 10, 30, 100, 250, 500, and 1000 years increasing from right to left on the plot.
        The signal frequency decreases slowly with $t$ (inverse chirp) while the amplitude remains at the same order of magnitude for up to $t=1000$ years.
        These results were computed for a galactic supernova at a distance $D=10$ kpc from the Earth.
        \label{fig:SNR_s13}
        }
\end{figure}

Core collapse in massive scalar tensor (MST) gravity can lead to the emission of large quantities of scalar radiation which becomes highly stretched out in time, during the dispersive propagation to Earth.
As observed from a detector on Earth, the GW signal is quasimonochromatic with slowly evolving frequency and amplitude given by Eqs.~(\ref{eq:omegastar}) and (\ref{eq:intstatamp4}), respectively.
In this section we discuss the detectability of these signals by ground-based GW detectors such as LIGO \cite{TheLIGOScientific:2014jea} and Virgo \cite{TheVirgo:2014hva}.
This should help guide future efforts to search for such signals thereby testing MST gravity.
Additionally, and as will be shown below, the absence of any current detection may already be sufficient to place more stringent constraints on the parameters of MST gravity than existing techniques.
(For a discussion of existing constraints see, for example, Ref.~\cite{Gerosa:2016fri} and references therein.)
However, a detailed analysis of the constraints implied by existing measurements is deferred to a future study.

Ground-based GW detectors routinely search for quasimonochromatic, continuous GW signals (for a recent review, see \cite{Riles:2017evm}).
The primary motivation for such searches is the possibility of detecting GWs from rapidly rotating, asymmetric neutron stars.
Here we hope to leverage these efforts for another purpose, to test a specific class of modified theories of gravity, namely MST gravity.
Continuous GW searches fall into three broad classes:
(i) \emph{all-sky} searches (see e.g.\ \cite{Pisarski:2019vxw, Abbott:2018bwn, Abbott:2017pqa, Abbott:2018bwn}), (ii) \emph{directed} searches, fixing the sky location to that of a known source (see e.g.\ \cite{Abbott:2017hbu, Abbott:2017mwl, Abbott:2018qee}), and (iii) \emph{targeted} searches fixing the sky location, the frequency, and possibly its time derivative to the corresponding values of a known source (see e.g.\ \cite{Abbott:2017ylp}).
All of (i), (ii), or (iii) can be adapted to search for scalar polarized GWs instead of the usual tensorial polarizations (see e.g.\ \cite{Abbott:2017tlp, Isi:2017equ}).
However, only methods (i) or (ii) can be used for our present purpose; we could either search the whole sky for or target the location of an historical supernova in the hope that the signal has been dispersively stretched to such an extent that it still retains a detectable amplitude.
Method (ii) is computationally cheaper than (i) and can be sensitive to quieter signals, although method (i) has the obvious advantage of covering the whole sky.
As for method (iii), fixing the signal frequency is not applicable here without further theoretical assumptions [this is because the frequency $\Omega(t)$ depends on the unknown mass of the scalar field; see Eq.~(\ref{eq:omegastar})], however, one may instead fix the relation between $\Omega$ and $\dot{\Omega}$ so as to increase the sensitivity of the search.

In this section we calculate the single-detector optimal signal-to-noise ratio (SNR) of our highly dispersed inverse chirp signals.
To estimate the SNR for a network of detectors, the individual SNRs can be added in quadrature.
We point out the significance of a multidetector network for being able to distinguish between the polarizations of a scalar signal and a standard tensorial GW.
The dispersed scalar field signal at the detector is modeled as a simple sine wave,
\begin{equation}
  \varphi(t)=A\sin(\Omega t+\phi_{0})\,.
  \label{eq:varphioft}
\end{equation}
Any evolution in the amplitude and frequency is neglected in our SNR estimates as such changes typically occur on timescales much longer than a typical LIGO/Virgo observation run, and (save for strong resonances in the noise spectrum)  variations of the noise spectral density over a short frequency interval are smaller than temporal variations due to nonstationarity of the instrument.
The scalar field is coupled to the physical metric $g_{\mu\nu}$ via Eq.~(\ref{eq:conformal_metric_Ffactor}); therefore, oscillations in the scalar field source oscillations in $g_{\mu\nu}$, i.e. GWs.
In massless ST theory these GWs are transverse, scalar-polarized, with strain amplitude
\begin{equation}
h_\mathrm{B}(t) = 2\alpha_0 \varphi(t)\,,
\end{equation}
sometimes called a breathing mode.
In MST theory, there is an additional longitudinal polarization with a smaller amplitude, \begin{equation}
h_\mathrm{L}(t) =\left(\frac{\omega_*}{\Omega}\right)^{2}2\alpha_0\varphi(t)\,.
\end{equation}
The response of a GW interferometer is given by \begin{align}
    h(t)=F\big(\theta(t), \phi(t)\big)[h_\mathrm{B}(t)-h_\mathrm{L}(t)] \,,
\end{align}
where $F(\theta,\phi)=\textstyle\frac{-1}{2}\sin^2\theta\cos2\phi$ is the interferometer antenna pattern which depends on the sky location $(\theta,\phi)$ of the source in a coordinate system attached to the detector \cite{PhysRevD.64.062001,Will:2014kxa}.
The antenna pattern is identical (up to a sign) for both polarizations implying that they cannot be distinguished.
As the detector rotates diurnally due to the motion of the Earth, the coordinates $(\theta,\phi)$, and hence the antenna response, change with time.
This periodic dependence of the antenna pattern tends to have an averaging effect; sometimes the source is in a favorable location while later it may cross a zero in the antenna pattern.
Therefore, for our simple SNR estimates we use the constant, sky averaged rms value for the antenna pattern,
\begin{align}
    \bar{F}=\sqrt{\iint\mathrm{d}\theta\,\mathrm{d}\phi\;\sin\theta\, F^{2}(\theta,\phi)}=\sqrt{4\pi/15}\,.
    \label{eq:Faverage}
\end{align}
Combining Eqs.~(\ref{eq:varphioft})-(\ref{eq:Faverage}),
%With these assumptions,
the effective strain $h(t)$ appearing in the interferometer's output is given by
\begin{align}
    h(t)=2A\alpha_0\bar{F}[1-(\omega_*/\Omega)^{2}]\sin(\Omega t+\phi)\,.
\end{align}
Here we neglect any Doppler shift in the source frequency caused by the motion of the Earth as this has a negligible effect on the SNR.

The noise in the instrument (commonly assumed to be stationary and Gaussian) is described by the (one-sided) noise power spectral density $S_n(f)$.
The optimal SNR $\rho$ is defined in the Fourier domain by the following integral over frequency $f$ \cite{Moore:2014lga}:
\begin{align} \label{eq:SNR_1_x}
    \rho^2 = 4\int_{0}^{\infty}\mathrm{d}f\;\frac{|\tilde{h}(f)|^2}{S_{n}(f)} \,.
\end{align}
For an (approximately) sinusoidal signal $h(t)$, the integrand
in this equation has support only at $f=\Omega/(2\pi)$,
so that the denominator can be pulled out of the integral
as a constant $S_n(\Omega/(2\pi)$.
In the limit $T\gg 1/\Omega$, the integral in Eq.~(\ref{eq:SNR_1_x}) can be approximated by a time domain integral (using Parseval's theorem) and evaluated to give
\begin{align} \label{eq:SNR_eqreq}
    \rho \!\approx \!\sqrt{\frac{S_o}{S_n(\frac{\Omega}{2\pi})}}, \,~\mathrm{where}~ \, S_o\! =\! T(A\alpha_0 \bar{F})^{2}\left[1-\left(\frac{\omega_*}{\Omega}\right)^{2}\right]^{2}\!.
\end{align}

In Fig.~\ref{fig:SNR_s13} we plot the quantity $\sqrt{S_o}$
(cross symbols) at specific frequencies as a measure of the signal amplitude for two months of observation and the quantity $\sqrt{S_n(f)}$
(solid curves)
as a measure of the instrumental noise; the height of the cross above the curve gives a visual measure of the SNR
[cf. Eq.~(\ref{eq:SNR_eqreq}) with $f=\Omega/(2\pi)$].
For each simulation a sequence of crosses are plotted corresponding to the same source observed at different (retarded) times $t$ after the original supernova; results are shown for $t=$1, 3, 10, 30, 100, 250, 500, and 1000 years.
Results are shown for several core collapse simulations using the $s39$ progenitor for different values of the MST theory parameters $\alpha_0$ and $\beta_0$ and for two different choices for the equation of state (EOS1 and EOS3).
The general trend is that as time passes the frequency slowly decreases following the inverse chirp formula in Eq.~(\ref{eq:omegastar}) while the amplitude can remain at the same order of magnitude for a very long time after the original supernova.
This trend is extremely robust to changes in the properties of the progenitor star; additional results for the progenitors $u39$ and $z39$ (both with EOS1 and EOS3) are shown in Appendix~\ref{app:App_SNR_estimates_LIGO}.

The results in Fig.~\ref{fig:SNR_s13} and Appendix~\ref{app:App_SNR_estimates_LIGO} show that if, for example, $(\alpha_0,\beta_0)=(10^{-2}, -20)$, then with the current LIGO capabilities a galactic supernova at $D=10\,{\rm kpc}$ could have a SNR of $\rho\sim 30$ at $\sim 200\,\mathrm{Hz}$ in 2 months of observation if observed $t=3$ years after core collapse.
Furthermore, such a source remains detectable in LIGO continuous wave searches for $t\sim 300$ years after the original supernova.
With the Einstein Telescope or Cosmic Explorer some signals may reach SNRs of $\sim 1000$ in just 2 months of observation and remain observable for up to 1000 years after the original supernova.
Note that the SNR scales with the duration of observation as $\sqrt{T}$ and with distance to the source as $1/D$.

These results are obviously promising for the prospects of making a detection or constraining $\alpha_0$, $\beta_0$, and $\mu$.
Because the signals remain detectable for such a long time, it will be worthwhile carrying out directed searches for continuous, scalar-polarized, inverse-chirp signals at the locations of historical supernovae.
If such searches yielded no detection, it seems likely that this could be used to place the tightest current constraints on the ($\alpha_0$, $\beta_0$, $\mu$) parameter space of MST gravity.
Supernova 1987A in the large Magellanic cloud is an example of a recent, nearby core-collapse supernova.
A detailed projection of the possible constraints are complicated by the $\mu$ dependence of the inverse-chirp profile in Eq.~(\ref{eq:omegastar}); we defer a careful analysis of this question to a future study.
\medskip
%
%================================================================
\section{Conclusions}
\label{concl}

We have performed the first extensive study of spherically symmetric
core collapse in MST
theory in which we cover a wide range of
equations of state and progenitor models, as well as a vast section
of the scalar parameter space centered around the threshold
for hyperscalarization. A stronger scalar field delays gravitational
collapse to the point of impeding BH formation.

For mildly negative values of the quadratic coefficient $\beta_0$
in the conformal factor, we recover the two well-known collapse
scenarios in GR, the formation of a NS and the
formation of a BH resulting from continued accretion
onto a proto-NS.  For sufficiently negative values of $\beta_0$,
we encounter three collapse scenarios qualitatively different from those in GR: the formation of a BH following multiple NS stages,
the multi-stage formation of a strongly scalarized NS, and the single-stage
formation of a strongly scalarized NS.

The fate of a progenitor (with a fixed equation
of state) in GR dictates the distribution of these five collapse
scenarios as
we vary the scalar parameters. As we change $\beta_0$ from
zero toward negative values,
only two possible successions of collapse scenarios are possible.
The first sequence is the following: two-stage BH formation,
multi-stage BH formation, multi-stage formation of a strongly
scalarized NS, single-stage formation of a strongly scalarized NS.
The second sequence is
single-stage formation of a low-compactness weakly scalarized NS,
multi-stage formation of a strongly scalarized NS, single-stage formation
of a strongly scalarized NS. The boundaries between the different
classes can vary with the equation of state, the metallicity, or the
mass of the progenitor, but for every progenitor we encounter
either one or the other sequence, depending on whether the star forms
a BH or a NS in GR.

The different scenarios are reflected in the scalar field (which
mirrors the matter density evolution) and, as a consequence, in the
scalar radiation. The scalar mass causes the GW signal to disperse
as it propagates, and by the time it would reach a detector the
signal will retain little information with regard to its source,
but it carries a highly characteristic imprint of the MST theory.
Over a wide range of MST parameters, we find that the resulting
gravitational-wave signals will be strong enough to
reach SNRs $\gtrsim 20$ over long periods of time, even up to
several centuries. This implies potential detection through the
study of historical supernovae or, through nondetection,
the most stringent constraints
on the $\left(\alpha_0,\beta_0\right)$ parameter space of MST theory.

\medskip
\begin{acknowledgments}
We made use of presupernova models by S. Woosley and A. Heger available at \href{https://2sn.org/stellarevolution/}
{2sn.org/stellarevolution}.
We thank Max Isi for helpful comments through the internal
LIGO-Virgo review.
U.S. is supported by the European Union’s H2020 ERC Consolidator
Grant “Matter and strong-field gravity: New frontiers in Einstein’s
theory” Grant No. MaGRaTh–646597 and the STFC Consolidator Grant
No. ST/P000673/1.
M.A. is supported by the Kavli Foundation.
D.G. is supported by Leverhulme Trust Grant No. RPG-2019-350.
%.
This work was supported by the GWverse COST Action Grant No. CA16104,
“Black holes, gravitational waves and fundamental physics.”
Computational work was performed on the SDSC Comet and TACC Stampede2
clusters through NSF-XSEDE Grant No. PHY-090003; the Cambridge CSD3
system through STFC capital Grants No. ST/P002307/1 and No.
ST/R002452/1, and STFC operations Grant No. ST/R00689X/1; the
University of Birmingham BlueBEAR cluster; the Athena cluster at
HPC Midlands+ funded by EPSRC Grant No.~EP/P020232/1; and the
Maryland Advanced Research Computing Center (MARCC).

\end{acknowledgments}

%=============================================================================
\bibliographystyle{apsrev4-1}
\bibliography{bibliography}

\clearpage
\appendix

%=============================================================================
\section{Collapse scenarios}
\label{app:scenarios}
In this appendix, we discuss in more detail the five qualitatively
different collapse scenarios listed in Sec.~\ref{sec:classification}
by analyzing for each case a prototypical example.
For all configurations discussed in this section, we use a scalar
mass $\mu=10^{-14}\,{\rm eV}$.

%=============================================================================
\subsection{Single-stage collapse to a weakly scalarized neutron star}
The formation of a weakly scalarized neutron star is the scenario
realized for weakly (or non-) negative values of $\beta_0$ and for
equations of state and progenitor models that result in a neutron
star in GR. The dynamics of this scenario barely differ from the
corresponding collapse in GR and result in a weak GW signal
as long as $\alpha_0 \ll 1$.
\begin{figure*}[p]
  \includegraphics[width=0.87\textwidth]{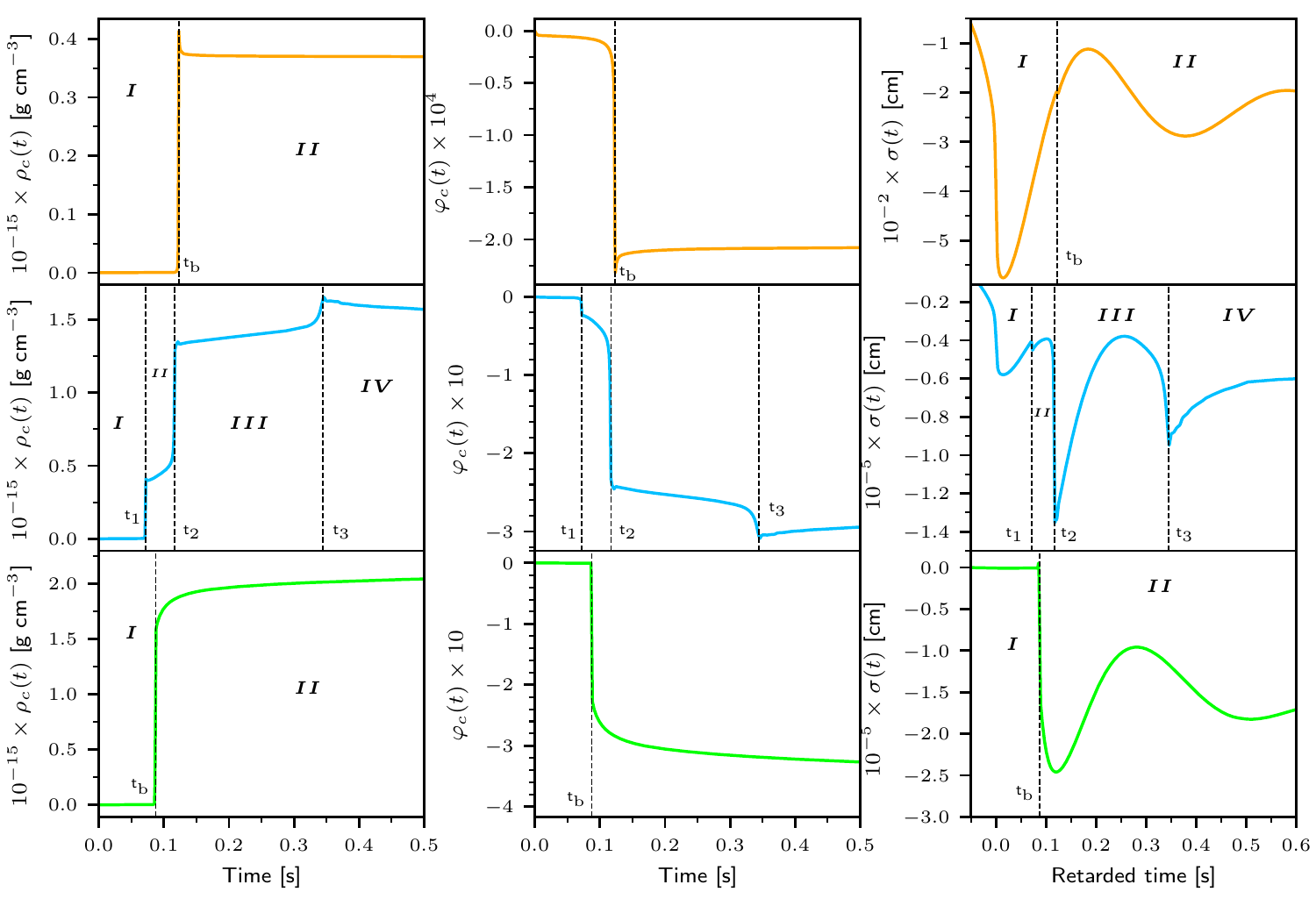}
  \caption{
           The central baryon density $\rho_c$ (left), the central scalar field
           value $\varphi_c$ (middle), and the wave signal
           $\sigma=r_{\rm ex}\varphi$ extracted at
           $r_{\rm ex}=3\times 10^4\,{\rm km}$ (right column) are shown
           as a function of time for three configurations.
           {\em Top}: Progenitor s39 with EOS3 and scalar parameters
           $\alpha_0=10^{-3},~\beta_0=-2$ promptly forms a weakly
           scalarized NS. {\em Center}: Progenitor s39 with EOS1
           and $\alpha_0=10^{-1},~\beta_0=-7$ undergoes a multi-stage
           collapse to a strongly scalarized NS. {\em Bottom}:
           Progenitor z39 with EOS1 and $\alpha_0=10^{-3},~\beta_0=-20$
           promptly collapses into a strongly scalarized NS. The Roman
           numerals label separate stages in the time evolution.
          }
  \label{fig:plotNS}
\bigskip
\bigskip
  \includegraphics[width=0.87\textwidth]{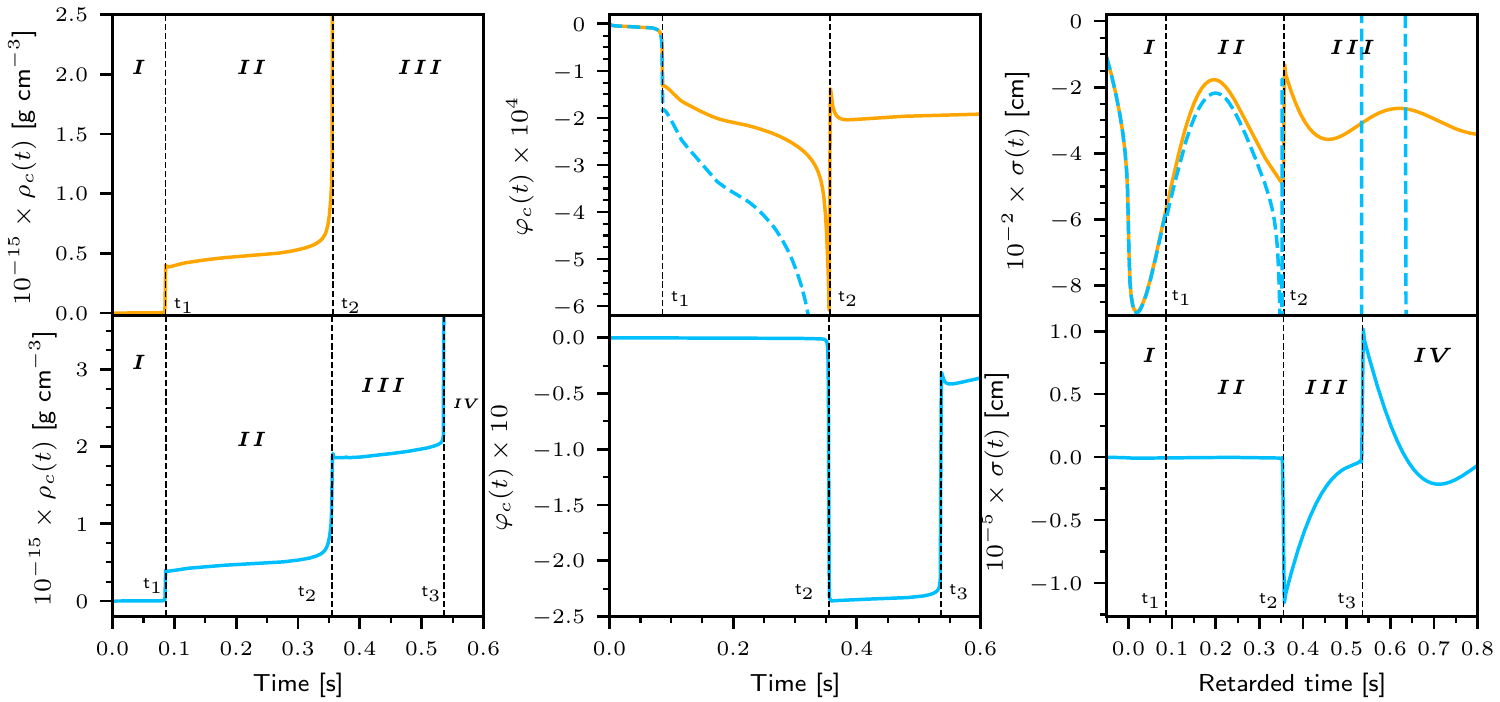}
  \caption{The central baryon density $\rho_c$ (left), the central scalar field
           value $\varphi_c$ (middle), and the wave signal
           $\sigma=r_{\rm ex}\varphi$ extracted at
           $r_{\rm ex}=3\times 10^4\,{\rm km}$ (right column) are shown
           as a function of time for two configurations.
           {\em Top}: The progenitor u39 with EOS1 and scalar parameters
           $\alpha_0=10^{-3},~\beta_0=-2$ temporarily forms a weakly
           scalarized NS before it collapses to BH at $t\approx 0.35\,{\rm s}$.
           {\em Bottom}: The same configuration with $\beta_0=-5$ also
           collapses into a BH eventually, but not before briefly
           settling down into a strongly scalarized NS phase between
           $t\approx 0.35\,{\rm s}$ and $t\approx 0.54\,{\rm s}$.
           For comparison, the scalar field and wave signal of the
           second configuration are also displayed as dashed lines
           in the upper panels. Note how the additional strongly
           scalarized NS stage leads to an increase in the GW signal
           by several orders of magnitude.
          }
  \label{fig:plotBH}
\end{figure*}

\begin{figure*} [hb!]
	\centering
  \includegraphics[width=0.9\textwidth]{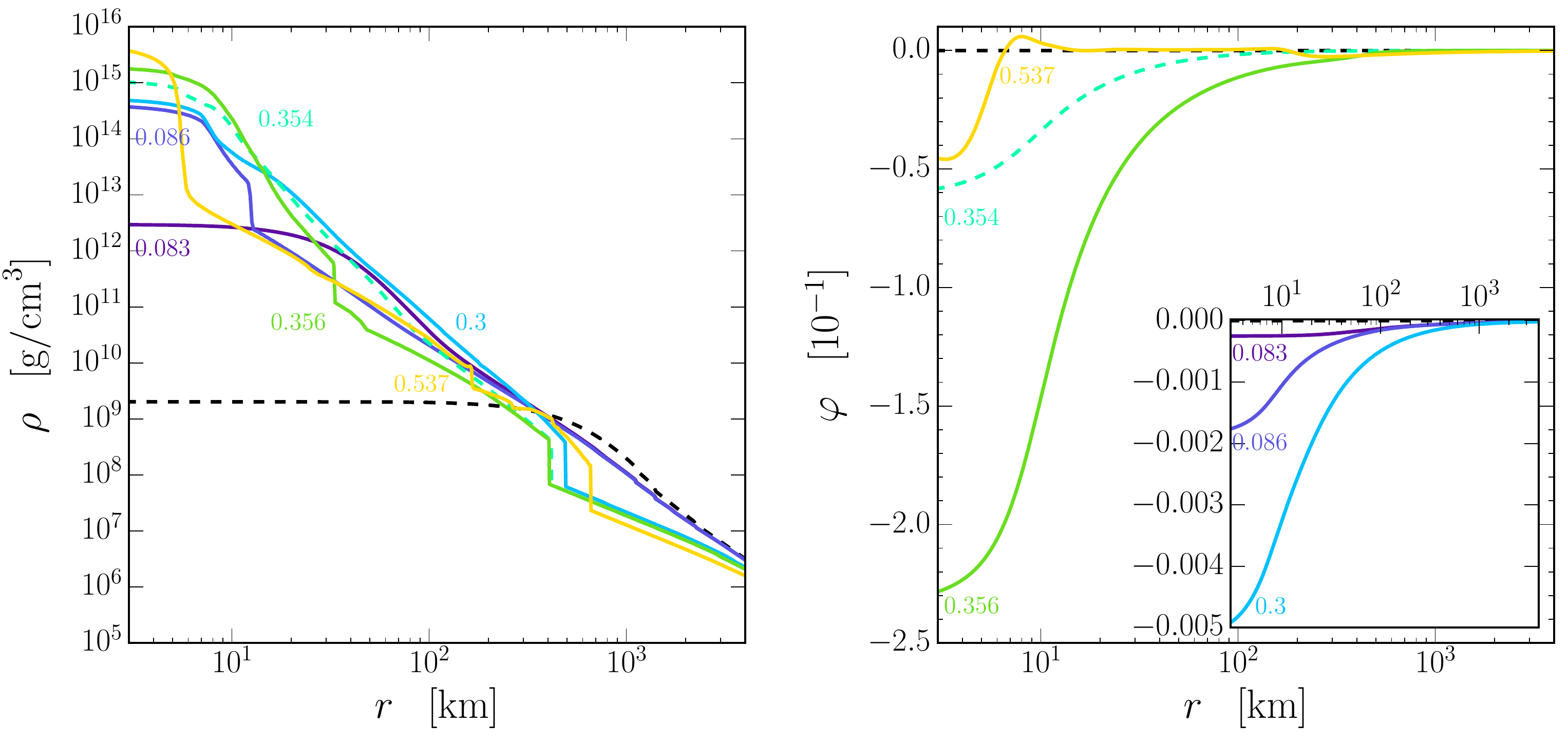}
  \caption{Snapshots of the baryon density and scalar field profiles
           in the collapse of the progenitor u39 with EOS1 and
           ST parameters $\alpha_0=10^{-3},~\beta_0=-5$. The dashed
           curves show the initial data. A first outgoing shock results
           from the core bounce at $t\approx 0.086\,{\rm s}$.
           The second contraction leads to a second core bounce
           at $t\approx 0.356\,{\rm s}$, and this time the scalar
           field also increases in amplitude (left panel), signaling
           the temporary formation of a strongly scalarized NS.
           At $t\approx 0.537\,{\rm s}$, the baryon density once again
           starts increasing sharply, this time leading to the formation
           of a BH and the corresponding descalarization.
          }
  \label{fig:snapshots_msBH}
\bigskip \bigskip
  \includegraphics[width=.9\textwidth]{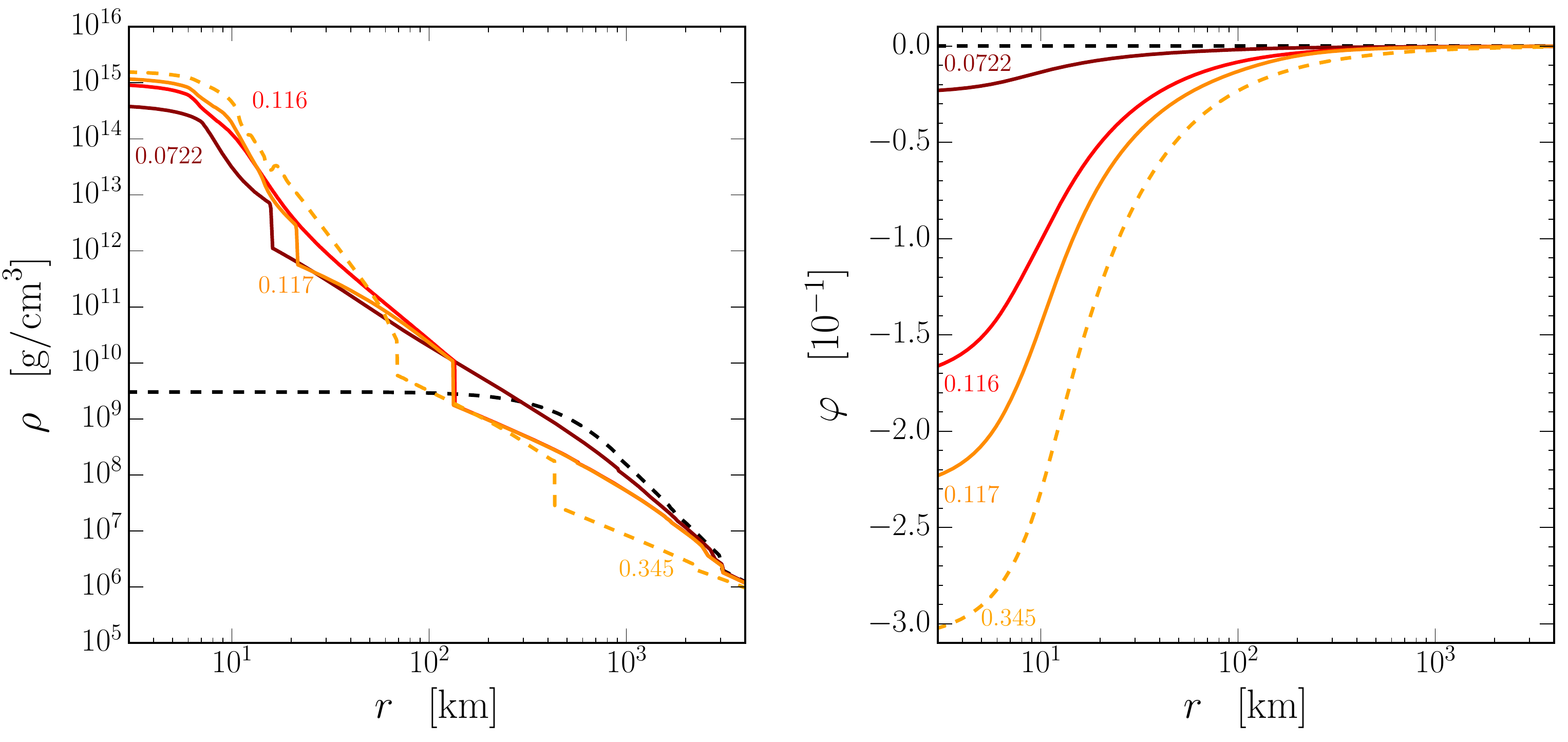}
  \caption{Snapshots of the baryon density and scalar field profiles
           in the collapse of the progenitor s39 with EOS1 and
           ST parameters $\alpha_0=10^{-1},~\beta_0=-7$. The dashed
           curves show the initial data. A first outgoing shock results
           from the core bounce at $t\approx 0.072\,{\rm s}$.
           The second contraction leads to a second core bounce
           at $t\approx 0.116\,{\rm s}$, and this time the scalar
           field also increases in amplitude (left panel), signaling
           the temporary formation of a strongly scalarized NS.
           At $t\approx 0.345\,{\rm s}$, both the baryon density
           and the scalar field amplitude once again
           jump, but by lesser margins. Close inspection of the data
           shows a mild shock that is barely perceptible in the
           density profile at $t=0.345\,{\rm s}$ around
           $r\approx 15\,{\rm km}$.
          }
  \label{fig:snapshots_msNS}
\end{figure*}

\begin{figure*} [t!]
	\centering
   \includegraphics[width=.9\textwidth]{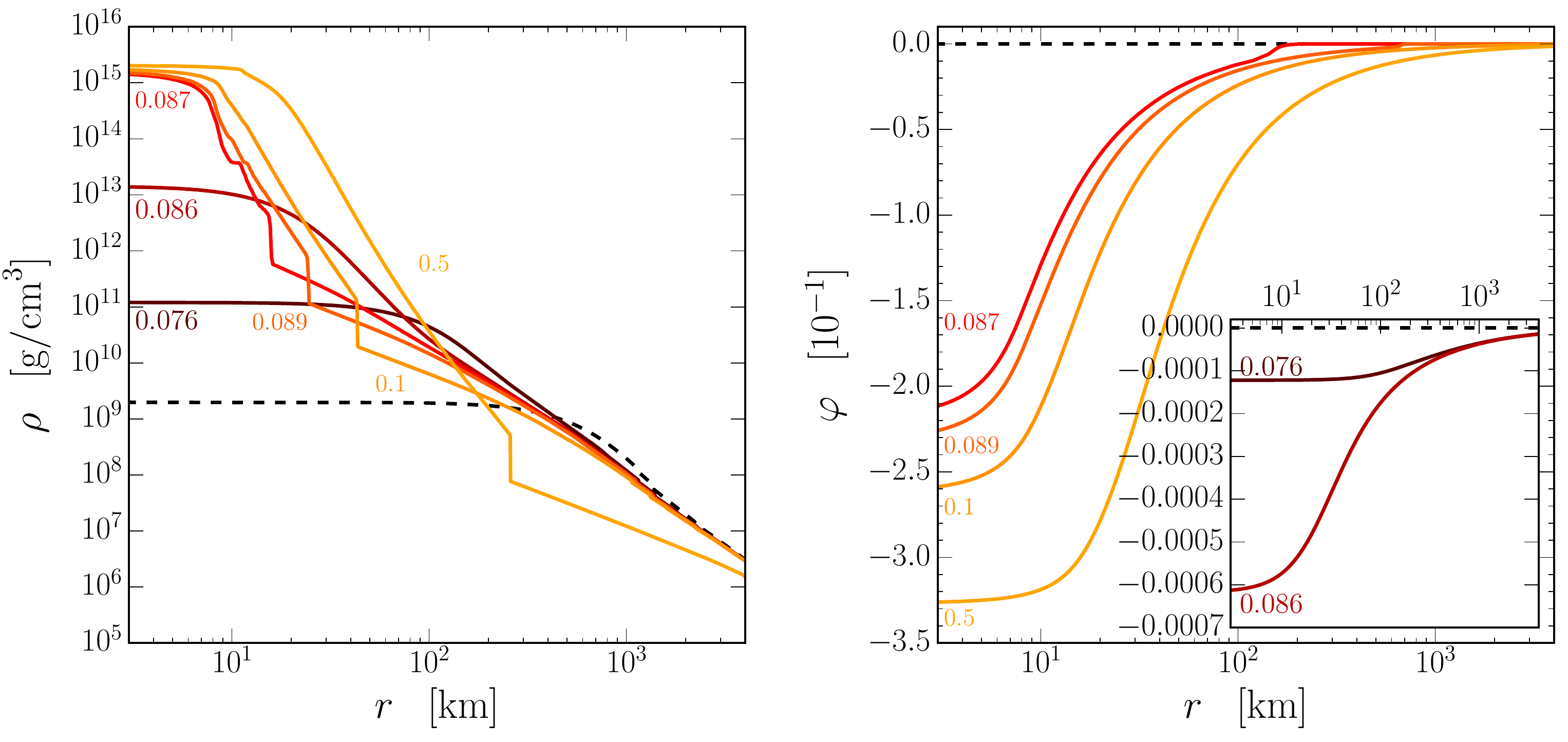}
	\caption[Snapshots of the baryon density and scalar field profiles during the simulation of a single-stage high-compactness NS.]{Snapshots of the baryon density (left) and scalar field (right) profiles during the simulation of a single-stage high-compactness NS. The progenitor is $z39$ with EOS1 and the scalar parameters are $\alpha_0=10^{-3}$, $\beta_0=-20$, and $\mu=10^{-14}$ eV. The dotted black line represents the initial profile.
	In this case, core bounce occurs at $t\approx 0.087\,{\rm s}$
    which leads to a shock propagating outwards. Over the remaining
    duration of the simulation no further shocks appear and the
    central density barely changes.}
	\label{fig_frame_prompt_high}
\end{figure*}

As an example, we plot in the top row of Fig.~\ref{fig:plotNS}
as functions of time the central
baryon density $\rho_c$, the central scalar field value $\varphi_c$,
and the wave signal $\sigma=r_{\rm ex}\varphi$ at $r_{\rm ex}
=3\times 10^4\,{\rm km}$ for the collapse of an s39 progenitor
with EOS3 and ST parameters $\alpha_0=10^{-3}$, $\beta_0=-2$.
This example displays all the characteristics we observe in
configurations collapsing in a single stage into a weakly scalarized
NS. The central density abruptly increases in one jump up to
a few times $10^{14}\,{\rm g/cm}^3$. For nonzero $\alpha_0$
the jump in density is accompanied by
a sudden change in the central scalar field away from zero, but
the scalar field only reaches an amplitude $\varphi_c = \mathcal{O}(\alpha)$;
cf.~the top center panel in Fig.~\ref{fig:plotNS}.
This weak scalarization leads to a correspondingly weak GW signal
as shown in the top right panel of the figure.

%=============================================================================
\subsection{Two-stage formation of a black hole}
In the GR limit, a larger ZAMS mass, a lower metallicity, or a softer equation
of state may result in the formation of a BH instead of a NS. For
non-negative or mildly negative values of $\beta_0$, this occurs in two
stages; the configuration briefly settles down into a weakly scalarized
NS before a second contraction phase results in the final BH (as in GR). In the top
row of Fig.~\ref{fig:plotBH}, we show as an example the progenitor
u39 with EOS1 and ST parameters $\alpha_0=10^{-3},\,\beta_0=-2$.
The upper left panel illustrates that the central density first
jumps to nuclear values $\mathcal{O}(10^{14})\,{\rm g/cm}^3$
and briefly levels off before a second jump signals the formation of a BH
at $t\approx 0.35\,{\rm s}$. The first contraction phase only leads
to a weak scalarization and a correspondingly weak GW signal in the
center and right panels. The scalarization in the second contraction
phase is more complicated; as the stellar compactness increases, the
scalar field rapidly strengthens. This increase is halted, however,
once a horizon forms and the BH descalarizes in accordance with
the no-hair theorems. The maximal degree of the scalarization critically
depends on how rapidly a BH forms and, thus, exhibits sensitive
dependence on the configuration's parameters. In our set of simulations,
we have found that all degrees from weak to strong scalarization
and GW emission are possible in the two-stage BH formation category and
that even tiny changes in a parameter can drastically modify the
ensuing GW signal; see, for example, the right panel in
Fig.~\ref{fig:heatmaps} where the dark ``two-stage BH'' region
in the $(\alpha_0,\beta_0)$ plane of the bottom plot covers the
entire range of scalarization displayed in the center plot.
Among the five qualitatively different collapse
scenarios, the two-stage BH formation is the only one that exhibits
such a sensitive dependence on the parameters.

%=============================================================================
\subsection{Multi-stage collapse to a black hole}
This scenario also leads to the formation of a BH, but the collapsing
star settles down into at least two temporarily stationary neutron-star
configurations with increasing central density. Furthermore, all but the
first neutron-star stages are strongly scalarized, so that this
scenario always generates a strong GW signal. As an example,
we show in the bottom row of Fig.~\ref{fig:plotBH}
for a progenitor u39 with EOS1 and ST parameters
$\alpha_0=10^{-3},~\beta_0=-5$ the central density
$\rho_c$, the central scalar field $\varphi_c$, and the
wave signal $\sigma=r_{\rm ex}\varphi$ at $r_{\rm ex}=3\times 10^4\,{\rm km}$
as functions of time. Note the similarity at early times to the
otherwise identical configuration with $\beta_0=-2$ shown in the upper
panel of the same figure. The key difference is that the second
contraction phase around $t\approx 0.35\,{\rm s}$ promptly results
in a BH if $\beta_0=-2$ but leads to an intermittent strongly scalarized
NS phase if $\beta_0=-5$.

In Fig.~\ref{fig:snapshots_msBH} we show snapshots of the radial profiles
of the baryon density $\rho_c$ and the scalar field $\varphi$ for this
model with $\beta_0=-5$. Each contraction to a temporarily NS stage
is accompanied by the formation of an outgoing shock through core
bounce; these are visible at times $t\approx 0.086\,{\rm s}$
and $t\approx 0.356\,{\rm s}$ in the profiles $\rho(r)$ in the
left panel of the figure. The first NS is weakly scalarized,
and we only see a significant increase in the scalar field amplitude
in the right panel
following the second contraction phase at $t\approx 0.356\,{\rm s}$.
The third and final contraction at $t\approx 0.537\,{\rm s}$
leads to a BH and, in accordance with the no-hair theorems,
the descalarization of the compact star. This strong scalarization and
ensuing descalarization results in the two peaks in the GW signal
of this configuration in the bottom right panel
of Fig.~\ref{fig:plotBH}.

%=============================================================================
\subsection{Multi-stage collapse to a neutron star}
This scenario resembles in many ways the multi-stage formation of a
BH discussed in the preceding subsection. Again, we observe a
first contraction phase resulting in a weakly scalarized NS followed
by one or more further contraction stages. The key difference is that
the end product is a highly compact, strongly scalarized
NS rather than a BH. An example of this scenario is given by the
collapse of the s39 progenitor with EOS1 and ST parameters
$\alpha_0=10^{-1},~\beta_0=-7$ in the center row of Fig.~\ref{fig:plotNS}.
This configuration reveals three contraction phases that are
also visible in the snapshots of the radial profiles of
the baryon density $\rho$ and the scalar field $\varphi$ in
Fig.~\ref{fig:snapshots_msNS}.
Again, we observe each contraction phase to result in a core bounce
and an outgoing shock visible in the left panel
of Fig.~\ref{fig:snapshots_msNS}: The first shock forms at
$t\approx 0.072\,{\rm s}$, the second at $t\approx 0.117\,{\rm s}$,
while the third discontinuity is weak and barely visible at
$t=0.345\,{\rm s}$ around $r\approx 15\,{\rm km}$. As in the case
of a multi-stage BH formation, the significant jumps in the scalar
field may result in multiple peaks in the wave signal
as shown in the center-right panel of Fig.~\ref{fig:plotNS}.

%=============================================================================
\subsection{Single-stage collapse to a strongly scalarized neutron star}
The single-stage formation of a strongly scalarized NS can be regarded as
the limit of the preceding multi-stage NS formation with the duration
of all intermediate quasistationary NS configurations shrinking to
zero. This is indeed what is observed if we start with a given
multi-stage NS model, such as the one discussed in the previous
subsection, and then amplify $\beta_0$ to increasingly negative values;
the lifetime of the intermittent stages decreases, and we approach a
single contraction phase to a strongly scalarized NS. Over the parameter
range we have considered, this scenario ubiquitously represents
the limiting scenario for highly negative values of $\beta_0$;
cf.~Fig.~\ref{fig:heatmaps}. The wave signal always consists of a single
strong peak for these configurations.

%=============================================================================

\section{Additional classification for several equations of state and progenitor models}\label{appeos}

\begin{figure*}
  \includegraphics[width=0.49\textwidth,clip=true]{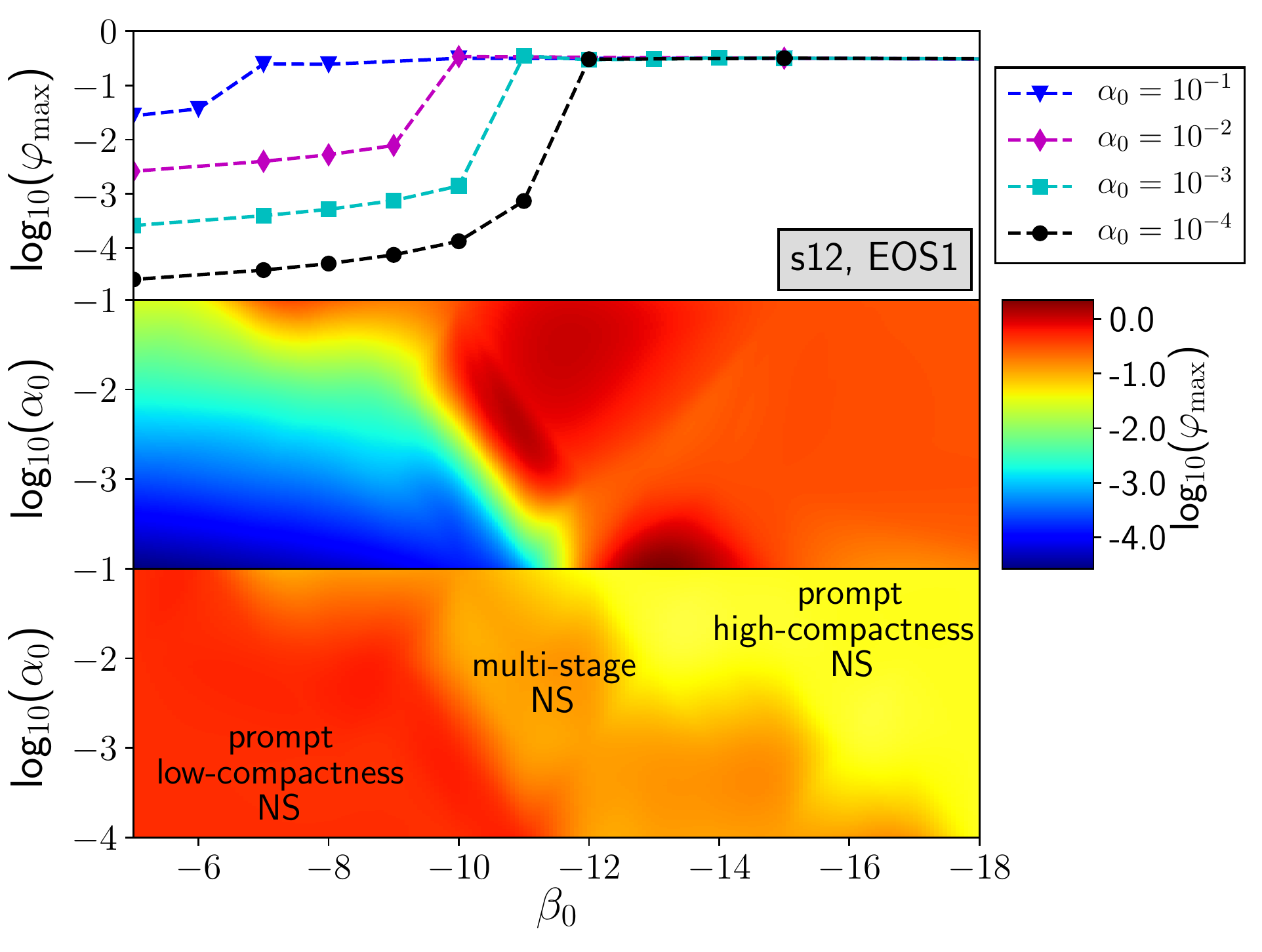}
  \includegraphics[width=0.49\textwidth,clip=true]{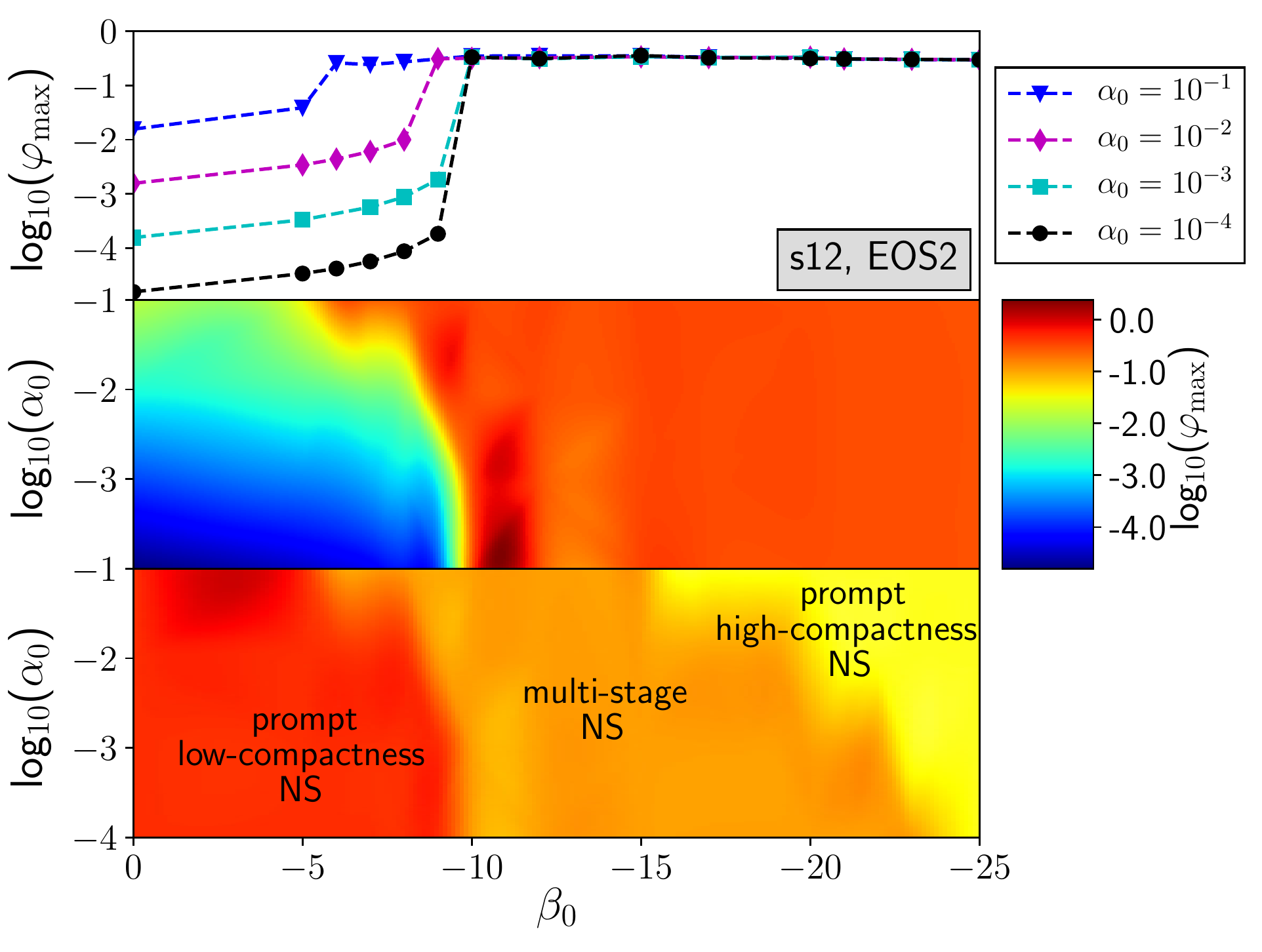}\\[10pt]
  \includegraphics[width=0.49\textwidth,clip=true]{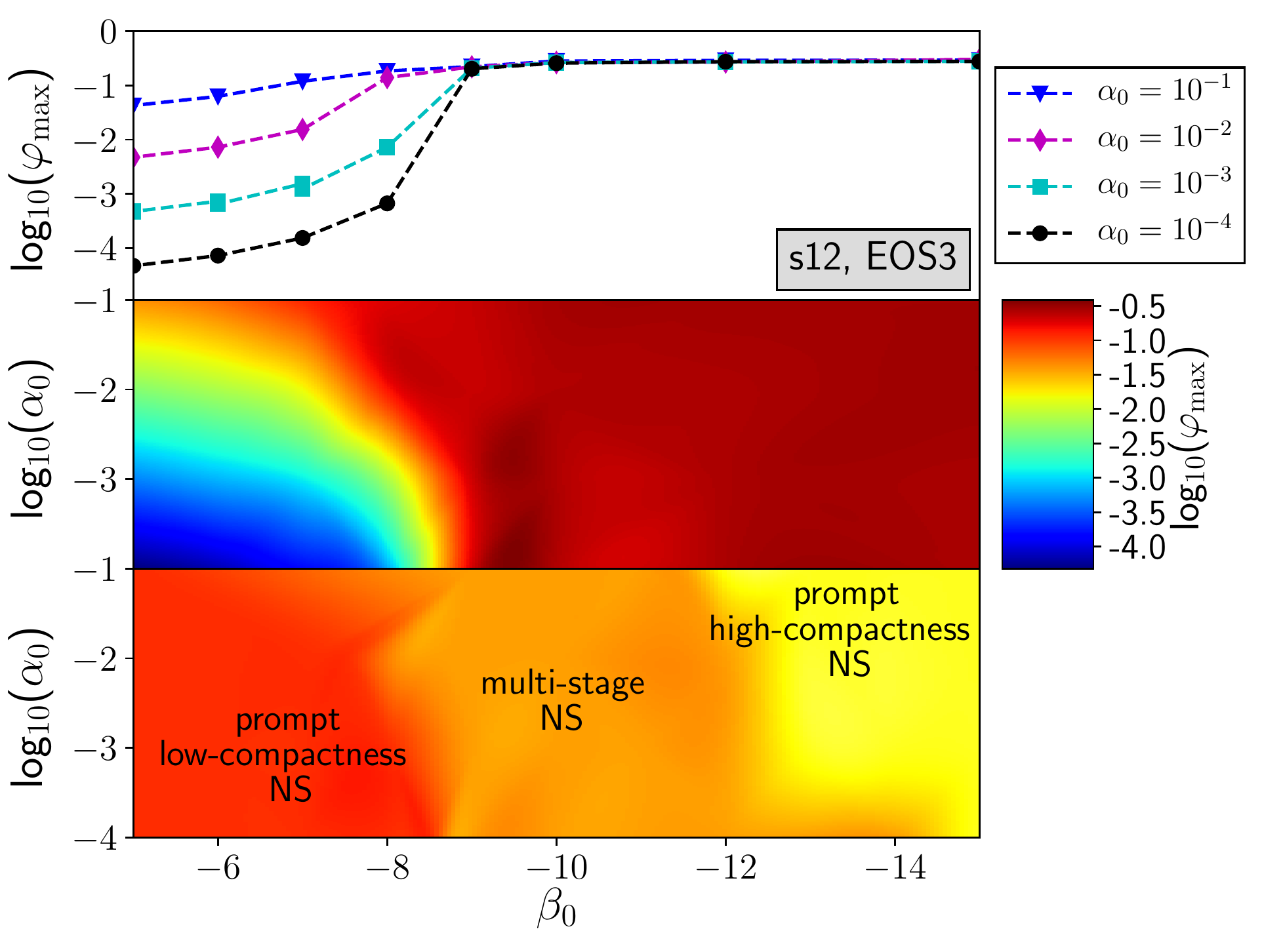}
  \includegraphics[width=0.49\textwidth,clip=true]{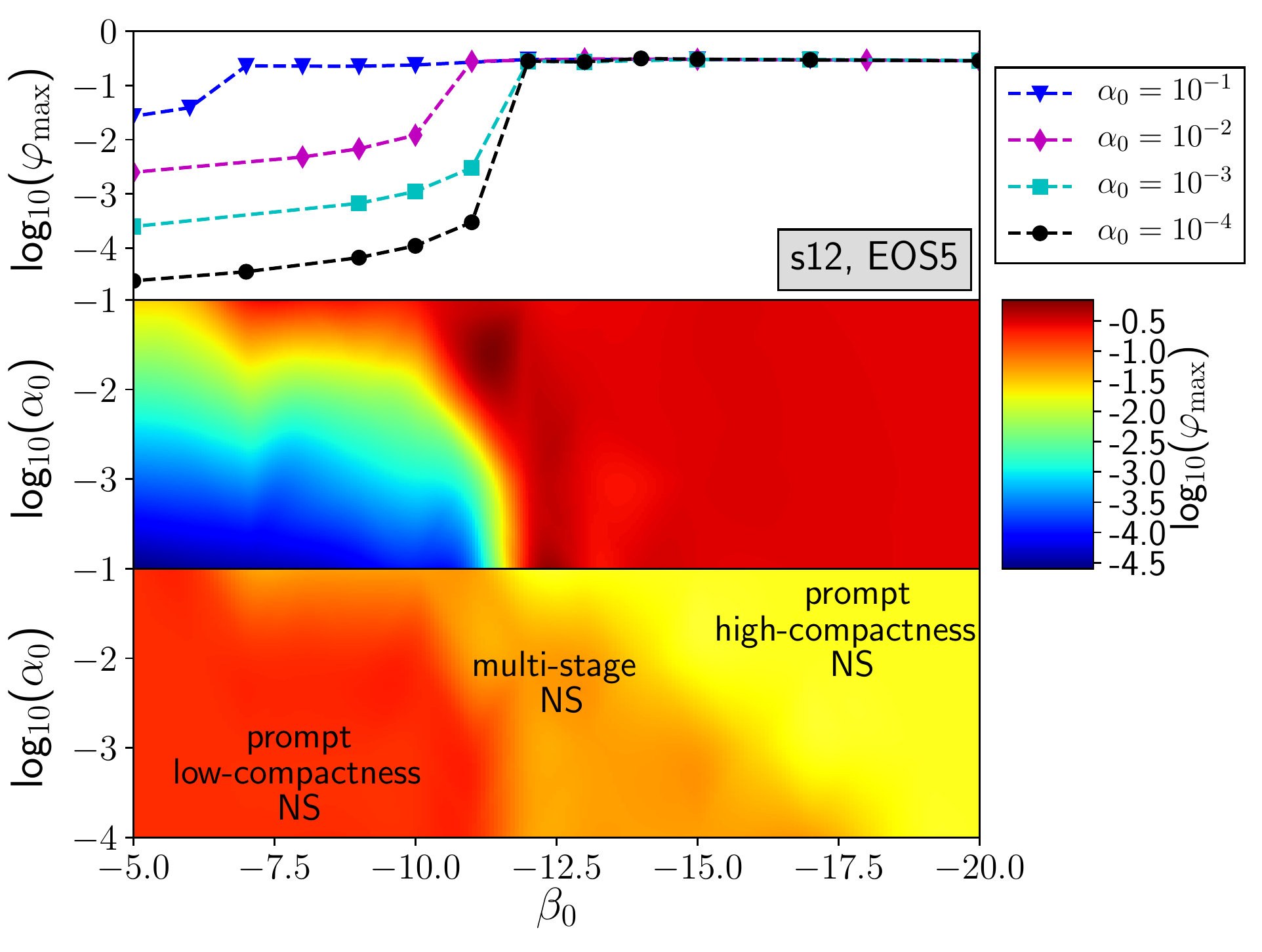}\\[10pt]
  \includegraphics[width=0.49\textwidth,clip=true]{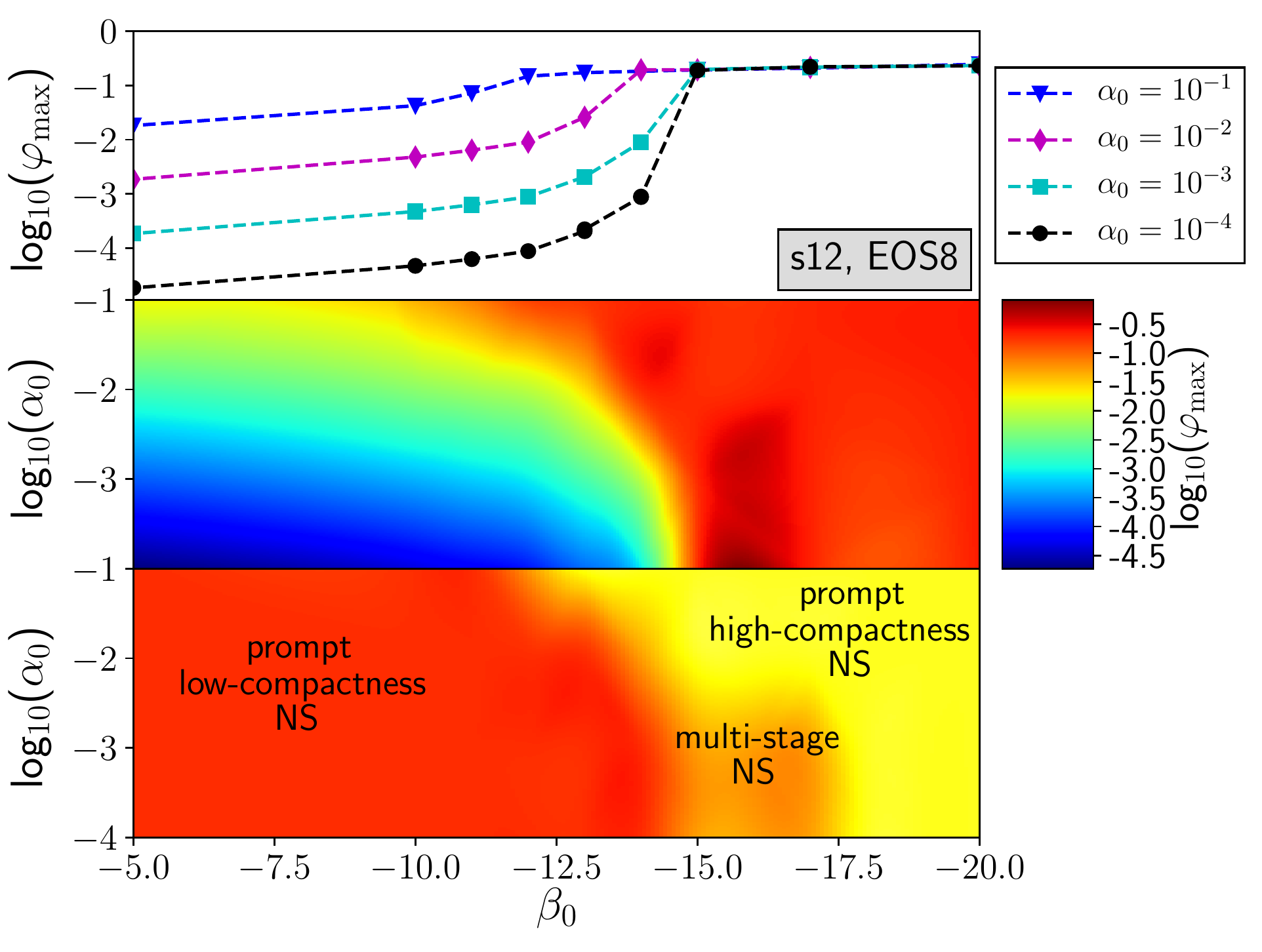}
  \includegraphics[width=0.49\textwidth,clip=true]{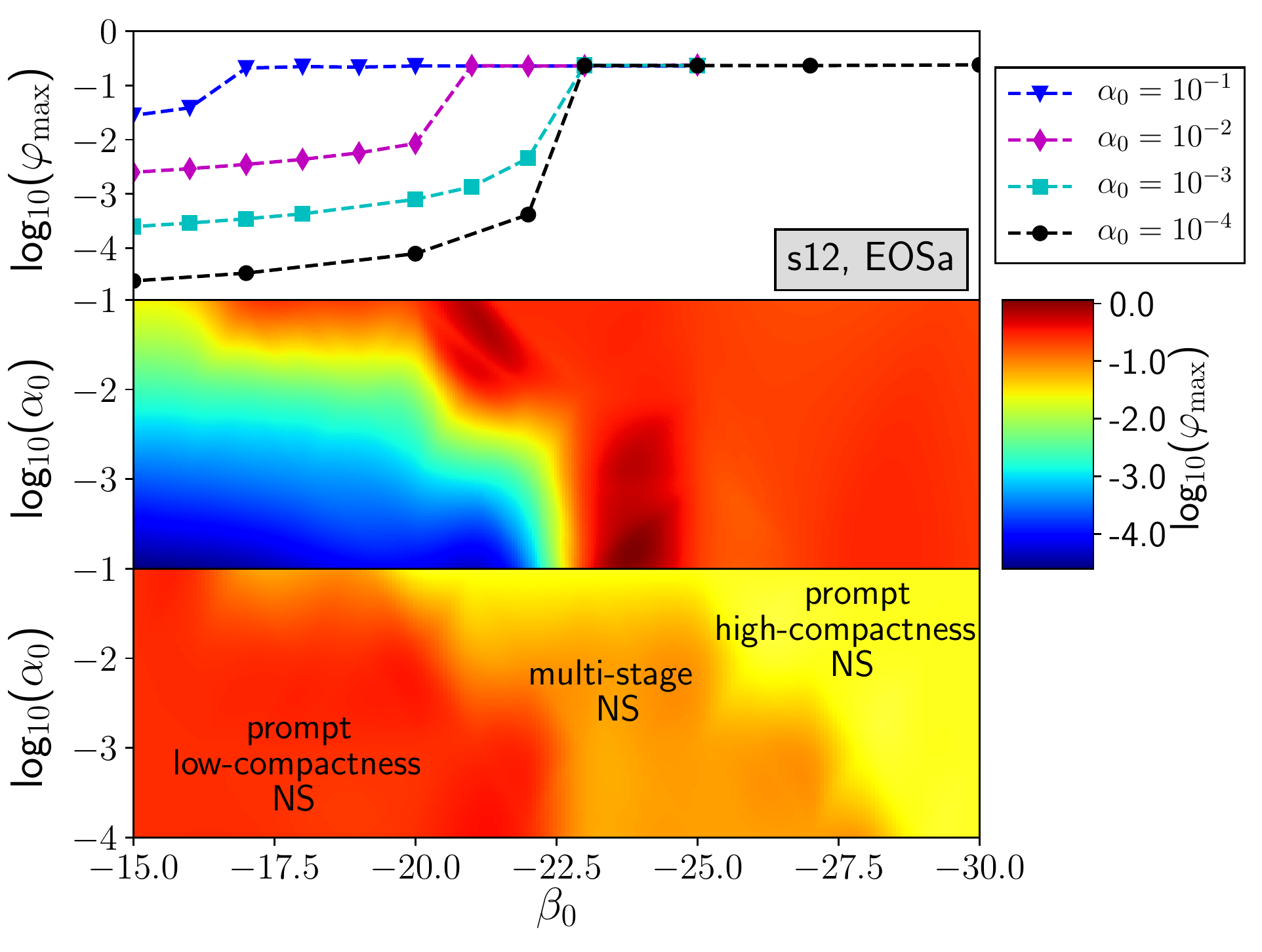}
    \caption{Similarly to Fig.~\ref{fig:heatmaps}, for each panel we consider a fixed progenitor star with ZAMS mass $12\,M_{\odot}$, solar metallicity, and several equations of state of Table \ref{tab:EOS}.
           Top rows: For selected values of $\alpha_0$, we plot the
           maximal scalarization of the star as a function of $\beta_0$.
           The middle rows provide a color (or ``heat'') map of the same
           quantity in the $(\alpha_0,\beta_0)$ plane: ``Red'' = strong
           scalarization, and ``Blue'' = weak scalarization. The bottom
           rows present a color code of the five qualitatively different
           collapse scenarios listed in Sec.~\ref{sec:classification}. Note that all progenitor models displayed here
           result in heat maps in the $(\alpha_0,\beta_0)$ plane qualitatively
           equal to that on the left side of Fig.~\ref{fig:heatmaps} (the neutron star case).}
            \label{fig:appendix_heatmaps_NS}
\end{figure*}

\begin{figure*}
  \includegraphics[width=0.49\textwidth,clip=true]{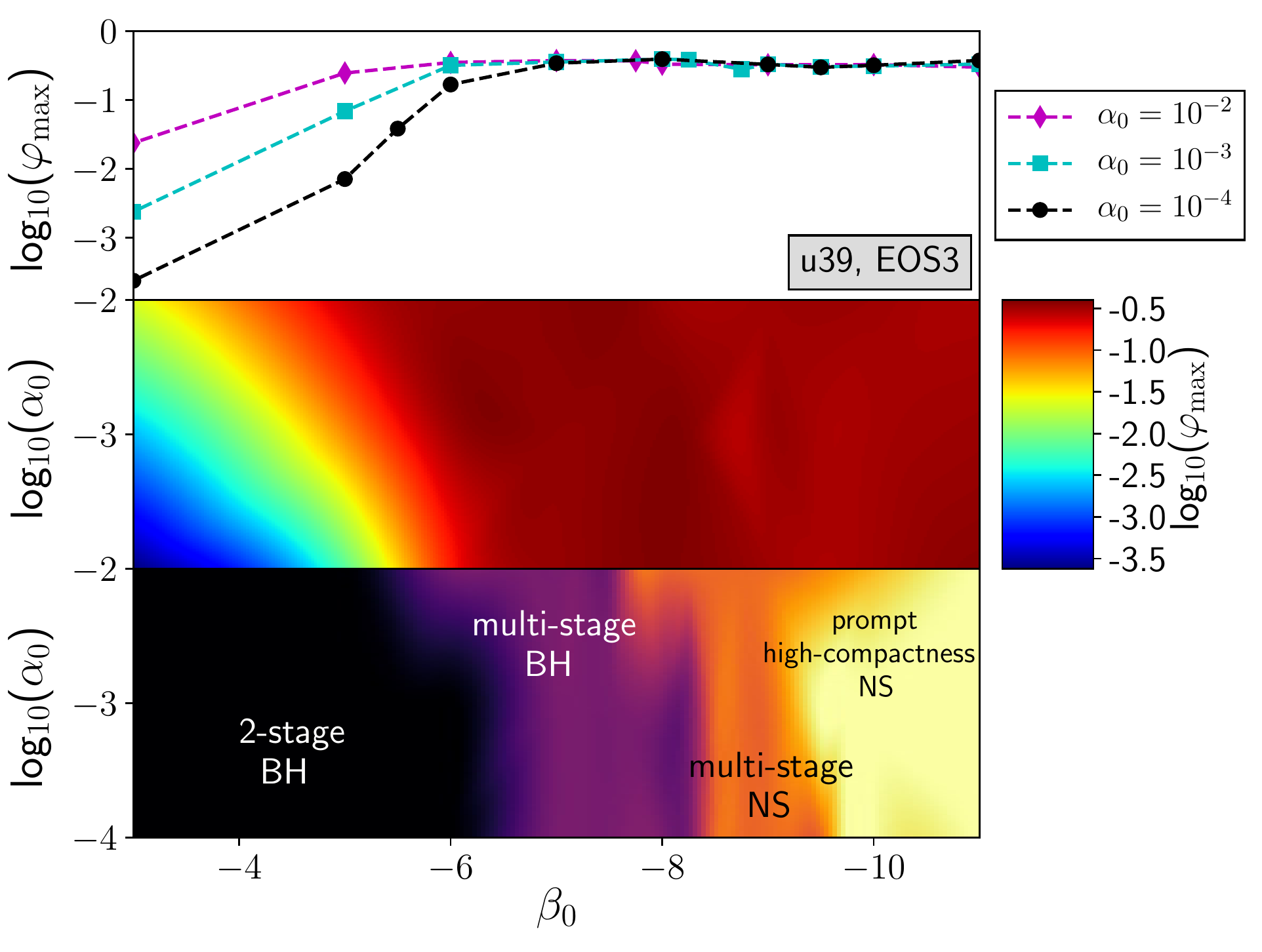}
  \includegraphics[width=0.49\textwidth,clip=true]{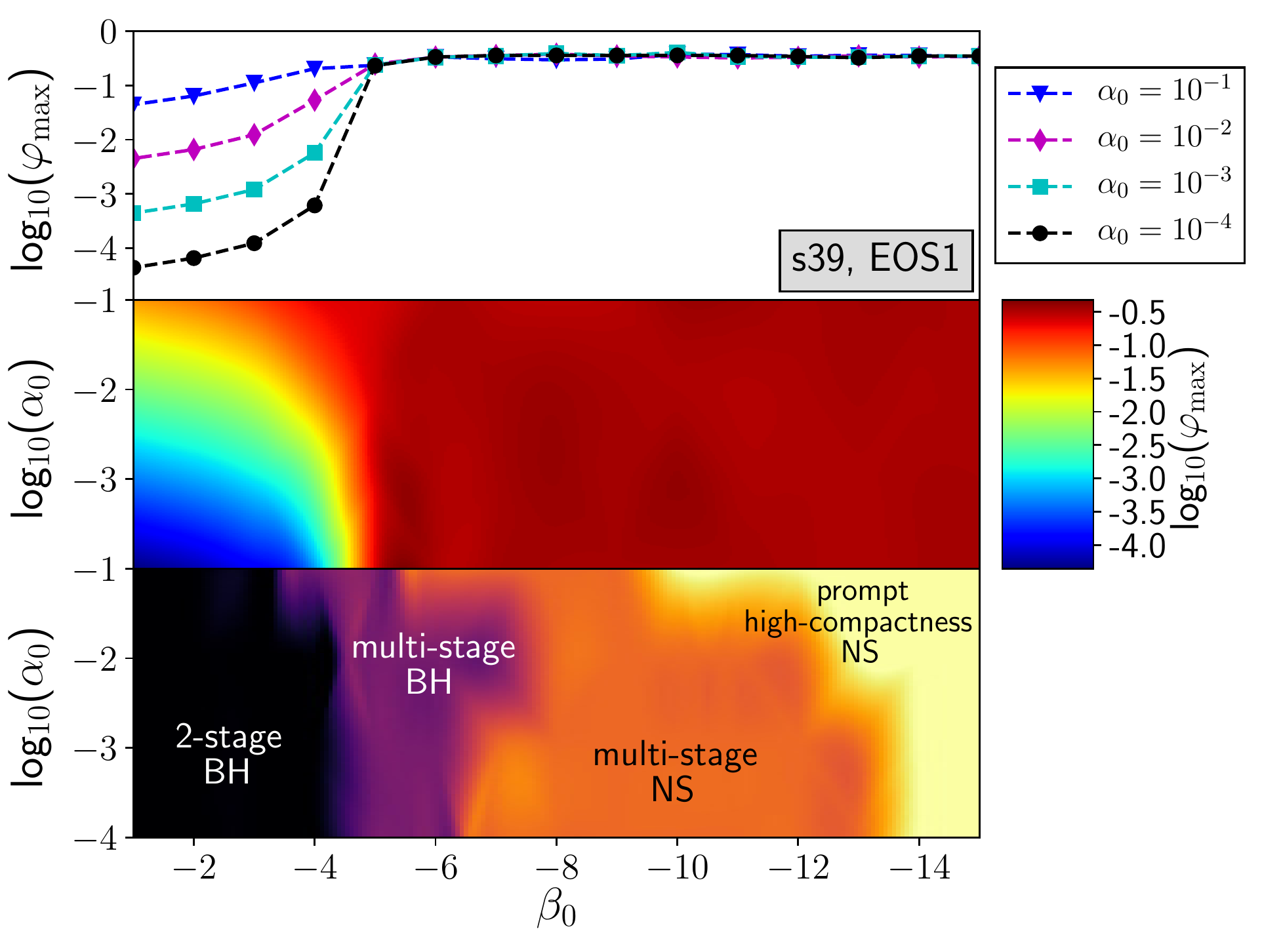}
  \includegraphics[width=0.49\textwidth,clip=true]{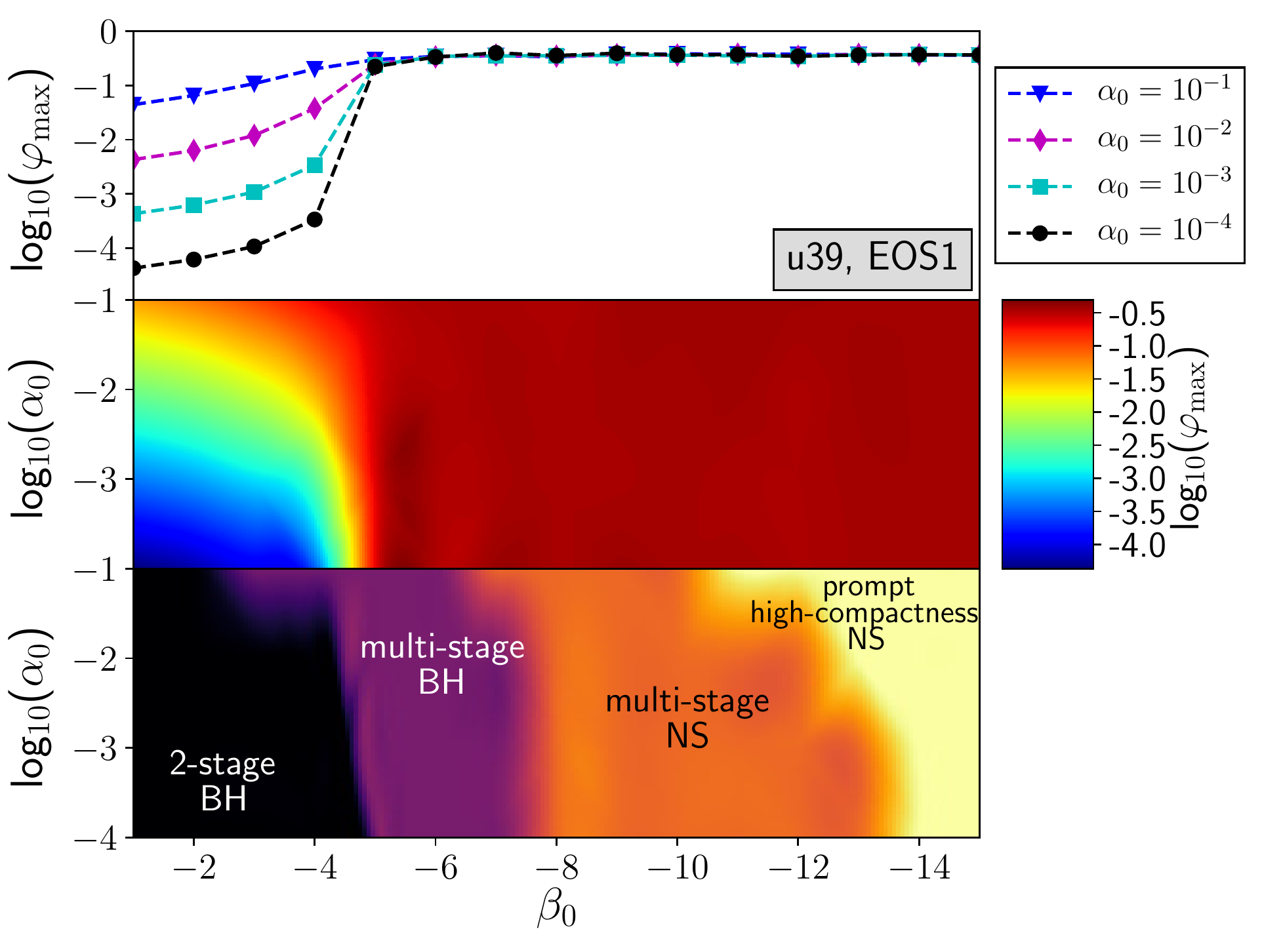}
  \includegraphics[width=0.49\textwidth,clip=true]{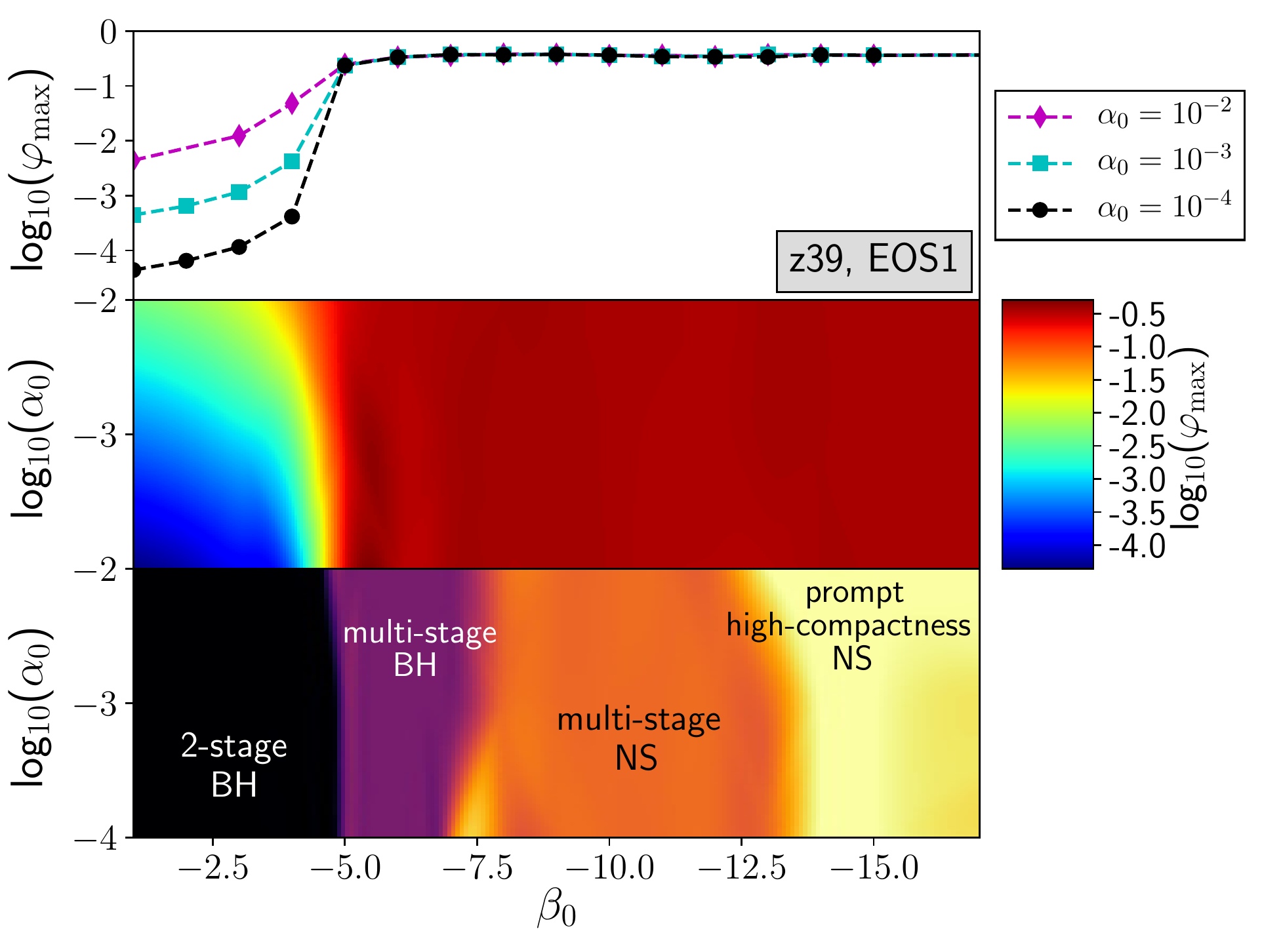}
  \caption{Similarly to Fig.~\ref{fig:heatmaps}, for each panel we consider a fixed progenitor star with ZAMS mass $39\,M_{\odot}$,all three metallicities and equations of state EOS1 and EOS3 of Table \ref{tab:EOS}.
           Top rows: For selected values of $\alpha_0$, we plot the
           maximal scalarization of the star as a function of $\beta_0$.
           The middle rows provides a color (or ``heat'') map of the same
           quantity in the $(\alpha_0,\beta_0)$ plane: ``Red'' = strong
           scalarization, and ``Blue'' = weak scalarization. The bottom
           rows present a color code of the five qualitatively different
           collapse scenarios listed in Sec.~\ref{sec:classification}. Note that all progenitor models displayed here
           result in heat maps in the $(\alpha_0,\beta_0)$ plane qualitatively
           equal to that on the right side of Fig.~\ref{fig:heatmaps} (the black hole case).}
           \label{fig:appendix_heatmaps_BH}
\end{figure*}

Figures \ref{fig:appendix_heatmaps_NS} and \ref{fig:appendix_heatmaps_BH} show additional results to accompany the discussion in Sec.~\ref{sec:classification} obtained with different stellar progenitors and equations of state. The main conclusion is that every progenitor model
           results in heat maps in the $(\alpha_0,\beta_0)$ plane qualitatively
           equal to that of the neutron star case (left panel of Fig.~\ref{fig:heatmaps}) or that of the black hole case (right panel of Fig.~\ref{fig:heatmaps}).

\begin{figure}
\centering
\includegraphics[width=0.42\textwidth]{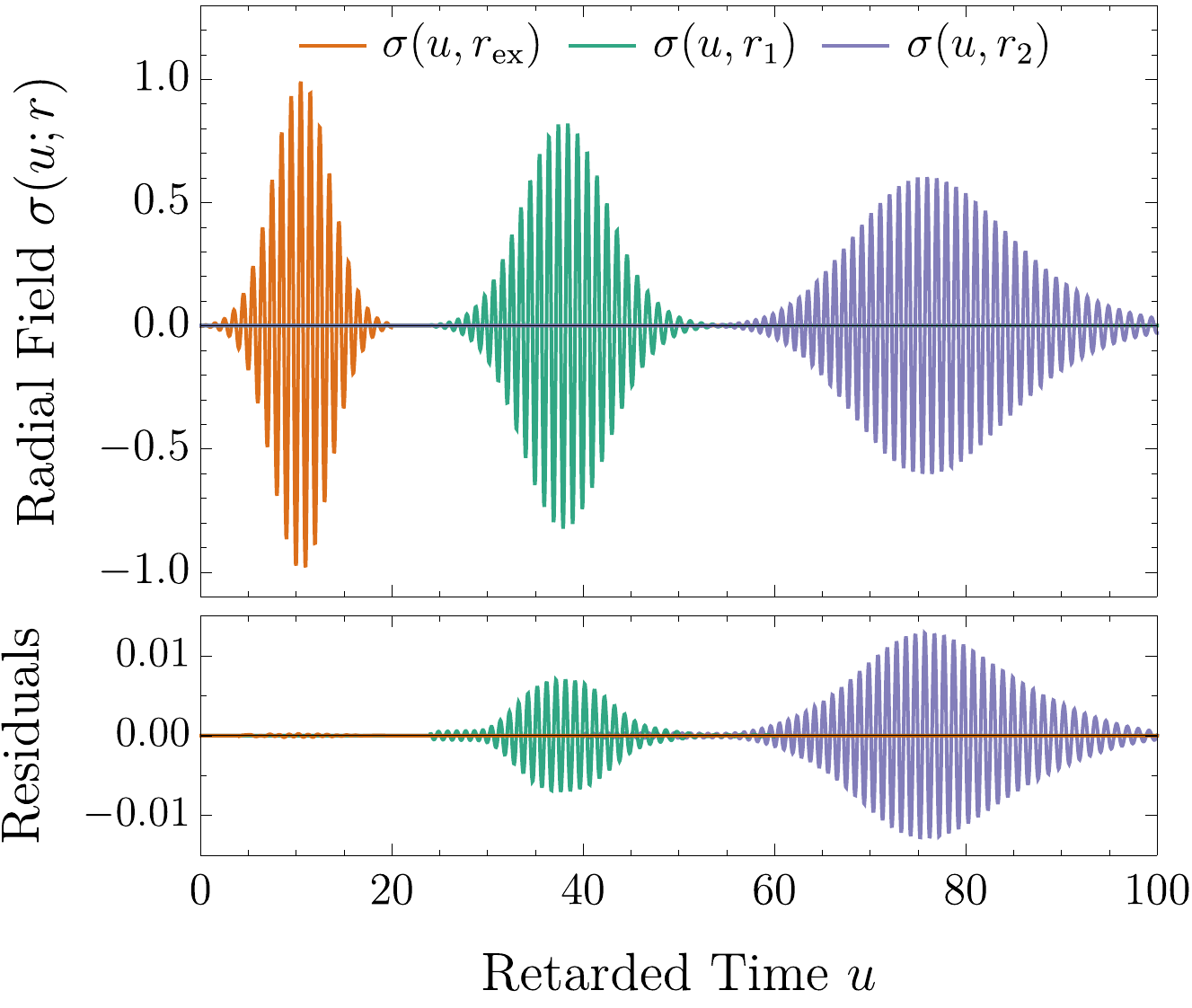}
\caption{ \label{fig:TestWave}
        The top panel shows the results of evolving an initially
        sin-Gaussian waveform out to radii $r_{1}=r_{\rm ex}+500cT$
        and $r_{2}=r_{\rm ex}+1200cT$ using the time domain numerical
        evolution of the wave equation (see Sec.~\ref{sec:1+1}).
        The evolution to large radii was also performed using the
        analytic Fourier domain approach (see Sec.~\ref{sec:FTprop}),
        and the bottom panel shows the differences, or residuals,
        between the two methods.
        }
\bigskip
        \centering
        \includegraphics[width=0.49\textwidth]{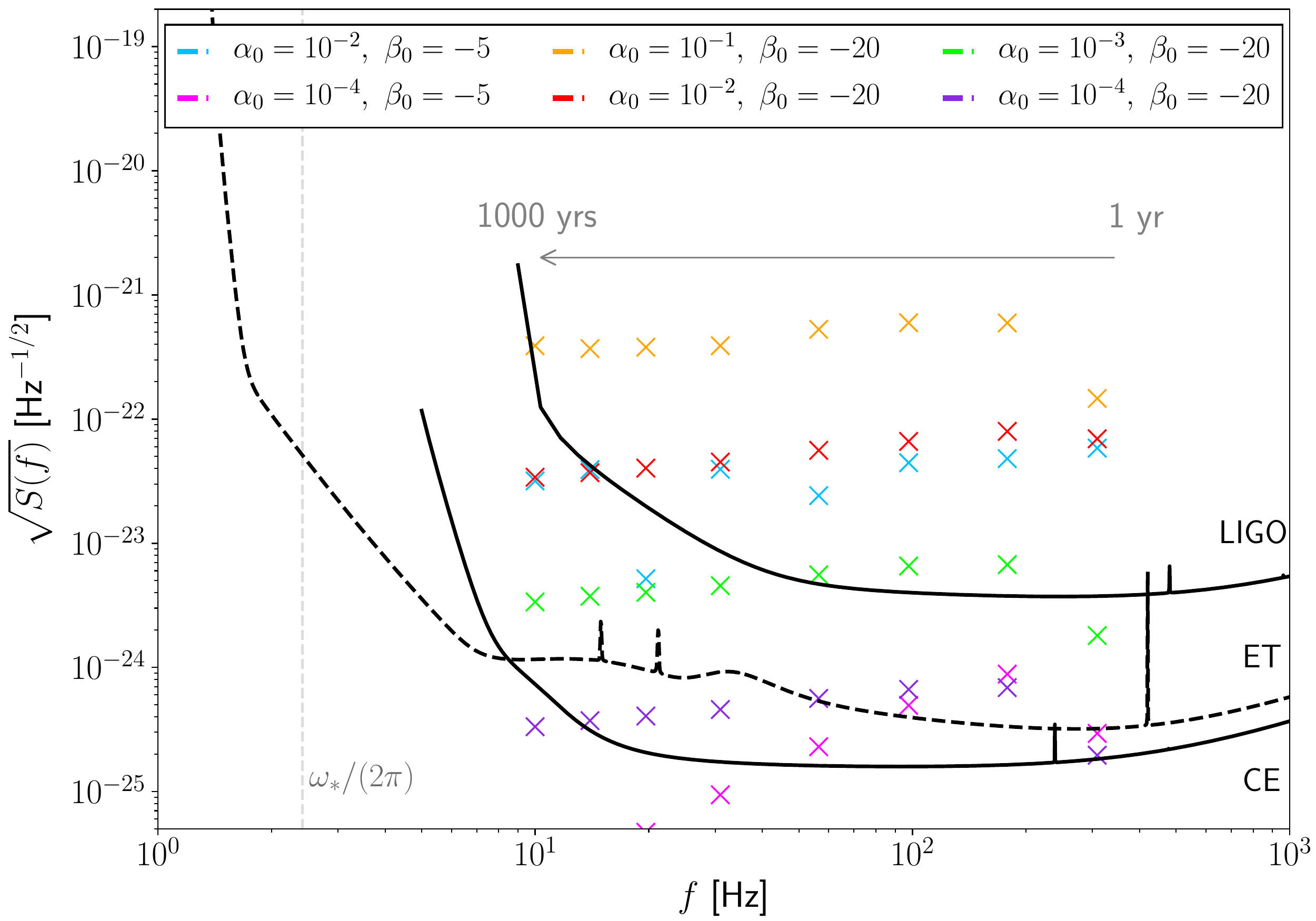}\\
        \medskip
        \includegraphics[width=0.49\textwidth]{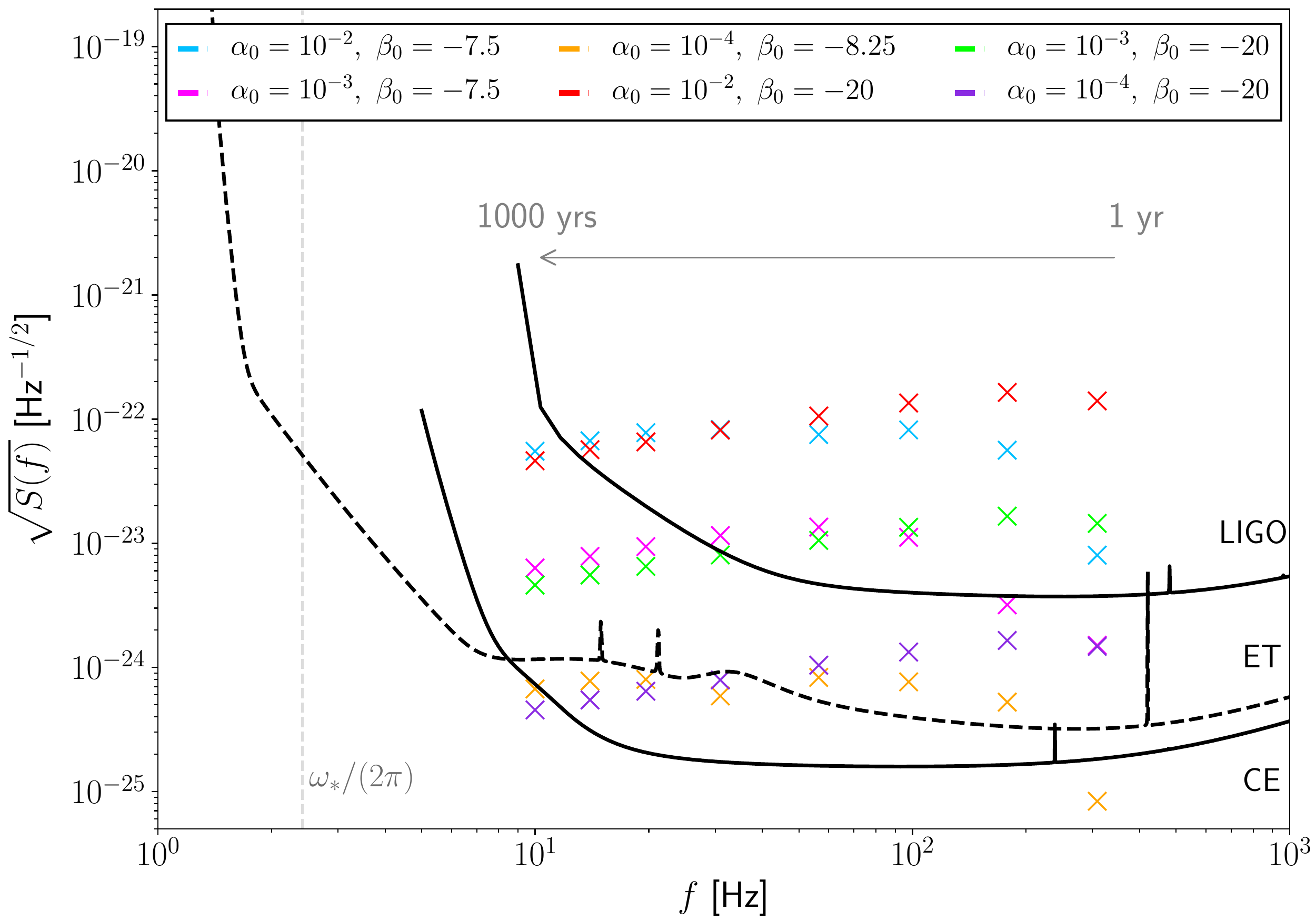}
        \caption{
        Similar to Fig.~\ref{fig:SNR_s13}, but for stellar collapse of the $u39$ progenitor model.
        \label{fig:SNR_u13}
        }
\end{figure}

%===============================================================
\section{Comparing the methods} \label{app:comparing_the_methods}

This appendix contains a test and comparison of the two methods described in Secs.~\ref{sec:1+1} and \ref{sec:analytic_propagation} for propagating signals from the extraction sphere to larger radii.
For this test, consider a simple signal which, on the extraction sphere,
is a cosine-Gaussian wave packet:
\begin{equation}
    \sigma(u;r_{\rm ex})\!=\!\cos\!\left(\frac{ 2\pi (u\!-\!r_{\rm ex})}{T}\right)\exp\!\left(\frac{-(u\!-\!r_{\rm ex})^{2}}{18T^{2}}\right).
\end{equation}
The parameter $T$ is an overall timescale which is set to unity
without loss of generality and the scalar field mass was chosen to
be $\omega_{*}=2/T$.  The signal was propagated to larger radii
using both of the methods described in Secs.~\ref{sec:1+1} and \ref{sec:analytic_propagation}, and the results are
summarized in Fig.~\ref{fig:TestWave}.

As can be seen from Fig.~\ref{fig:TestWave}, there is excellent
qualitative agreement between the two methods.  At the quantitative
level there are small errors (generally $\lesssim 1\%$, as can be
seen from the lower panel) which are due to numerical errors in the
$1+1$ time domain evolution (this has been checked by verifying the
scaling of the errors with grid resolution).  As the signals propagate
to larger radii, the peak lags at later retarded times due to the
subluminal wave propagation.  Additionally, the variation in the
group velocity between the different Fourier components of the
wave packet leads to a broadening of the peak; careful inspection
of the $\sigma(u;r_{2})$ profile reveals the beginnings of an
\emph{inverse chirp} profile (see Sec.~\ref{sec:asymp}) where the
high frequencies arrive first, followed by the low frequencies.

%=============================================================================
\section{Additional SNR results }\label{app:App_SNR_estimates_LIGO}

\begin{figure}[b]
        \centering
        \includegraphics[width=0.49\textwidth]{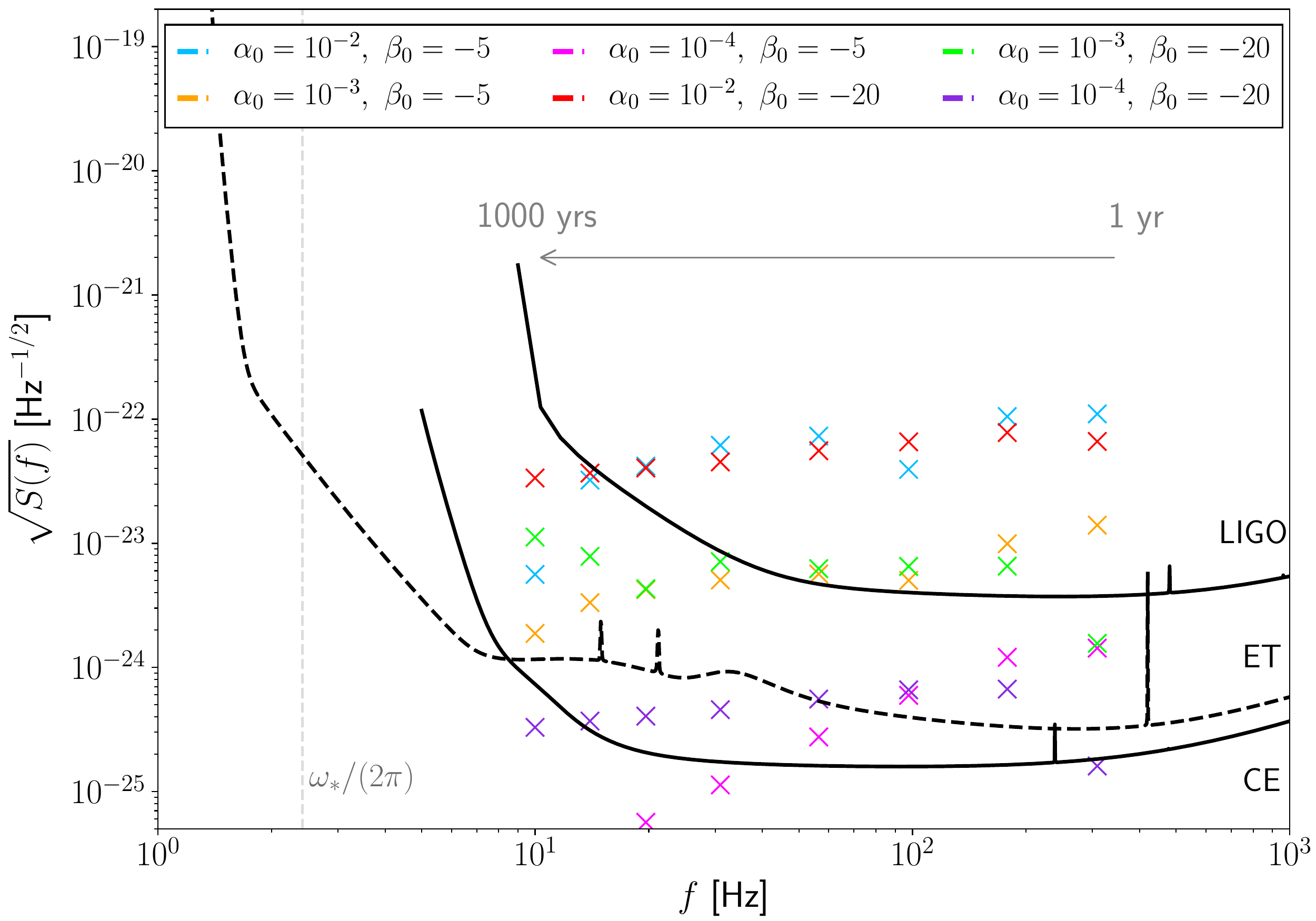}\\
        \medskip
        \includegraphics[width=0.49\textwidth]{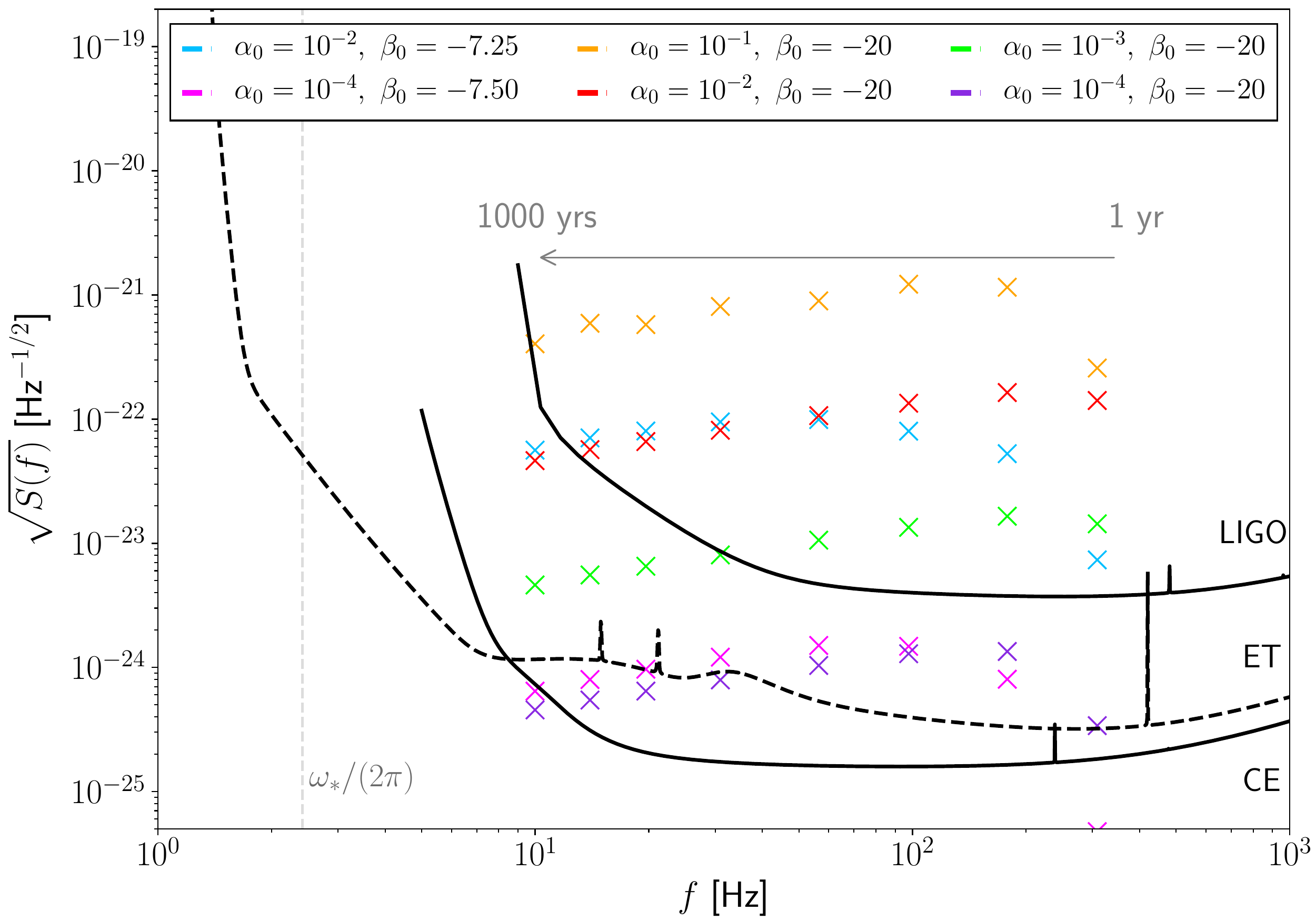}
        \caption{
        Similar to Fig.~\ref{fig:SNR_s13}, but for stellar collapse of the $z39$ progenitor model.
        \label{fig:SNR_z13}
        }
\end{figure}

Figures \ref{fig:SNR_u13} and \ref{fig:SNR_z13} show additional results to accompany the discussion in Sec.~\ref{sec:LIGO_obs} obtained with different stellar progenitors. The main conclusion is that the properties of the progenitor have only a mild effect on the SNR.

\vspace{-1cm} $\;$

\end{document}